\documentclass[preprint2, twocolumn, times, tighten]{aastex631}

\newcommand{\Paa}{\ensuremath{\mathrm{Pa\alpha}}}
\newcommand{\Ha}{\ensuremath{\mathrm{H\alpha}}}

\newcommand{\SFRPaa}{\ensuremath{\mathrm{SFR}_{\mathrm{Pa\alpha}}}}
\newcommand{\APaa}{A\ensuremath{_{\mathrm{Pa\alpha}}}}

\newcommand{\Msun}{\ensuremath{{M_{\odot}}}}
\newcommand{\Lsun}{\ensuremath{{L_{\odot}}}}
\newcommand{\Zsun}{\ensuremath{{\mathrm{Z}_{\odot}}}}
\newcommand{\logM}{\ensuremath{{M_{\star}/M_{\odot}}}}
\newcommand{\LIR}{\ensuremath{L_{\rm IR}}}

\providecommand{\e}[1]{\ensuremath{\times 10^{#1}}}

\providecommand{\metals}[1]{\ensuremath{12+\log{(\mathrm{O/H})}}}

\usepackage{amsmath}
\usepackage{CJK}

\begin{document}

\title{Calibrating Photometric Mid-Infrared Star Formation Rates for JWST}

\author[0000-0002-8909-8782]{Stacey Alberts}
\affiliation{AURA for the European Space Agency (ESA), Space Telescope Science Institute, 3700 San Martin Dr., Baltimore, MD 21218, USA}
\affiliation{Steward Observatory, University of Arizona,
933 North Cherry Avenue, Tucson, AZ 85719, USA}
\email{salberts@stsci.edu}

\author[0000-0003-2303-6519]{George H. Rieke}
\affiliation{Steward Observatory, University of Arizona, 933 North Cherry Avenue, Tucson, AZ 85719, USA}

\author[0000-0003-4702-7561]{Irene Shivaei} 
\affiliation{Centro de Astrobiolog\'ia (CAB), CSIC-INTA, Ctra. de Ajalvir km 4, Torrej\'on de Ardoz, E-28850, Madrid, Spain}

\author[0000-0001-7673-2257]{Zhiyuan Ji}
\affiliation{Steward Observatory, University of Arizona, 933 North Cherry Avenue, Tucson, AZ 85719, USA}
% \email{zhiyuanji@arizona.edu}

\author{Pascal Oesch}
\affiliation{Department of Astronomy, University of Geneva, Chemin Pegasi 51, 1290 Versoix, Switzerland}
\affiliation{Niels Bohr Institute, University of Copenhagen, Lyngbyvej 2, DK2100 Copenhagen, Denmark}
\affiliation{Cosmic Dawn Center (DAWN), Copenhagen, Denmark}

\author{Gabriel Brammer}
\affiliation{Niels Bohr Institute, University of Copenhagen, Lyngbyvej 2, DK2100 Copenhagen, Denmark}
\affiliation{Cosmic Dawn Center (DAWN), Copenhagen, Denmark}

\author[0000-0003-4337-6211]{Jakob M.\ Helton}
\affiliation{Department of Astronomy \& Astrophysics, The Pennsylvania State University, University Park, PA 16802, USA}
%\email{jakobhelton@psu.edu}

\author[0000-0002-6221-1829]{Jianwei Lyu (\begin{CJK}{UTF8}{gbsn}吕建伟\end{CJK})}
\affiliation{Steward Observatory, University of Arizona, 933 North Cherry Avenue, Tucson, AZ 85719, USA}

\author{Erica J. Nelson}
\affiliation{Department for Astrophysical and Planetary Science, University of Colorado, Boulder, CO 80309, USA}

\author[0000-0001-9687-4973]{Naveen Reddy}
\affiliation{Department of Physics and Astronomy, University of California, Riverside, 900 University Avenue, Riverside, CA 92521, USA}

\author[0000-0002-5104-8245]{Pierluigi Rinaldi}
\affiliation{Space Telescope Science Institute, 3700 San Martin Drive, Baltimore, Maryland 21218, USA}
% \email{prinaldi@stsci.edu}

\author[0000-0001-6561-9443]{Yang Sun}
\affiliation{Steward Observatory, University of Arizona, 933 North Cherry Avenue, Tucson, AZ 85719, USA}

\author[0000-0001-7160-3632]{Katherine E. Whitaker}
\affiliation{Department of Astronomy, University of Massachusetts, Amherst, MA 01003, USA}
\affiliation{Cosmic Dawn Center (DAWN), Denmark}

\author[0000-0003-2919-7495]{Christina C.\ Williams}
\affiliation{NSF National Optical-Infrared Astronomy Research Laboratory, 950 North Cherry Avenue, Tucson, AZ 85719, USA}
% \email{christina.williams@noirlab.edu}

\author[0000-0001-9262-9997]{Christopher N.\ A.\ Willmer}
\affiliation{Steward Observatory, University of Arizona, 933 North Cherry Avenue, Tucson, AZ 85719, USA}
% \email{cnaw@as.arizona.edu}

\author{Stijn Wuyts}
\affiliation{Department of Physics, University of Bath, Claverton Down, Bath, BA2 7AY, UK}

%\collaboration{20}{(AAS Journals Data Editors)}

%% Note that the \and command from previous versions of AASTeX is now
%% depreciated in this version as it is no longer necessary. AASTeX 
%% automatically takes care of all commas and "and"s between authors names.

%% AASTeX 6.31 has the new \collaboration and \nocollaboration commands to
%% provide the collaboration status of a group of authors. These commands 
%% can be used either before or after the list of corresponding authors. The
%% argument for \collaboration is the collaboration identifier. Authors are
%% encouraged to surround collaboration identifiers with ()s. The 
%% \nocollaboration command takes no argument and exists to indicate that
%% the nearby authors are not part of surrounding collaborations.

%% Mark off the abstract in the ``abstract'' environment. 
\begin{abstract}

The mid-infrared (IR) spectrum of galaxies has a long history as a valuable proxy for the dust-obscured star formation rate (SFR) in massive galaxies.  Now, with JWST, we can explore the mid-IR's full potential as a SFR tracer over four orders~of~magnitude in total infrared luminosity ($9\lesssim\mathrm{log}~L_{\rm IR}/\Lsun\lesssim13$). 
First, combining the SMILES and FRESCO surveys, we evaluate MIRI photometry against 
the \Paa~emission line $-$ a gold standard SFR indicator $-$ in Main Sequence (MS) galaxies at cosmic noon.  We find the rest-frame $6-8\,\mu$m luminosity has a steeply superlinear relation with SFR$_{\Paa}$ below $\sim8\,\Msun$ yr$^{-1}$, in contrast with the unity slope seen in coeval massive galaxies. We derive broken power-law SFR indicators from single-band MIRI photometry plus a representative dust template, with a scatter typical of IR SFRs ($\sim0.2-0.3$ dex). Despite the break in the mid-IR behavior and our simplifying assumption of a single dust SED, we next successfully formulate a UV+IR composite relation (scatter $\sim0.15$ dex) under the usual assumption of energy balance. This implies that the rest-frame $6-8\,\mu$m primarily tracks the global dust-obscuration fraction $-$ which decreases rapidly at log $\logM\lesssim10$ $-$ rather than reflecting a deficit in PAH abundances at low mass. Our results thus support MIRI photometry as a robust SFR proxy at log $\logM\gtrsim9$ up to $z\sim3$.  Finally, extending to local and $z\gtrsim1$ ultraluminous infrared galaxies not represented in SMILES, we examine when \Paa\ and the IR reliably track SFR in the bright regime.

\end{abstract}

%% Keywords should appear after the \end{abstract} command. 
%% The AAS Journals now uses Unified Astronomy Thesaurus concepts:
%% https://astrothesaurus.org
%% You will be asked to selected these concepts during the submission process
%% but this old "keyword" functionality is maintained in case authors want
%% to include these concepts in their preprints.

%% From the front matter, we move on to the body of the paper.
%% Sections are demarcated by \section and \subsection, respectively.
%% Observe the use of the LaTeX \label
%% command after the \subsection to give a symbolic KEY to the
%% subsection for cross-referencing in a \ref command.
%% You can use LaTeX's \ref and \label commands to keep track of
%% cross-references to sections, equations, tables, and figures.
%% That way, if you change the order of any elements, LaTeX will
%% automatically renumber them.
%%
%% We recommend that authors also use the natbib \citep
%% and \citet commands to identify citations.  The citations are
%% tied to the reference list via symbolic KEYs. The KEY corresponds
%% to the KEY in the \bibitem in the reference list below. 

\section{Introduction} \label{sec:intro}

Our understanding of galaxy evolution hinges on accurately quantifying the rate of formation of new stars in galaxies across cosmic time.  As such, a substantial amount of time and effort has gone into assessing observables from the X-ray to the radio as star formation rate (SFR) indicators \citep[see][for a review]{kennicutt2012}.  Direct tracers of high energy ionizing or non-ionizing continuum photons such as the Balmer lines or the ultraviolet (UV), respectively, are directly proportional to the emission from young stars.  However, their use is often compromised by absorption or scattering from cosmic dust \citep[e.g.,][]{cardelli1989, gordon1997, calzetti2000, reddy2008}.  This has spurred the use of indirect tracers such as the infrared to obtain a complete and unbiased accounting of recent star formation.

The dominant absorbing medium, cosmic dust, is an important constituent in a galaxy's ecosystem; it both influences and is influenced by the physical and chemical state of the interstellar medium (ISM).  Early work quantifying the cosmic optical and infrared backgrounds (COB and CIB, respectively) found that half of starlight is reprocessed by dust into the infrared \citep[e.g.,][]{lagache2005, dole2006}, highlighting the importance of combining direct and indirect tracers to measure the total SFR.  Though the COB and CIB are averaged across cosmic time, we now know that dust obscuration peaks along with the general SFR density of the Universe during the epoch of cosmic noon \citep[$1\lesssim z\lesssim3$; see][for a review]{madau2014}.  During this critical era, luminous infrared galaxies\footnote{Commonly termed LIRGs ($10^{11}<L_{\rm IR}/\Lsun<10^{12}$) and ultra-LIRGs or ULIRGs ($10^{12}<L_{\rm IR}/\Lsun<10^{13}$).} dominate the total SFR budget in galaxies, with the obscured star formation component accounting for $\gtrsim80\%$ of the total SFR in massive galaxies\citep[log $M_{\star}/\Msun\gtrsim10-10.5$;][]{whitaker2017, zavala2021}. 

Accounting for this obscured SFR component has historically relied on sensitive survey facilities such as Infrared Space Observatory \citep[ISO;][]{Kessler1996}, Spitzer Space Telescope \citep{Werner2004} and Herschel Space Observatory \citep{Pilbratt2010}, spanning the wide wavelength range and spectral features that encompass the mid- to far-infrared ($\sim5-1000\mu$m).  Studies with these facilities faced two major challenges.  First, the poor to moderate spatial resolution at IR wavelengths meant we were confusion limited \citep{dole2006}, limiting us to relatively shallow sensitives compared to the UV and optical.  Analyses of individual galaxies were limited to the bright end and stacking \citep{whitaker2014a, whitaker2017} was required to reach galaxies on the main sequence \citep[MS;][]{elbaz2011, popesso2023}, which averaged over galaxy-to-galaxy variations.  Pushing past the confusion limit requires larger apertures, such as we now have for the mid-IR with JWST \citep{Gardner2023, rigby2023}.

The second challenge is more fundamental.  Both the fraction of recent star formation absorbed by dust and the fraction of dust emission coming from young stars vary as a function of galaxy population, cosmic time, and wavelength observed.  The relationship between the IR and SFR cannot be easily extrapolated from massive to dwarf galaxies, nor from local to high redshift.  For example, in heavily obscured galaxies, the ratio of SFRs derived from the infrared to those derived from hydrogen recombination lines like \Paa\ at 1.876$\,\mu$m increases with increasing infrared luminosity \citep{alonso-herrero2006, calzetti2007a}.

Observing dust in a wide range of populations and redshifts to disentangle these complexities can now be advanced substantially by JWST.  In particular, the Mid-Infrared Instrument \citep[MIRI;][]{rieke2015, wright2023} provides sensitive access to the mid-infrared (observed $\sim5-26\,\mu$m) up to $z\sim3$ through both imaging and spectroscopic modes. In star forming galaxies (SFGs), the mid-IR is dominated by broad emission features from polycyclic aromatic hydrocarbons \citep[PAHs;][]{smith2007, tielens2008, li2020, lai2020}. The role of PAHs in the ISM is not yet fully known, but it may be substantial: PAHs regulate the chemistry and ionization balance in the ISM \citep[e.g.,][]{tielens1985, bakes1994, helou2001} and are tightly correlated with molecular gas \citep[e.g.,][]{pope2013, whitcomb2022, leroy2022}, which may indicate they act as catalysts for the formation of H$_2$ molecules \citep[e.g.,][]{bauschlicher1998, foley2018, barrera2023}.

PAHs are also tightly correlated with star formation on scales from H{\sc II} regions to entire galaxies \citep{calzetti2007a}.  As bright features, PAHs in aggregate typically contribute $\sim10\%$ of the total infrared emission \citep[$L_{\rm IR}\equiv L[8-1000\mu$m][]{wu2010, shipley2016} and up to $\sim20\%$ in extreme cases \citep{smith2007}. As such, PAHs $-$ along with the rest-frame mid-IR continuum $-$ have a prolific history as a robust measure of the obscured SFR component \citep{lefloch2005,  Reddyt2006, elbaz2011, magnelli2011, wuyts2011, whitaker2012, whitaker2017, rujopakarn2013, shipley2016, cluver2017, mahajan2019, kovacs2019}.  This has been especially vital at higher redshifts, as previous instruments such as Spitzer/MIPS at $24\mu$m \citep{rieke2004} directly probed the rest-frame $8\mu$m spectral region during cosmic noon.  

The catch, however, is that these studies have largely focused on massive SFGs with roughly solar metallicities, where PAHs are ubiquitous \citep{elbaz2005, yan2005, sajina2007} and relatively well behaved.  In these galaxies, stellar energy is largely absorbed and re-radiated by dust, such that the total infrared luminosity is closely representative of the bolometric luminosity of young stars, making it an excellent proxy not only for the obscured SFR but also the total SFR \citep{kennicutt2012}.  On the other hand, in the most heavily obscured SFGs, the ratio of the SFR derived from the IR to that derived from e.g. \Paa\ is known to increase with increasing $L_{\rm IR}$.  This trend might arise because in these environments the dust competes for a fraction of the ionizing photons \citep{alonso-herrero2006}.  Or alternatively, that the dust is hotter, increasing the IR output relative to the true SFR \citep{calzetti2007a}.  This discrepancy has not yet been solved and needs to be considered when evaluating IR- (and optical-)based SFRs. 

\begin{figure*}[htb!]
    \centering
    \includegraphics[width=0.8\textwidth]{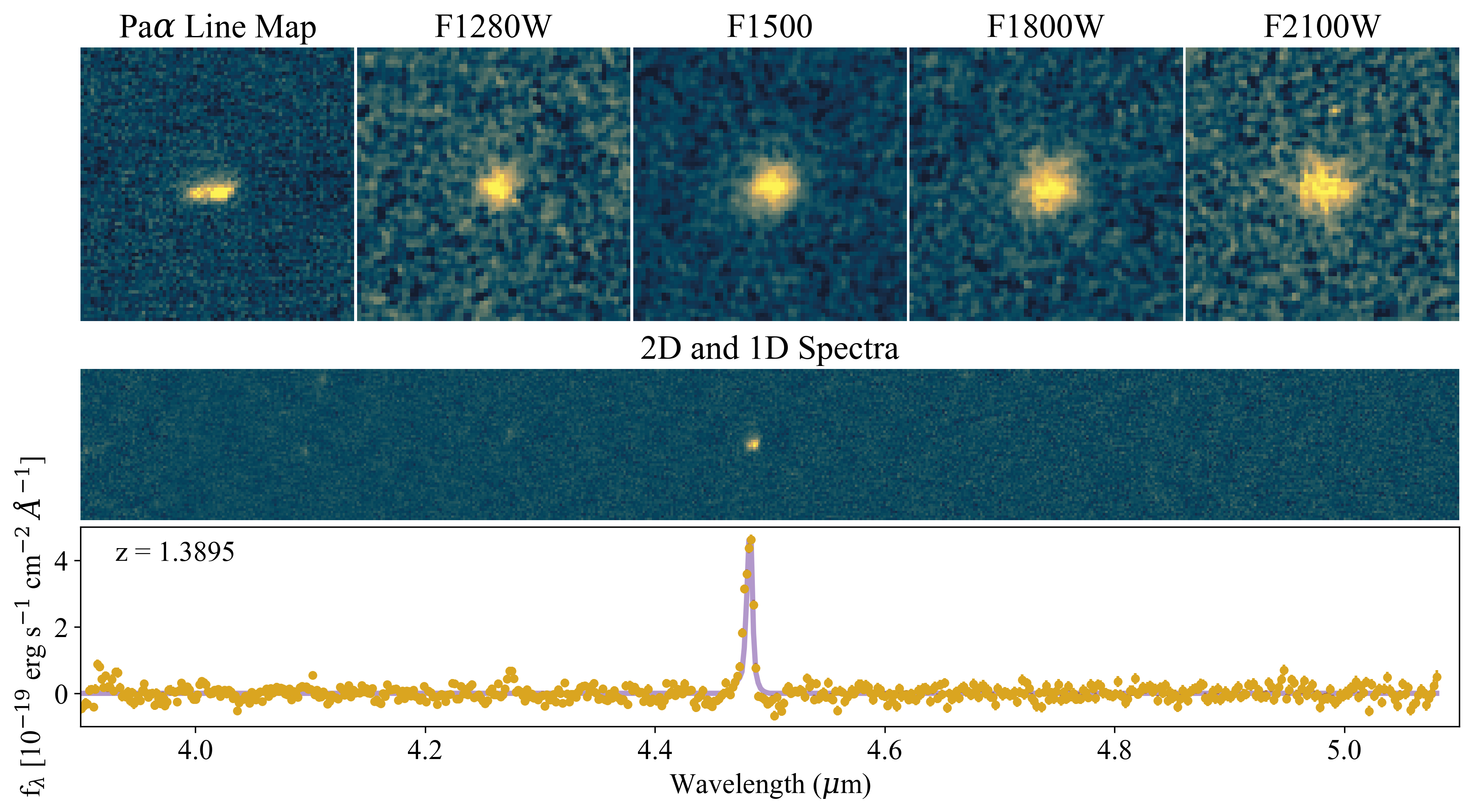}
    \caption{An example of a \Paa ~emitter at $z=1.3895$.  The top row shows image cutouts of the \Paa ~line map and the F1280W, F1500W, F1800W, and F2100W bands, which contain PAH features at this redshift. The middle and bottom rows show the NIRCam grism F444W 2D and 1D spectra, respectively.  The best fit is shown via the purple line. }
    \label{fig:paa}
\end{figure*}

With MIRI, we can now go beyond the most massive galaxies, to more fully understand the mid-IR spectrum and its utility as an indirect SFR tracer in main sequence galaxies \citep[e.g.,][]{kirkpatrick2023, ronayne2023}  during cosmic noon.  A fresh calibration is necessary: in this regime, we expect galaxies to be progressively less metal-rich \citep{maiolino2019}, with lower dust masses and evolving dust properties as conditions change in the ISM \citep[e.g.,][]{remy-ruyer2015, chastenet2025}.  PAHs in particular are known to be sensitive to metallicity and the local radiation field, with lower metallicity galaxies having lower PAH abundances both locally \citep{engelbracht2005, madden2006, draine2007, marble2010, remy-ruyer2015, aniano2020, chastenet2025} and at cosmic noon \citep{shivaei2024}. 

At the same time, the global dust properties are also changing: the obscured fraction of star formation, driven moreso by far-IR emitting, large grains in thermal equilibrium, is strongly correlated with stellar mass \citep[e.g.,][]{whitaker2017}.  This leads to the unobscured SFR component becoming substantial. Finally, IR emission is also known to arise from stellar populations with a range of ages, such that in local, low SFR regions and galaxies, up to $\sim30-80\%$ of the emission at $\sim8\,\mu$m can come from evolved stars \textit{not} associated with recent star formation.  This contribution decreases with increasing SFR and is considered negligible in SFGs at cosmic noon as the MS evolves to higher typical SFRs.  However, it is a critical point in more quiescent populations \citep{leja2019a, hayward2014, fumagalli2014}. 

Given all of these factors, the mid-infrared contains substantial information about the conditions of the ISM and recent star formation. In this work, we examine the behavior of the mid-IR spectrum as a SFR proxy, primarily the rest-frame $8\mu$m regions dominated by the luminous $7.7\mu$m PAH, which contributes up to $50\%$ of the total PAH emission and $\sim70\%$ of the emission at $\sim8\mu$m \citep{smith2007}.  We calibrate this behavior against a fiducial SFR indicator; namely, we take advantage of JWST's access to the \Paa\ emission line, a near-infrared hydrogen recombination line previously established as a ``gold standard'' SFR indicator in local studies \citep[e.g.][]{alonso-herrero2006, calzetti2007a}.  \Paa\ directly probes the ionizing photons from very young stars \citep[with timescales $\lesssim10$ Myr; e.g.,][]{leitherer1990, rigby2004} and is relatively insensitive to dust attenuation compared to the more commonly-used Balmer emission lines \citep{rieke2009, cleri2022, gimenez-arteaga2022, reddy2023, neufeld2024, calzetti2025}, which can fail to account for optical thick star formation \citep{reddy2026}.  A combination of MIRI imaging and NIRCam grism spectroscopy via the SMILES and FRESCO surveys is used to build a sample of MS SFGs, free of the selection bias of previous analyses using targeted spectroscopy.  With this, we formulate MIRI-based and composite UV+IR SFR calibrations down to MS galaxies with log $\logM\sim9$ and SFRs$\,\sim\!1$ and assess their quality and applicability over the redshift range $0.3<z<3$. 

\begin{figure*}[htb!]
    \centering
    \includegraphics[width=0.7\textwidth]{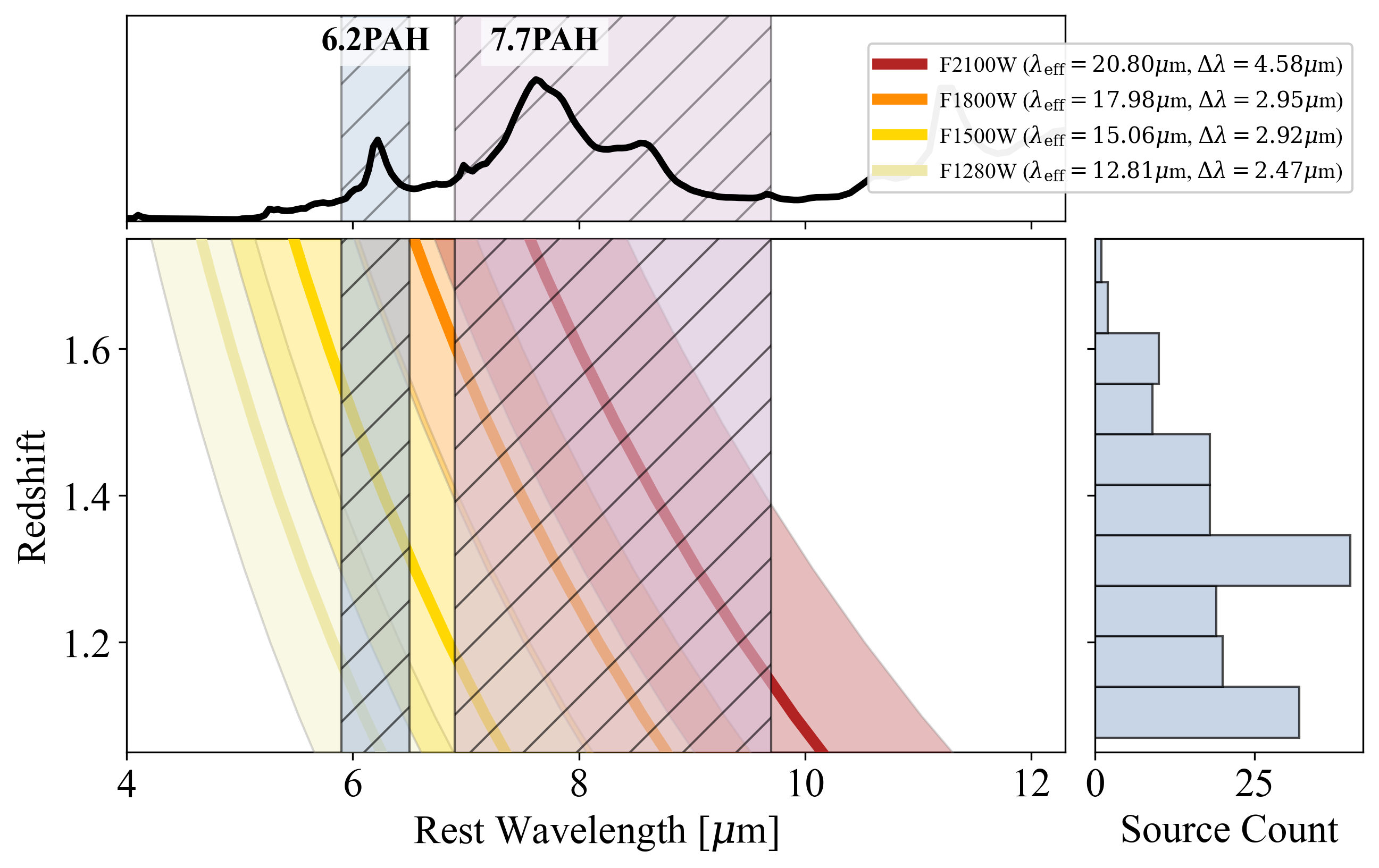}
    \caption{The rest wavelengths and bandwidths of four MIRI filters (F1280W, F1500W, F1800W, F2100W) over the redshift range $1<z<1.75$.  The F1800W and F2100W filters are dominated by the 7.7$\,\mu$m PAH emission complex (top panel).  The narrower 6.2$\,\mu$m PAH emission line (top panel) falls mostly in the F1500W, with partial coverage in the F1280W.  The width of the PAHs features \citep{draine2021} is shown via the hatched regions. (top) The mid-infrared regions of a representative log $L_{\rm IR}/\Lsun=11.25$ SFG template from \citet{rieke2009}.  (right) A histogram of the redshifts of the sources used in this study.
    }
    \label{fig:dist}
\end{figure*}

As the area of these surveys limits the number of luminous galaxies in our main calibration sample, we also examine the use of SFR indicators at the bright end using data from the literature and a small sample of local ULIRGs with JWST/NIRSpec \citep{Jakobsen2022,  Boker2023} spectroscopy \citep{perna2024, rieke2025}.  Specifically, we address two issues: 1) whether the PAH deficits at high $L_{\rm IR}$ observed in local galaxies apply to IR-based SFRs at higher redshift.  And 2) we use multiple hydrogen recombination lines detected in the local ULIRG sample to measure accurate extinction-corrected \Paa\ SFRs in very luminous sources.  We compare this to IR-based SFRs, addressing whether Pa $\alpha$ can give accurate SFR determinations in these extreme galaxies, or whether infrared approaches are preferred.

This paper is organized as follows: in Sections~\ref{sec:data} we present the datasets used in this study and in \ref{sec:sample} we describe the selection of our \Paa\ emitter sample and SED fitting.  In Section~\ref{sec:paa_sfrs}, we derive our fiducial SFRs from the \Paa\ emission line corrected for attenuation via both hydrogen recombination line ratios and SED fitting.  Our main analysis is presented in Section~\ref{sec:mir_sfrs}, which uses the fiducial \Paa\ SFRs to examine the behavior of the rest-frame $8\mu$m spectral region via MIRI photometry and calibrate SFR indicators via monochromatic flux densities, total infrared luminosities, and composite UV+IR relations.  Section~\ref{sec:disc} discusses the general robustness and applicability of our calibrations, including causes of scatter and outliers. Finally, in Section~\ref{sec:ulirgs},  we examine SFR estimation for very luminous infrared galaxies.  Our conclusions are presented in Section~\ref{sec:conclusions}.   In this work, we use the standard flat $\Lambda$CDM cosmology from \citep{planckcollaboration2020}. Any magnitudes quoted are AB mag \citep{oke1983}. Scatter is quoted using the normalized median absolute deviation (NMAD), $\sigma_{\rm NMAD} \approx 1.4826\times\,$MAD.

\section{Data}\label{sec:data}

\subsection{SMILES MIRI: Data Reduction and Photometric Extraction}\label{sec:miri_data}

The Systematic Mid-infrared Instrument Legacy Extragalactic Survey \citep[SMILES;][]{rieke2024, alberts2024a} imaged $\sim34$ arcmin$^2$ of the GOODS-S field in 8 MIRI imaging filters.  This work focuses primarily on the four mosaics at $\sim12-21\,\mu$m, F1280W, F1500W, F1800W, and F2100W, which reached $5\sigma$ point source sensitivities of 0.62, 0.75, 1.8, and 2.8 $\mu$Jy, respectively\footnote{We note that we do not use the F2550W imaging from SMILES due to its poorer sensitivity limit of $17\mu$Jy \citep[$5\sigma$;][]{alberts2024a}.}. We use aperture photometry based on Kron \citep{kron1980} apertures, with photometric uncertainties determined by placing random apertures across the source-masked image. For a full description of the data reduction and photometric catalog, see \citet{alberts2024a}.

\subsection{FRESCO: Spectral Extraction and Line Properties}

The First Reionization Epoch Spectroscopically Complete Observations survey \citep[FRESCO;][]{oesch2023} obtained NIRCam wide field slitless spectroscopy (WFSS) and associated direct imaging in GOODS-N and GOODS-S, with the latter overlapping the SMILES MIRI imaging described in Section~\ref{sec:miri_data}.  Spectra were obtained using grismR in a single dispersion direction with the F444W filter, providing continuous wavelength coverage over $\sim3.8-5\mu$m. Direct imaging was obtained in the F182M, F210M, and F444W filters to $5\sigma$ depths of 28.4, 28.2, and 28.2 mag, respectively. The FRESCO data reduction was performed using {\tt grizli} \citep{brammer2018} and will be presented in G. Brammer et al. (in prep).  Source extraction is based a stacked detection image, created from the FRESCO direct imaging. 

With 7 ks exposure times per pointing with the grism, FRESCO reaches a $5\sigma$ line sensitivity of $2\e{-18}$ ergs s$^{-1}$ cm$^{-2}$ with $R\sim1600$. As slitless spectroscopy, this provides a complete, flux-limited census of line emitters, free of the selection effects associated with slit spectroscopy. In this work, we focus on the \Paa\ line, which falls in the F444W filter at $1<z<1.75$. In Figure~\ref{fig:paa}, we show NIRCam and MIRI cutouts and the 2- and 1D spectra for an example \Paa\ emitter in the SMILES field.

\subsection{Other Data} \label{sec:other_data}

NIRCam kron photometry is adopted from the JWST Advanced Deep Extragalactic Survey \citep[JADES;][]{eisenstein2023, rieke2023} DR1 release which includes the following: 1) HST ACS photometry covering 0.4-0.85$\mu$m with 5 filters from the Hubble Legacy Field \citep{illingworth2016, whitaker2019}, NIRCam 0.9-4.4$\mu$m photometry in 6 filters from JADES and NIRCam medium band coverage (F182M, F210M, F430M, F460, F480M) from the JWST Extragalactic Medium-band Survey \citep[JEMS;][]{williams2023}.  The NIRCam data reduction and photometry catalogs are described in \citet{rieke2023}. The 5$\sigma$ point source sensitivities reached are on order 2-22 nJy (28.3-30 mag).

The GOODS-S field contains extensive spectroscopic coverage, mostly from targeted campaigns. For verification of a subset of the FRESCO grism-based redshifts, we use spectroscopic redshifts from JADES NIRSpec \citep{bunker2023, deugenio2023}, MUSE \citep{urrutia2019, bacon2023}, ASPECS \citep{walter2016}, and compilations of various sources in the literature for the CANDELS fields \citep{kodra2023}.
Finally, H$\alpha$+[N{\sc II}] emission line fluxes are obtained from the publicly available CANDELS Ly$\alpha$ Emission at Reionization (CLEAR) catalog from \citet{simons2023}, which were derived based on HST G102 and G141 grism spectroscopy.  
At the HST grism resolution ($R\sim100-200$) H$\alpha$ is blended with [N{\sc II}] and so we extract the H$\alpha$ line fluxes from the blended catalog fluxes by adopting the [N{\sc II}]/H$\alpha$ line ratio - stellar mass relation at $z\sim2$ from \citet{sanders2021}. The line ratio ranges from $0.06-0.23$ over log $\logM\sim9-10.6$ and we deblend the lines as follows:

\begin{equation}
    f_{\mathrm{H}\alpha} = f_{\mathrm{H}\alpha+[\mathrm{NII]}} \times \frac{1}{1+\mathrm{[NII]}/\mathrm{H}\alpha}
\end{equation}

We note the ratios have been corrected for the contribution from diffuse ionized gas to the line emission \citep{sanders2017, zhang2017, sanders2021} and that we don't account for any dependence of the ratio on SFR \citep[as motivated by e.g.,][]{wuyts2016}.    

\section{Sample and SED fitting}\label{sec:sample}

\subsection{Sample Selection}

To build our sample, we start with the 271 \Paa\ emitters with a $>4\sigma$ line detection that are within the SMILES footprint.  We match this catalog to the SMILES MIRI and JADES NIRCam detection catalogs with a search radius of $0.3\arcsec$, finding 194 matches.  
We then perform an inspection and further remove 15 matches based on their NIRCam grism spectra (i.e. visual inspection reveals the \Paa\ line is not at the redshift of existing spec-$z$s or we identify double-peaked lines that may not yield a robust \Paa\ flux) and an additional 10 which are blended or near the mosaic edge in MIRI. This leaves a robust sample of 169 \Paa\ emitters with MIRI+NIRCam photometric counterparts.  The redshift distribution of this parent sample, spanning $z\sim1-1.7$ as set by the F444W grism filter, is shown in Figure~\ref{fig:dist} compared with the central (rest) wavelength of the F1280W, F1500W, F1800W, and F2100W filters over this range.  The width of each filter is indicated by the shaded bands. The hatched regions show the width of the 6.2$\mu$m and 7.7$\mu$m PAH complexes, as defined in \citet{draine2021}.  The $6.2\mu$m PAH enters the F1280W, F1500W, F1800W and F2100W filters at $z\sim[0.8, 1.1, 1.55, 1.85]$ and exits at $z\sim[1.38, 1.8, 2.3, 2.9]$, respectively.  For the $7.7\mu$m PAH, the $z$ bounds are $z\sim[0.2, 0.4, 0.7, 0.9]$ and $z\sim[1, 1.37, 1.79, 2.3]$.   

Given the strength and width of the 7.7$\mu$m PAH feature, our focus will be on the F1800W and F2100W fluxes of our sample.  The SMILES catalog provides forced photometry in all bands 10$\mu$m and longer, regardless of SNR, based on a F560W+F770W detection image (see Section~\ref{sec:data}).  From our parent sample of 169, 128 ($79\%$) have a $\geq3\sigma$ detection in either F1800W or F2100W; 48 ($28\%$) have $1-3\sigma$ photometric measurements in F1800W and/or F2100W; and 15 ($8\%$) are completely undetected with $<1\sigma$ forced photometry. The marginally detected and undetected sub-samples will be incorporated using stacking.

In addition, we visually examine the 77 \Paa\ sources with no MIRI counterpart, finding that 34/77 are blended with a neighbor and the remaining 44 are low mass (median log $\logM\sim8.5$), below the approximate mass completeness limit estimated in \citet{shivaei2024}.
These \Paa\ emitters with no MIRI counterpart are combined with the 15 emitters with $<1\sigma$ in F1800W and F2100W and will be incorporated into the main analysis by stacking their MIRI cutouts.

Finally, to identify AGN, we match these to the catalogs from \citet{lyu2022a} and \citet{lyu2024}; 18 have evidence for AGN from the X-ray and/or mid-infrared.  These sources will be excluded from the main analysis but are discussed in Section~\ref{sec:agn}. 

\subsection{SED Fitting}\label{sec:sed_fitting}

SED fitting is done with full HST, NIRCam, and MIRI photometry using the Bayesian fitting code {\tt Prospector} \citep{leja2019b, johnson2021} and the Flexible Stellar Population Synthesis (FSPS) code \citep{conroy2009, conroy2010}.  We direct the reader to \citet{ji2023} for a detailed description of our {\tt Prospector} setup, which we summarize here.  The star formation history (SFH) assumed is non-parametric with the continuity prior \citep{leja2019b} and is composed of 9 lookback time bins, where the SFR is constant in each bin\footnote{As shown in \citet{leja2019b} using simulated galaxy catalogs, the best-fit properties are insensitive to the number of bins used above five bins.}.  The first two time bins are fixed to be $0-30$ and $30-100$ Myr to capture recent star formation activity with a relatively high time resolution. SED fitting with a nonparametric SFH using {\tt Prospector} produces high quality reconstructions of the SEDs and SFHs of synthetic galaxies in cosmological simulations, with low systematic biases \citep{leja2019b, johnson2021, tacchella2022, ji2023}.

Nebular continuum and line emission modeling is based on \citet{byler2017}.  For dust attenuation, we assume a two-component model \citep{tacchella2022}
where the attenuation of nebular emission and young populations is treated differently from that of old stellar populations \citep{charlot2000}.  Stellar populations older than 10 Myr are parameterized using a modified Calzetti Law \citep[via a variable slope;][]{noll2009} with the UV dust bump at 2175 \AA\ tied to the slope of the attenuation following \citet{kriek2013}.  The attenuation of the young population is assumed to be an inverse function of wavelength and tied to the attenuation of the old population via a clipped normal prior on their ratio, centered at 1, with a width of 0.3 and a range of $0-2$ \citep{ji2023}.
%such that their ratio is assumed to have a . 
We allow the stellar and gas-phase metallicity as well as the ionization parameter $U$ to vary.  We additionally allow an AGN component and check that the addition of this component does not significantly alter our fits. From our fits, we obtain stellar masses, SED-fitted SFRs, mass-weighted ages, and a dust attenuation models.

\begin{figure}[t!]
    \centering
    \includegraphics[width=0.45\textwidth]{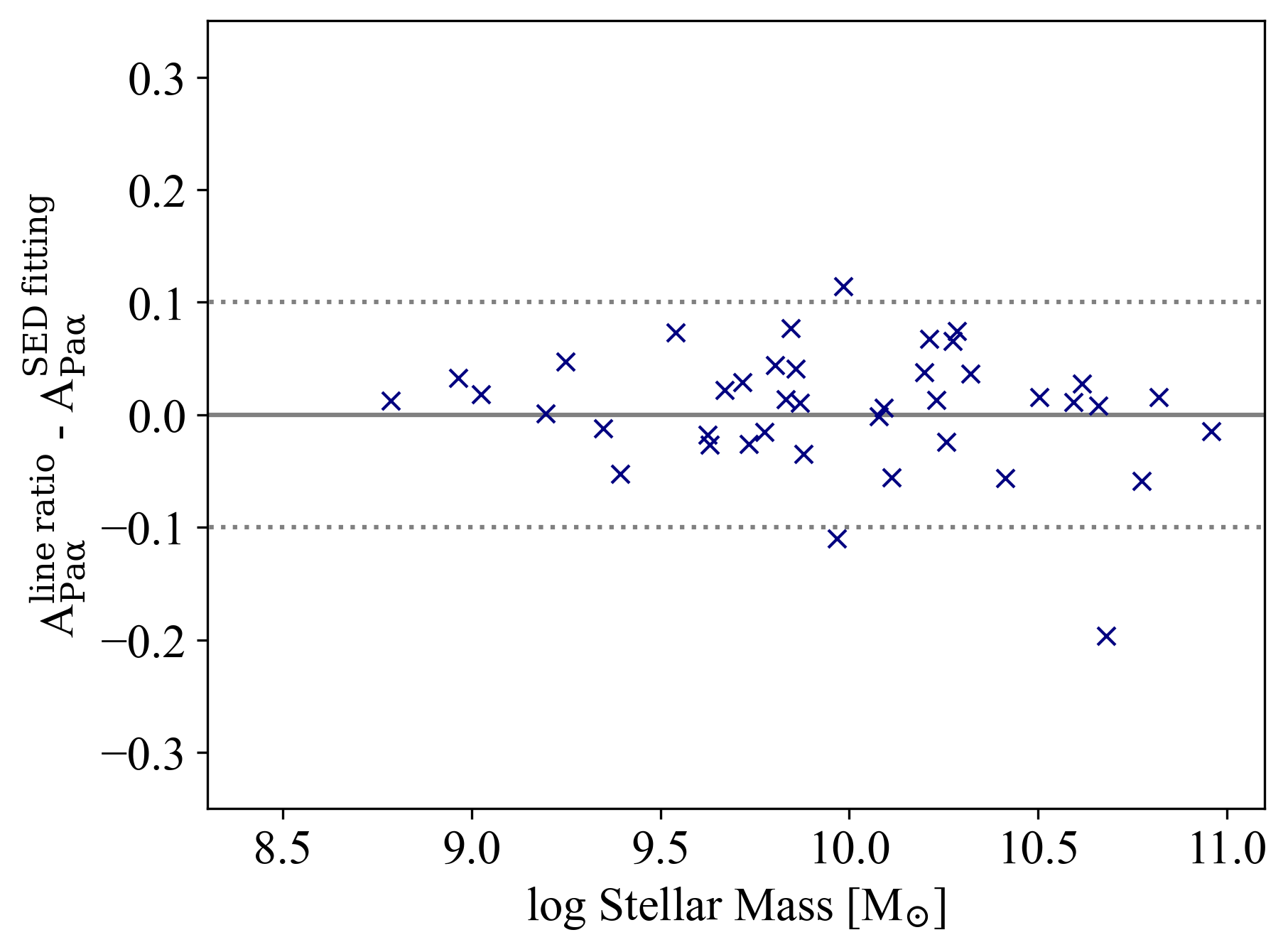}
    \caption{The difference in \APaa\ derived from the \Paa/\Ha\ line ratio and from SED fitting as a function of stellar mass.  The two measurements are in good agreement within 0.1 mag (dotted lines), corresponding to a difference in the final \Paa\ flux of $\lesssim10\%$.}
    \label{fig:apaa_comparison}
\end{figure}

\section{Fiducial Star Formation Rates from Pa$\alpha$}\label{sec:paa_sfrs}

For our goal of calibrating SFRs using MIRI imaging, we obtain fiducial SFRs from the Pa$\alpha$ emission line.  Compared to the commonly used stronger Balmer lines, the Paschen lines emit in the near-infrared and therefore are less susceptible to dust extinction; in extreme examples of dusty regions in local SFGs with $A_{\rm V}\sim10-30$, $A_{\rm Pa\alpha}$ is found to be $<2$ mag and optically thin \citep{murphy2001, dannerbauer2005, armus2007}.  With JWST, we are just now starting to expand the use of \Paa\ to higher redshifts; recent works with NIRSpec and NIRCam found higher SFRs implied by \Paa\ as compared to \Ha\ and spatial offsets in their emission that may indicate optically thick star formation is being missed by \Ha\ in MS galaxies at cosmic noon \citep{reddy2023a, reddy2025, lorenz2025}.  

To determine the attenuation of \Paa\ in our sample and obtain total SFRs, we first derive \APaa\ using the ratio of \Paa\ to \Ha\ for sources where \Ha\ is available from CLEAR (Section~\ref{sec:other_data}). 
As intrinsic hydrogen recombination line ratios can be predicted given reasonable assumptions on electron density and temperature
\citep{osterbrock1989}, they are often used as a direct measure of nebular line attenuation. From the CLEAR catalog, we find 56 matches to our sample with \Ha\ +[N{\sc II}] fluxes detected at greater than $5\sigma$ and no evidence for AGN activity.  We remove the [N{\sc II}] contribution as described in Section~\ref{sec:other_data}. The nebular extinction is then

\begin{equation}
    \mathrm{E(B-V)}_{\mathrm{neb}} = \frac{2.5}{\kappa_{\mathrm{H}\alpha} - \kappa_{\mathrm{Pa}\alpha}} \times \frac{\mathrm{log}(f_{\mathrm{Pa}\alpha}/f_{\mathrm{H}\alpha})}{0.109}
\end{equation}

\noindent
where 0.109 is the intrinsic (dust-free) line ratio under Case B recombination \citep{hummer1987, reddy2023}. We adopt the Galactic extinction curve from \citet{cardelli1989} with $R_V=3.1$, which has been shown to well approximate the nebular attenuation curve in the optical at $z\sim2$ \citep{reddy2015, reddy2020}.
We find 13 measurements of \APaa\ that are negative, likely indicating uncertainty in removing [NII] from \Ha\ . Of the remainder, the median \APaa\ from the \Paa/\Ha\, line ratio is $0.06$ mag with a scatter of $\sigma_{\rm NMAD}=0.06$. 

\begin{figure}[b!]
    \centering
    \includegraphics[width=0.45\textwidth]{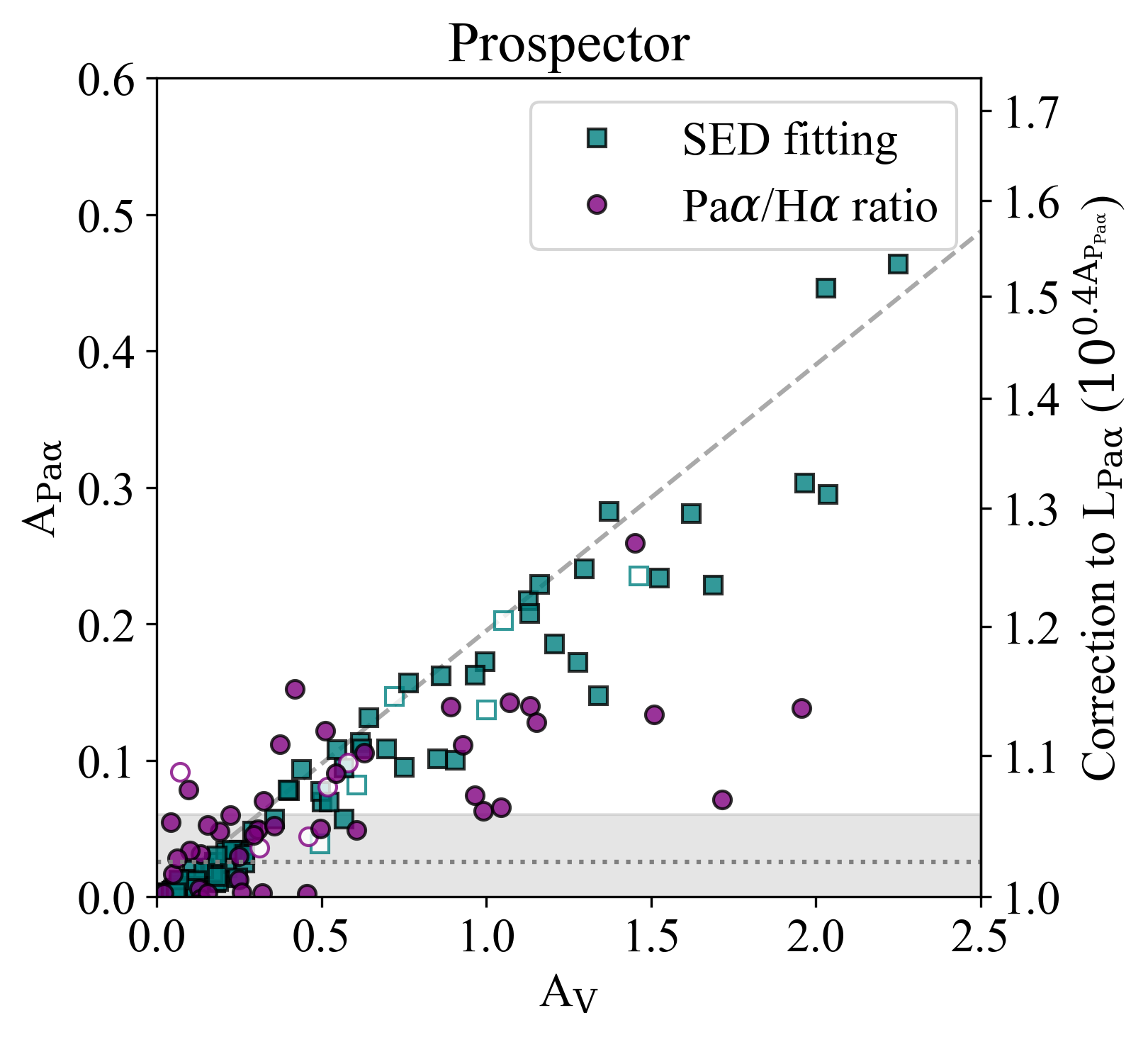}
    \caption{A$_{\rm Pa\alpha}$ and the correction factor 10$^{0.4 A_{\rm Pa\alpha}}$ (right axis) as a function of A$_{V}$ for values derived from SED fitting (blue squares) and from the \Paa/\Ha line ratios (purple circles). Open symbols denote AGN. The dotted line and shaded region show the median value 0.03 with a scatter of 0.04 dex.  The dashed line shows the relation A$_{V}$/A$_{\rm Pa\alpha}=6$ \citep{calzetti2007a}.}
    \label{fig:apaa}
\end{figure}

\begin{figure*}[th!]
    \centering
    \includegraphics[width=0.7\textwidth]{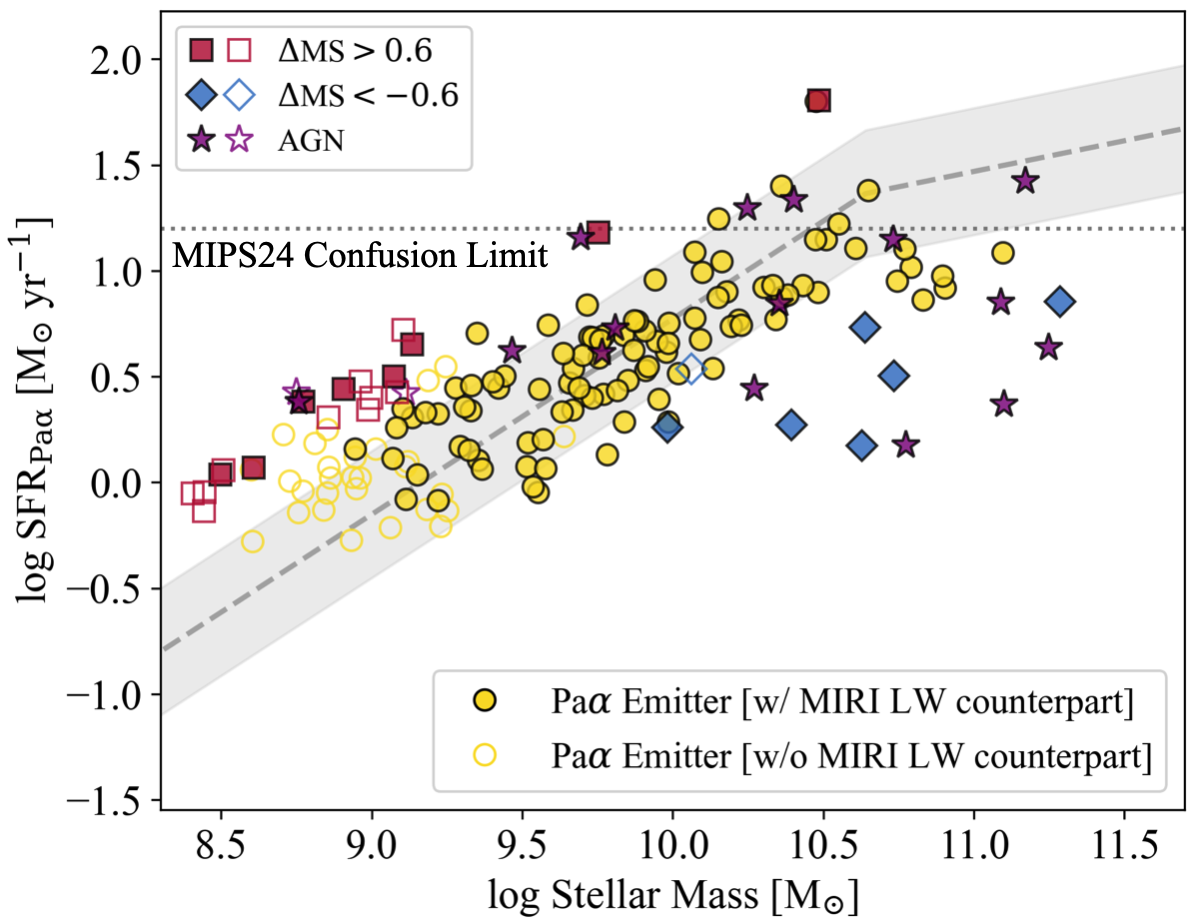}
    \caption{SFR versus stellar mass for the \Paa\ emitters in our sample.  Solid (open) circles show \Paa\ emitters on the MS detected (undetected) in MIRI long wavelength filters (F1800W or F2100W). The dashed line and shaded region shows the MS at $z\sim1.3$, the median redshift in our sample. \Paa\ emitters above the MS ($\Delta$MS$\,>0.6$), below the MS ($\Delta$MS$\,<-0.6$), and hosting AGN are shown as red squares, blue diamonds, and purple stars, respectively.  The MIPS 24$\mu$m confusion limit at $z\sim1.3$ is shown as a dotted line. 
    }
    \label{fig:sfr_ms}
\end{figure*}

For the portion of our sample without an \Ha\ match or a negative \APaa\ derived from the \Paa\ to \Ha\ ratio\footnote{The majority of the sources with negative \APaa\ are log $\logM<9.5$ and have attenuation consistent with zero from the SED fitting.}, we derive the correction from our SED fitting with {\tt Prospector}. The description of the two-component dust attenuation model is described in Section~\ref{sec:sed_fitting}.  As our fits were done with photometry only, we assume that the attenuation curve of the older stellar population ($>10$ Myr) is more robustly constrained, which gives us the stellar attenuation. To obtain \APaa\, we then assume that the stellar and nebular attenuation are equal.  We show that this results in good agreement between the \APaa\ derived from the \Paa\--\Ha\ line ratios vs the SED fitting in Figure~\ref{fig:apaa_comparison} for the sub-sample where we can measure both. Our assumption is likely appropriate for our high mass \Paa\ emitters \citep{reddy2020, shivaei2020}.  For dwarf galaxies at $z\sim2$, the relationship between nebular and stellar reddening is found to be closer to that originally derived for local starbursts \citep[$\sim2$;][]{calzetti2000, reddy2020, shivaei2020}; however, this relationship is also known to have significant scatter and as the attenuation of \Paa\ is predicted to be especially low in dwarf galaxies, we expect our assumption to have negligible impact on our results.

Our final \APaa\ values are shown as a function of the $A_{\rm V}$ derived from SED fitting in Figure~\ref{fig:apaa}.   We display the relation expected for a Calzetti attenuation curve ($A_{\rm V}/\APaa=6$) for reference \citep{calzetti2007a}. We correct the \Paa\ line luminosities as   

\begin{equation}
L_{\mathrm{Pa\alpha, corr}} = L_{\mathrm{Pa\alpha}} 10^{0.4A_{\mathrm{Pa\alpha}}}
\end{equation}

\noindent Overall, we find \APaa\ is largely $<0.2$ mag, corresponding to a less than $20\%$ correction for dust attenuation, in good agreement with previous studies \citep{reddy2018, reddy2023a, neufeld2024}.  

From the attenuation corrected \Paa\ luminosities, we then calculate our fiducial \SFRPaa\ using the calibration presented in \citet{reddy2023} as

\begin{equation}
    \mathrm{SFR}^{\mathrm{R23}}_{\mathrm{Pa\alpha}}\, [\Msun\,\mathrm{yr^{-1}}] = C(\mathrm{Pa\alpha}) L_{\mathrm{Pa\alpha, corr}}\, [\mathrm{ergs\,s^{-1}}]
\end{equation}

\noindent
where $C$(\Paa) = $1.95\e{-41}$ and 3.9\e{-41} for sub-solar ($Z_{\star}=0.001$) and solar ($Z_{\star}=0.02$ or 12+log(O/H)=8.69) metallicities, respectively.  This calibration takes into account 1) that the ionizing photon rate increases with decreasing metallicity at fixed SFR \citep{chisholm2019} and 2) that stellar binarity has been shown as important in optical line calibrations \citep{steidel2016, reddy2022}. As in \citet{reddy2023a}, the solar metallicity calibration is adopted for all galaxies at $z\leq1.4$ and for high mass galaxies (log $\logM>10.45$) at $z\geq1.4$.  Sub-solar metallicity is adopted for low-mass galaxies at $z\geq1.4$. This \Paa\ SFR calibration will be adopted throughout the paper. As a check, we will also verify all of our results using a non-metallicity dependent calibration as in \citet{shipley2016}: log SFR [$\Msun$ yr$^{-1}$] = $-40.33 + \log L_{\rm Pa\alpha}$ [ergs s$^{-1}$].

\begin{figure*}[ht!]
    \centering
    \includegraphics[width=0.9\textwidth]{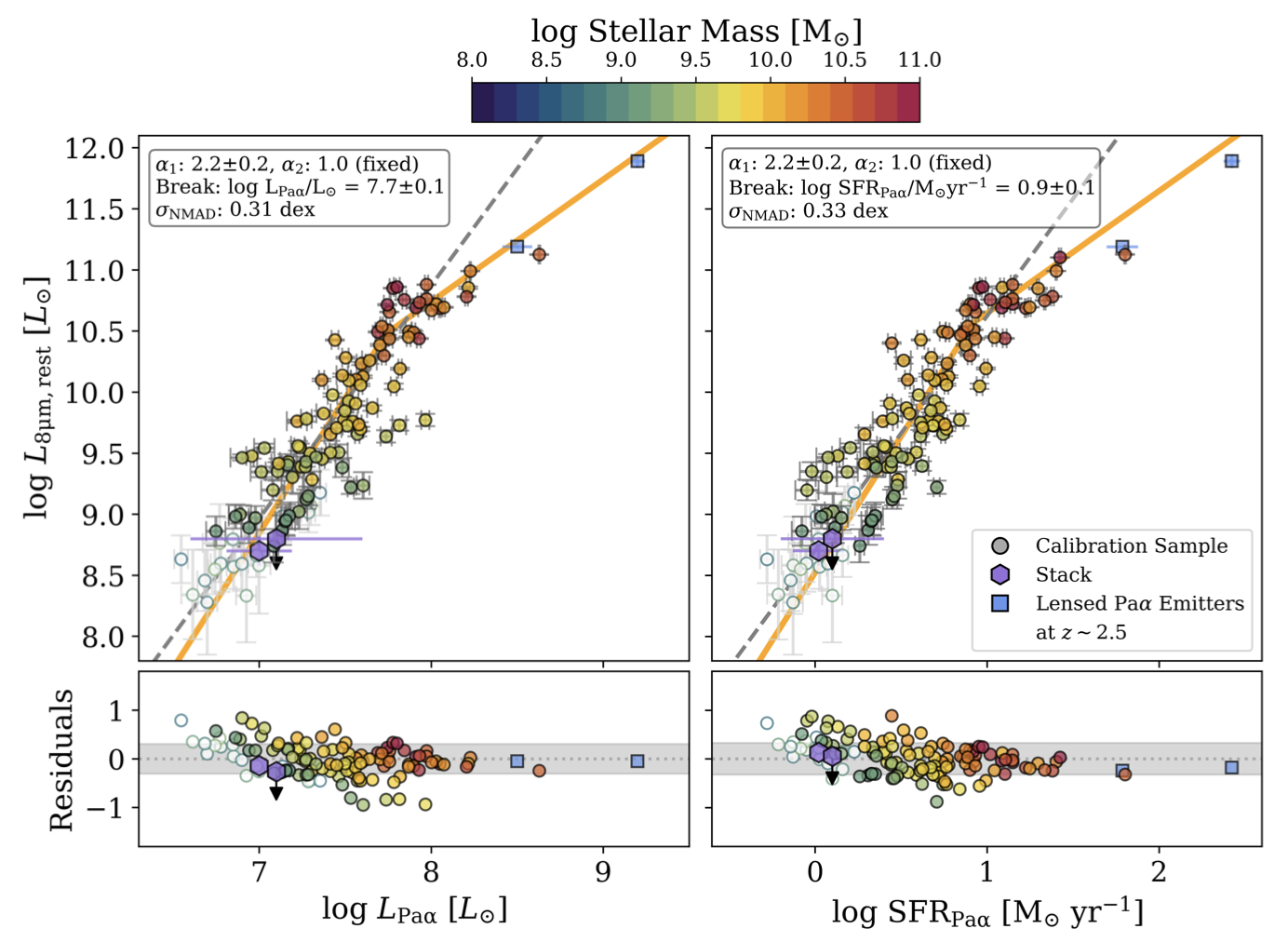}
    \caption{(left) The relation between the extinction-corrected \Paa\ luminosity and the observed $L_{\rm 8\mu m, rest}$ (Eqn~\ref{eqn:mir}) for MS galaxies with (closed circles) and without (open circles) a $>3\sigma$ detection in F1800W or F2100W.  The purple hexagons show the stack of the \Paa\ emitters with only marginal detections (SNR$=1-3$) in F1800W or F2100W and the upper limit on the stack of \Paa\ emitters with no MIRI detection. The blue squares show two lensed galaxies with measurements of \Paa\ and the $7.7\mu$m PAH feature at $z\sim2.5$ \citep{rigby2008,rujopakarn2011, shipley2016}. The gray dashed line is a single component linear fit.  A broken power law fit  
    (Eqn~\ref{eqn:pl}) with free parameters amplitude, breakpoint, and faint end slope is shown via the orange, solid line.  The bright end slope is fixed to unity following the literature \citep[e.g.,][]{shipley2016} The bottom panel shows the residuals from the broken power law fit.  (right) The relation between the extinction-corrected \Paa\ SFR using the \citet{reddy2023a} calibration and the observed $L_{\rm 8\mu m, rest}$ (Eqn~\ref{eqn:mir}).  Sample and symbols are as in the lefthand panel.  
    The colorbar corresponds to stellar mass for all panels.}
    \label{fig:lmir}
\end{figure*}

The final \Paa\ based fiducial SFRs, spanning a range of $\sim0.5-40$ $\Msun$ yr$^{-1}$, are shown as a function of stellar mass in Figure~\ref{fig:sfr_ms}.  We sample the full range of the MS ($-0.3<\Delta \mathrm{MS}<0.3$) over the range log $\logM\sim9-10.8$ and reach over an order of magnitude lower in SFR than the MIPS24 confusion limit \citep{dole2004} at $z\sim1.3$.

\section{Calibration of mid-infrared-based SFR indicators with MIRI}\label{sec:mir_sfrs}

In this section, we derive a SFR calibration for the MIRI broadband photometry using our fiducial \Paa\ SFRs.  In the redshift range of our sample ($1\lesssim z\lesssim1.75$), the MIRI broadband filters at $12-21\mu$m contain PAH emission as well as small grain dust and stellar continuum emission.  The F1800W and F2100W bands are dominated by the strong $7.7\,\mu$m PAH over the full redshift range and the F1280W and F1500W filters contain the narrower $6.2\,\mu$m PAH at $z\lesssim1.3$ and $1.2\lesssim z\lesssim1.6$, respectively (Figure~\ref{fig:dist}).  To use the cleanest calibration sample possible, we remove 1) 18 AGN, which may influence PAHs \citep[e.g.,][]{diamond-stanic2010, inami2018, garcia-bernete2022, lai2022} and \Paa; 2) 18 non-AGN starbursts \citep[$\Delta$MS$\,>0.6$; e.g.,][]{rodighiero2011}, where intense radiation fields may influence the PAHs \citep[e.g.,][]{kim2024a}; and 3) 6 \Paa\ emitters below the MS ($\Delta$MS$\,<-0.6$), where dust heating may not be related to young stars \citep[e.g.,][]{hayward2014}. Our clean calibration sample contains 126 \Paa\ emitters.  The removed outliers will be examined related to the clean sample and calibration in Section~\ref{sec:scatter}.

\subsection{The relationship between \Paa\ and the $8\mu$m mid-infrared spectral region}\label{sec:paa_lmir}

In Figure~\ref{fig:lmir}, we show the luminosity of the rest-frame 8$\,\mu$m spectral region as measured by MIRI longwave filters as a function of the extinction-corrected \Paa\ line luminosity (panel a) and \SFRPaa\ (panel b). To sample the same spectral features as much as possible across our redshift range, we adopt the observed F1800W luminosity for the lower half of our redshift range and F2100W for the upper half such that 

\begin{equation}\label{eqn:mir}
L_{\mathrm{8\mu m, rest}} =\begin{cases}
  L_{\mathrm{F1800W}} & \text{for } z<1.4\\   
  L_{\mathrm{F2100W}} & \text{for } z\geq1.4  
\end{cases}
\end{equation}

\noindent where $L_x \equiv \nu L_{\nu}$. We make no correction for the different filter transmission curves nor do we apply a k-correction here (see Section~\ref{sec:uv_ir}).  

\begin{deluxetable*}{lccccc|cc}\label{tbl:fit1}
\tablecaption{SFR$_{\Paa}-L_{x}$ Best-Fit Parameters and Monochromatic $L_{x}$-based SFR Calibration}
\tablewidth{0pt}
\tablehead{
\colhead{Parameter} & 
\multicolumn{5}{c|}{SFR$_{\Paa}-L_{x}$ Fits (Eqn~\ref{eqn:pl})} & 
\multicolumn{2}{c}{$L_{x}$-based SFR Calibration (Eqn~\ref{eqn:sfr_mir})} \\
\cline{2-6} \cline{7-8}
& \colhead{N$^{a}_{\mathrm{gal}}$} & \colhead{$\log$ A} & \colhead{$\log$ x$^{b}_{\rm break}$} & \colhead{$\alpha$} & \colhead{$\sigma_{\rm NMAD}$} & \colhead{$a_x$} & \colhead{$C_x$} 
}
\startdata
$L^c_{\mathrm{\rm 8\mu m, rest}}$ & & 10.6 & $0.9\pm0.1$ & & 0.33 & &  \\
\quad $<\mathrm{x_{\rm break}}$ & 74 & & & $2.2\pm0.2$ & & $0.45\pm0.07$ & $3.8\pm0.2$  \\
\quad $>\mathrm{x_{\rm break}}$ & 21 & & & $1.0$ (fixed) & & $-$ & $9.6\pm0.1$ \\ \\ 
$L_{\rm F1800W}$ & & 10.5 & $0.9\pm0.1$ & & 0.35 & &  \\
\quad $<\mathrm{x_{\rm break}}$ & 75 & & & $2.3\pm0.3$ & & $0.44\pm0.07$ & $3.7\pm0.2$  \\
\quad $>\mathrm{x_{\rm break}}$ & 21 & & & $1.0$ (fixed) & & $-$ & $9.6\pm0.1$ \\ \\ 
$L_{\rm F2100W}$ &  & 10.2 & $0.9\pm0.1$ & & 0.39 & &  \\
\quad $<\mathrm{x_{\rm break}}$ & 62 & & & $2.4\pm0.5$ & & $0.42\pm0.1$ & $3.4\pm0.3$  \\
\quad $>\mathrm{x_{\rm break}}$ & 23 & & & $1.0$ (fixed) & & $-$ & $9.3\pm0.2$ \\ \\ %\hline
\\
\enddata
\tablecomments{(a) Number of galaxies with SNR$\,>3$ in a given band.  (b) x$_{\rm break}$ is given in $\Msun$ yr$^{-1}$. See Section~\ref{sec:calibration} for discussion on adopting a redshift-independent break in stellar mass: log $\logM=10.1$. (c) as defined in Eqn~\ref{eqn:mir}.}
\end{deluxetable*}

For \Paa\ sources with marginal or no detection at F1800W or F2100W, we perform inverse-variance weighted and median stacking with bootstrapping in two categories: 1) clean samples\footnote{Meaning AGN, starbursts, and sources significantly below the MS have been removed.} with a marginal SNR$\,=1-3$ measurement in F1800W or F2100W and 2) a clean sample with SNR$\,<1$ in both red filters or no MIRI counterpart based on the F560W+F770W detection image (see Section~\ref{sec:miri_data}).  For the former, the stacks of 17 and 13 marginally detected sources in F1800W and F2100W, respectively, are detected at $\sim4-5\sigma$, while splitting this sample by redshift (Eqn~\ref{eqn:mir}) yields stacks\footnote{To create a combined stack, we first split by redshift at $z=1.4$ (Eqn~\ref{eqn:mir}) and stack separately in F1800W and F2100W and then combine the stacked fluxes weighted by the number in each stack 
} detected at $\sim6-7\sigma$.  We verify that the weighted average stacks are consistent with median stacks, ruling out a bias toward outliers.  These stacks will be used in the calibration of the MIRI SFRs.  The second, undetected sample (29 sources) is not detected via stacking in either filter and has significant differences between the weighted and median stack, indicating this subsample contains outliers.  From the median stacks and bootstrapped errors we derive $3\sigma$ upper limits, which we will check against our calibration for consistency.

In Figure~\ref{fig:lmir} (left) we show $L_{\rm 8\mu m, rest}$ versus $L_{\rm Pa\alpha}$ in log space. To quantify the behavior of $L_{\rm 8\mu m, rest}$, we perform fitting using orthogonal distance regression (ODR) from the {\tt scipy} software package \citep{scipy2020}, which takes into account uncertainties in both parameters\footnote{As only measurement uncertainties are used in the fits, we impose a ceiling of SNR$\,=20$, equivalent to assuming $\sim5\%$ systematic uncertainty, to keep very luminous sources from driving the fit.}.  Fitting is done to \Paa\ emitters with F1800W or F2100W SNR$\,\geq3$ and to the stacks of those with SNR$\,=1-3$.  We start with a single component linear fit, finding a superlinear slope with $\alpha=1.9\pm0.1$ with a scatter of 0.38 dex.  This is somewhat steeper than the slope derived in \citet{ronayne2023}, which used the dust-corrected UV as their fiducial SFR indicator.  We note, however, their use of SED fitting to isolate $L_{\rm 7.7PAH}$ differs from our approach of using the total broadband fluxes, which contain PAH plus continuum emission.

As discussed in the introduction, however, pre-JWST studies have firmly established that for massive, high SFR galaxies, there is a unity relation between the mid- and far-IR and SFR \citep{rujopakarn2013, shipley2016, alonso-herrero2006, papovich2007, wuyts2011, pope2008, siana2008, siana2009, rieke2009, zhu2008}, such that we expect our superlinear fit to turn over at some  threshold. This is visually demonstrated in Figure~\ref{fig:lmir} by two lensed galaxies at $z\sim2.5$ in which both \Paa\ and the $7.7\mu$m PAH were detected by Spitzer/IRS \citep[][]{rujopakarn2011, shipley2016}.  Given 1) the small areal coverage of SMILES relative to pre-JWST MIR surveys and 2) the typical SFRs of the MS galaxies in the redshift range where we have \Paa\ (i.e. tens of $\Msun$ yr$^{-1}$), we do not expect that our sample will have much overlap with the populations studied in the Spitzer era, but these previous works and the hint of a turn over in our data (Figure~\ref{fig:lmir}) motivate us to examine a two component fit. 

\begin{deluxetable*}{lccccc|cc}\label{tbl:fit2}
\tablecaption{SFR$_{\Paa}-L_{\rm IR}$ Best-Fit Parameters and $L_{\rm IR}$-based SFR Calibration}
\tablewidth{0pt}
\tablehead{
\colhead{Parameter} & 
\colhead{N$^{a}_{\mathrm{gal}}$} &
\multicolumn{4}{c|}{SFR$_{\Paa}-L_{\rm IR}$ Fits (Eqn~\ref{eqn:pl})} & 
\multicolumn{2}{c}{$L_{\rm IR}$-based SFR Calibration (Eqn~\ref{eqn:sfr_mir2})} \\
\cline{3-6} \cline{7-8}
 & & \colhead{$\log$ A} & \colhead{$\log$ x$^{b}_{\rm break}$} & \colhead{$\alpha$} & \colhead{$\sigma_{\rm NMAD}$} & \colhead{$a_x$} & \colhead{$C_x$} 
}
\startdata
$L_{\rm IR}^{\rm F1280W}$ & 63 & 11.3 & $0.9\pm0.1$ & & 0.21 & &  \\
\quad $<\mathrm{x_{\rm break}}$ & & & & $1.9\pm0.2$ & & $0.52\pm0.01$ & $5.0\pm0.1$  \\
\quad $>\mathrm{x_{\rm break}}$ & & & & $1.0$ (fixed) & & $-$ & $10.4\pm0.2$ \\ \\ \hline
$L_{\rm IR}^{\rm F1500W}$ & 126 & 11.1 & $0.9\pm0.1$ & & 0.27 & & \\
\quad $<\mathrm{x_{\rm break}}$ & & & & $2.1\pm0.2$ & & $0.47\pm0.04$ & $4.4\pm0.1$  \\
\quad $>\mathrm{x_{\rm break}}$ & & & & $1.0$ (fixed) & & $-$ & $10.4\pm0.1$ \\ \\ \hline
$L_{\rm IR}^{\rm F1800W}$ & 122 & 11.4 & $0.9\pm0.1$ & & 0.31 & & \\
\quad $<\mathrm{x_{\rm break}}$ & & & & $2.3\pm0.2$ & & $0.44\pm0.04$ & $4.1\pm0.1$  \\
\quad $>\mathrm{x_{\rm break}}$ & & & & $1.0$ (fixed) & & $-$ & $10.5\pm0.1$ \\ \\ \hline
$L_{\rm IR}^{\rm F2100W}$ & 107 & 11.4 & $0.9\pm0.1$ & & 0.31 & &   \\ 
\quad $<\mathrm{x_{\rm break}}$ & & & & $2.2\pm0.3$ & & $0.46\pm0.06$ & $4.3\pm0.1$  \\
\quad $>\mathrm{x_{\rm break}}$ & & & & $1.0$ (fixed) & & $-$ & $10.5\pm0.1$ \\
\enddata
\tablecomments{(a) Number of galaxies with SNR$\,>3$ in a given band.  (b) x$_{\rm break}$ is given in $\Msun$ yr$^{-1}$. See Section~\ref{sec:calibration} for discussion on adopting a redshift-independent break in stellar mass: log $\logM=10.1$.}
\end{deluxetable*}

As such, we fit a broken power law following 

\begin{equation}\label{eqn:pl}
f(\mathrm{x}) =\begin{cases}
  A\left(\frac{\mathrm{x}}{\mathrm{x}_{\mathrm{break}}}\right)^{\alpha_1} & \text{for } \mathrm{x}<\mathrm{x}_{\mathrm{break}}\\   
  A\left(\frac{\mathrm{x}}{\mathrm{x}_{\mathrm{break}}}\right)^{\alpha_2} & \text{for } \mathrm{x}>\mathrm{x}_{\mathrm{break}} . 
\end{cases}
\end{equation}
\vspace{10pt}

\noindent where the amplitude (A), break point ($x_{\rm break}$), and $\alpha_1$ are allowed to vary.  Following previous studies, we fix $\alpha_2=1$.  Our two component best-fit has a superlinear slope of $\alpha_1=2.2\pm0.2$ up to 
log $L_{\rm Pa\alpha}/\Lsun\,=7.7\pm0.1$, above which a linear slope fits the turn over in our data.  Similarly if we fit $L_{\rm 8\mu m, rest}-$SFR$_{\Paa}$ (Figure~\ref{fig:lmir}, [right]), we find the same superlinear slope up to SFR$_{\Paa}\sim8\,\Msun$ yr$^{-1}$.  
Our broken power law relation yields a scatter of $\sim0.3$ dex, about a factor of 2, comparable with previous studies \citep[e.g.,][]{calzetti2007a, reddy2010, wuyts2011, rujopakarn2013, derossi2018, ronayne2023}.  We confirm that we find a consistent broken power-law fit (with slightly increased scatter) if we instead use the SFR$_{\Paa}$ calibration from \citet{shipley2016}, which has no metallicity dependence.  We likewise find consistent fits within the uncertainties if we repeat this exercise with $L_{\rm F1800W}$ and $L_{\rm F2100W}$ separately, though we see increased scatter (up to $
\sim0.4$ dex), likely due to the lack of $k$-correction (see Section~\ref{sec:templates}).

Given that the broken power law provides a good description of the SFR$_{\Paa}-L_{\rm 8\mu m, rest}$ relation, we calibrate MIRI-based SFR relations at $z\sim1.3$ as 

\begin{equation}\label{eqn:sfr_mir}
\log\mathrm{SFR} =\begin{cases}
  a_x \log L_x - C_x & \text{for } <\,8\Msun\, \mathrm{yr^{-1}} \\  
  \log L_x - C_x & \text{for } >\,8\Msun\, \mathrm{yr^{-1}} 
\end{cases}
\end{equation}
\noindent
where luminosities are in $\Lsun$. As most studies will not have an independent measurement of the SFR, we propose an equivalent breakpoint of log $\logM=10.1$, based on the MS at $z\sim1.3$ \citep{leja2022}; we explore and justify this assumption in Section~\ref{sec:breakpoint}. Our best-fit parameters are provided in Table~\ref{tbl:fit1} for $L_{\rm 8\mu m, rest}$ (Eqn~\ref{eqn:mir}), $L_{\rm F1800W}$, and $L_{\rm F2100W}$. 

\subsection{Application to higher redshift}\label{sec:application}

So far we have focused on the rest-frame $8\mu$m region as it is well-studied and known to be dominated by the luminous $7.7\mu$m PAH in SFGs, contributing $\sim70\%$ of the flux in the IRAC ch4 band in local galaxies \citep{smith2007}.  Expanding the usefulness of MIRI-based SFRs to $z\gtrsim2$, however, will need to make use of dust features at shorter wavelengths.  We examine here the rest-$6\mu$m region, which contains the $6.2\,\mu$m PAH.  For F1800W and F2100W, $\gtrsim70\%$\footnote{Using the PAH clip points presented in \citet{draine2021}.}of the $7.7\mu$m PAH falls within the band at $0.9<z<1.5$ and $1.1<z<2.1$, respectively.  By comparison, $\gtrsim70\%$ of the $6.2\mu$m PAH falls into F1800W and F2100W at $1.6<z<2.2$ and $1.9<z<2.85$, respectively.  To test whether the rest-frame $6\mu$m region is viable as a SFR indicator, we employ the F1280W and F1500W bands over our redshift range.  For F1280W, coverage of the dust emission at $6.2\mu$m starts at $z\sim1.3$.  The F1500W on the other hand covers dust emission at rest $\sim6-8\mu$m over our full redshift range.  As in the previous section, we test these bands against the \Paa\ luminosities and our fiducial \Paa\ SFRs.  We find again that our data is well fit with a broken power law with a breakpoint at $\sim8\,\Msun$ yr$^{-1}$ and similar scatter ($\sim0.3$ dex), showing the same behavior through the rest $\sim6\,\mu$m spectral region.  This suggests that the $6.2\,\mu$m and $7.7\mu$m and their associated underlying continua behave similarly in MS galaxies at $z\sim1.3$, even in lower mass, lower metallicity galaxies.  From this, we conclude that the F1800W and F2100W filters can also be calibrated for the rest $6\mu$m region, extending the applicable range up to $z\sim3$, particularly if we control for the $k$-correction, as we discuss in the next section.  We note that we will not test the viability of the $3\mu$m region in this work as the $3.3\mu$m PAH is relatively weak and sits on continuum more likely unrelated to young stars.  Spectroscopy is therefore needed to characterize this feature beyond the local Universe, see \citet{lyu2025, mckinney2025b}.

\begin{figure*}[ht!]
    \centering
    \includegraphics[width=0.99\textwidth]{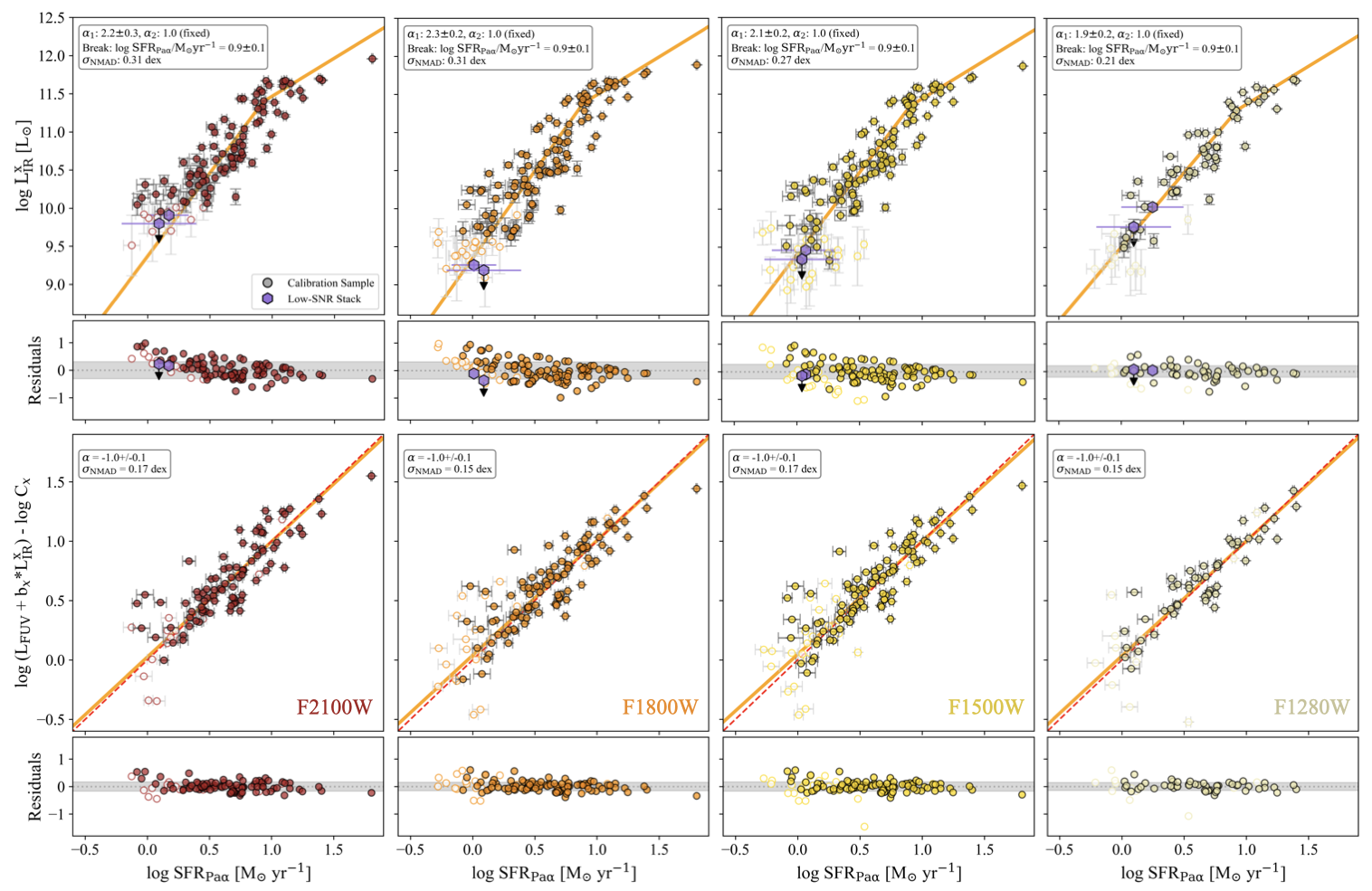}
    \caption{(top panels) $L_{\rm IR}$  derived using single-band MIRI photometry in F2100W, F1800W, F1500W, or F1280W ($z<1.3$ only) as a function of the fiducial \Paa\ SFRs. Symbols and orange, solid line are as in Figure~\ref{fig:lmir}.  (bottom panels) The UV+IR composite SFR calibration (Eqn~\ref{eqn:uv_mir_sfrs}).  The solid orange line is a single component linear fit.  The dashed red line is the one-to-one line. The fitted slope and scatter are given in the legend.  Closed symbols are detections in both UV and IR, open symbols have SNR$<3$ in either. }
    \label{fig:lir1}
\end{figure*}

\subsection{(UV+)L$_{\rm IR}$ SFR Calibration using MIRI}\label{sec:uv_ir}

In the previous section, we showed that the relation between the observed F1800W or F2100W luminosities from MIRI and our fiducial \Paa\ SFRs can be described as a broken power law with a break at log $\logM\sim10.1$.  Though this empirical approach is appealing in that it minimizes assumptions, it has drawbacks. First, it doesn't $k$-correct to account for redshift, which is particularly important given the large variations as a function of wavelength in the mid-IR spectral region. And, second,
for log $\logM\lesssim10$, we are entering the regime where the obscured component of star formation no longer dominates in cosmic noon galaxies \citep{whitaker2017}, increasing the uncertainty in extrapolating the obscured SFR to a total SFR.
As such, in the next two sections, we refine our MIRI-based SFR calibration by scaling from single-band MIRI photometry to the total dust luminosity $L_{\rm IR}\equiv L[8-1000\mu$m] and then accounting for the unobscured SFR component through composite SFR calibration \citep[e.g.,][]{bell2005, treyer2010, hao2011, calzetti2007a, kennicutt2007, kennicutt2009, wuyts2011, whitaker2014a, lee2015, popesso2023}.

\subsubsection{Template-based $L_{\rm IR}$}\label{sec:templates}

The mid-IR spectrum of galaxies undergoes large variations over a short wavelength range due to the broad emission and absorption features (see Figure~\ref{fig:dist}).  As such, small variations in the redshift of a source can correspond to large intrinsic photometric scatter for a given mid-IR filter.  To remove these variations, standard practices are to $k$-correct to the same rest wavelength \citep[e.g.,][]{rieke2009} or quantify the total infrared luminosity from one or more mid- to far-infrared bands using SED fitting and/or by scaling representative templates \citep[e.g.,][]{rujopakarn2013, boquien2021}.  This process accounts for the redshift and differences in the shape and width of the filters used. 

To derive $L_{\rm IR}$, we adopt the set of empirical templates from \citet{rieke2009}, which have been shown to represent the SEDs of IR-bright galaxies with a scatter of $~\sim0.2$ dex in derived IR SFR up to $z\sim3$ \citep[e.g.,][]{rujopakarn2013, derossi2018}.  For simplicity, we first scale a single template with log $L_{\rm IR}/\Lsun = 11.25$ to the MIRI F1800W and F2100W photometry $-$ separately and accounting for their respective transmission curves $-$ and repeat the calibration against SFR$_{\Paa}$ as describe in the previous section.  We find that the trends of $L_{\rm IR}^{\rm F1800W}$ and $L_{\rm IR}^{\rm F2100W}$ with \Paa\ can again be described with a broken power law (Eqn~\ref{eqn:pl}) with the best-fit $\alpha_1$ slope and breakpoint (Table~\ref{tbl:fit2}) consistent with what was found in Section~\ref{sec:paa_lmir}.  This reflects that using a single empirical template is assuming a fixed ratio between $L_{\rm MIR}$ and $L_{\rm IR}$.  However, we find that the scatter is somewhat improved by incorporating the $k$-correction (reduced by $\sim0.05$ dex). 
Our fits are shown in Figure~\ref{fig:lir1} and we note that our galaxies span the range of log $L_{\rm IR}/\Lsun\sim9-11.5$ and are mostly in the sub-LIRG regime.

Next we test the full range of \citet{rieke2009} templates, spanning log $\LIR\sim10-11.75$.  To choose a template for each galaxy, we start with the \LIR\ derived above and then iterate until convergence to adopt the template closest to the galaxy's actual \LIR.  We find, however, that this \textit{increases} the scatter when we then compare to SFR$_{\Paa}$.  This is not wholly unexpected.  Studies of highly star-forming ($>100\,\Msun$ yr$^{-1}$) galaxies at cosmic noon found that the \LIR\ derived using MIPS 24$\mu$m agreed better with the \LIR\ measured with longer Spitzer and Herschel bands when using an average template instead of a luminosity-dependent conversion \citep{papovich2007, elbaz2010, wuyts2008, wuyts2011}.  \citet{rujopakarn2013} likewise showed that the variations in the dust SED correlated with the IR surface density rather than total luminosity.  Though MIRI's spatial resolution will allow the measurement of the IR surface density at cosmic noon for some populations, we want to provide a calibration that does not require this information.  Even if we did include it, these studies of massive, IR luminous galaxies may not apply to our sample.  UV-selected samples of less massive galaxies, perhaps more analogous to ours, have found significant variation in the ratios of the mid- to far-IR emission \citep[e.g.,][]{reddy2012, shivaei2017, reddy2018}.  As such, we opt to proceed using the single dust template from above to derive \LIR\ for the rest of this work, to minimize scatter. In the next two sections, we show that the unobscured SFR component is a major source of scatter in our sample and perform a sanity check of the obscured SFR component implied by our choice of template.  Then in Section~\ref{sec:disc}, we further discuss the implications of using a single representative IR template in our calibration.  

Adopting the log $L_{\rm IR}/\Lsun = 11.25$ template for the remainder of this study, we calibrate model-dependent MIRI-based SFRs as 

\begin{equation}\label{eqn:sfr_mir2}
\log\mathrm{SFR} =\begin{cases}
  a_x \log L_{\mathrm{IR}} - C_x \\ \qquad\qquad \text{for } \mathrm{SFR}<8\,\Msun \mathrm{yr}^{-1} \\  
  \\
  \log L_{\mathrm{IR}} - C_x\\ \qquad\qquad \text{for } \mathrm{SFR}>8\,\Msun \mathrm{yr}^{-1}
\end{cases}
\end{equation}

\noindent In Table~\ref{tbl:fit2}, we include the best-fit parameters for SFRs based on $L_{\rm IR}$ scaled from the F1800W or F2100W photometry for reference.  As before, we suggest that the break at 8$\,\Msun$ yr$^{-1}$ can be substituted with log $\logM=10.1$.  This is discussed further in Section~\ref{sec:calibration}. 

\subsubsection{Composite SFR Indicators}

\begin{deluxetable*}{lccc|ccc}\label{tbl:fit3}
\tablecaption{ UV+IR SFR Calibration}
\tablewidth{0pt}
\tablehead{
\colhead{Parameter} & 
\multicolumn{3}{c|}{NUV+$L_{\rm IR}$ SFR Calibration} &
\multicolumn{3}{c}{FUV+$L_{\rm IR}$ SFR Calibration} \\
\cline{2-4} \cline{5-7}  
& \colhead{$b_x$} & \colhead{$C_x$} & \colhead{$\sigma_{\rm NMAD}$} & \colhead{$b_x$} & \colhead{$C_x$} & \colhead{$\sigma_{\rm NMAD}$}
}
\startdata
$L_{\rm IR}^{\rm F1280W}$ & $0.11\pm0.02$ & $9.5\pm0.04$ & 0.16 & $0.14\pm0.02$ & $9.6\pm0.05$ & 0.15 \\
$L_{\rm IR}^{\rm F1500W}$ & $0.10\pm0.01$ & $9.5\pm0.04$ & 0.15 & $0.13\pm0.02$ & $9.6\pm0.04$ & 0.17 \\
$L_{\rm IR}^{\rm F1800W}$ & $0.08\pm0.01$ & $9.5\pm0.03$ & 0.16 & $0.11\pm0.01$ & $9.6\pm0.04$ & 0.15 \\
$L_{\rm IR}^{\rm F2100W}$ & $0.12\pm0.02$ & $9.6\pm0.05$ & 0.17 & $0.17\pm0.03$ & $9.7\pm0.05$ & 0.17 \\
\enddata
% \tablecomments{}
\end{deluxetable*}

Composite SFR indicators combine a direct UV or optical tracer of unobscured emission from young stars with an infrared tracer to capture the missing component absorbed and re-radiated by dust.  This method is more robust against the systematic uncertainties introduced by using e.g. the UV or IR alone and so has a long history in the literature \citep[e.g.,][]{kennicutt2009, hao2011, kennicutt2012}.  It is particularly important in our sample, which spans the range in stellar mass where the obscured SFR component varies from $\sim20-90\%$ \citep{whitaker2017}.  We adopt the common prescription of the form 

\begin{equation}\label{eqn:uv_mir_sfrs}
    \log \mathrm{SFR} = \log (L_{\rm UV} + b_x L^x_{\rm IR}) - \log C_x
\end{equation}

\noindent which is valid under assumptions of energy balance for direct tracers where the attenuation for the continuum or line emission in the UV/optical is similar to the mean dust opacity of the starlight heating the dust \citep[see discussion in][] {kennicutt2009}.  As this is not the case for \Paa\ \citep{kennicutt2009} and the UV can be measured from e.g. HST photometry, we focus on UV+IR composite relations. We measure the rest-frame near-UV at $2300$ \AA  ~from our SED-fitting (Section~\ref{sec:sed_fitting}), which is bracketed across our redshift range by the F435W and F606W/F775W HST bands.  For easy reference to the literature, we will also look at the far-UV at $1600$ \AA, though this requires extrapolation.      

In Figure~\ref{fig:lir1} (bottom panels), we show the best-fit SFRs as a function of our fiducial SFR$_{\Paa}$ and in Table~\ref{tbl:fit2} we list our best-fit parameters.  Somewhat surprisingly, we find that Eqn~\ref{eqn:uv_mir_sfrs} provides an excellent description of our data over its full range, with significantly less scatter ($\sim0.15$) dex compare to the IR-only calibrations.  This indicates that the IR continues to be a robust tracer of the obscured star formation component down to log $\logM\sim9$ and total SFRs of $\sim1\,\Msun$ yr$^{-1}$.  We further discuss the implications of this in Section~\ref{sec:calibration}. 

The scale factor, $b_x$, describes the relation between the direct and indirect SFR tracers and in this case is dependent on both the fraction of the dust that is heated specifically by UV photons and the fraction of $L_{\rm IR}$ that is heated by emission related to young stars.  Using F1800W $-$ covering the rest-frame $8\mu$m $-$ and NUV we find $b_{\rm F1800W} = 0.08\pm0.01$, somewhat lower than calibrations based on local galaxies \citep{hao2011, kennicutt2012}.  Our best fit parameters for Eqn~\ref{eqn:uv_mir_sfrs} for all four MIRI filters and the NUV at $2300\,\AA$ and FUV at $1600\,\AA$ are found in Table~\ref{tbl:fit3}. 

\begin{figure*}[htb!]
    \centering
    \includegraphics[width=0.99\textwidth]{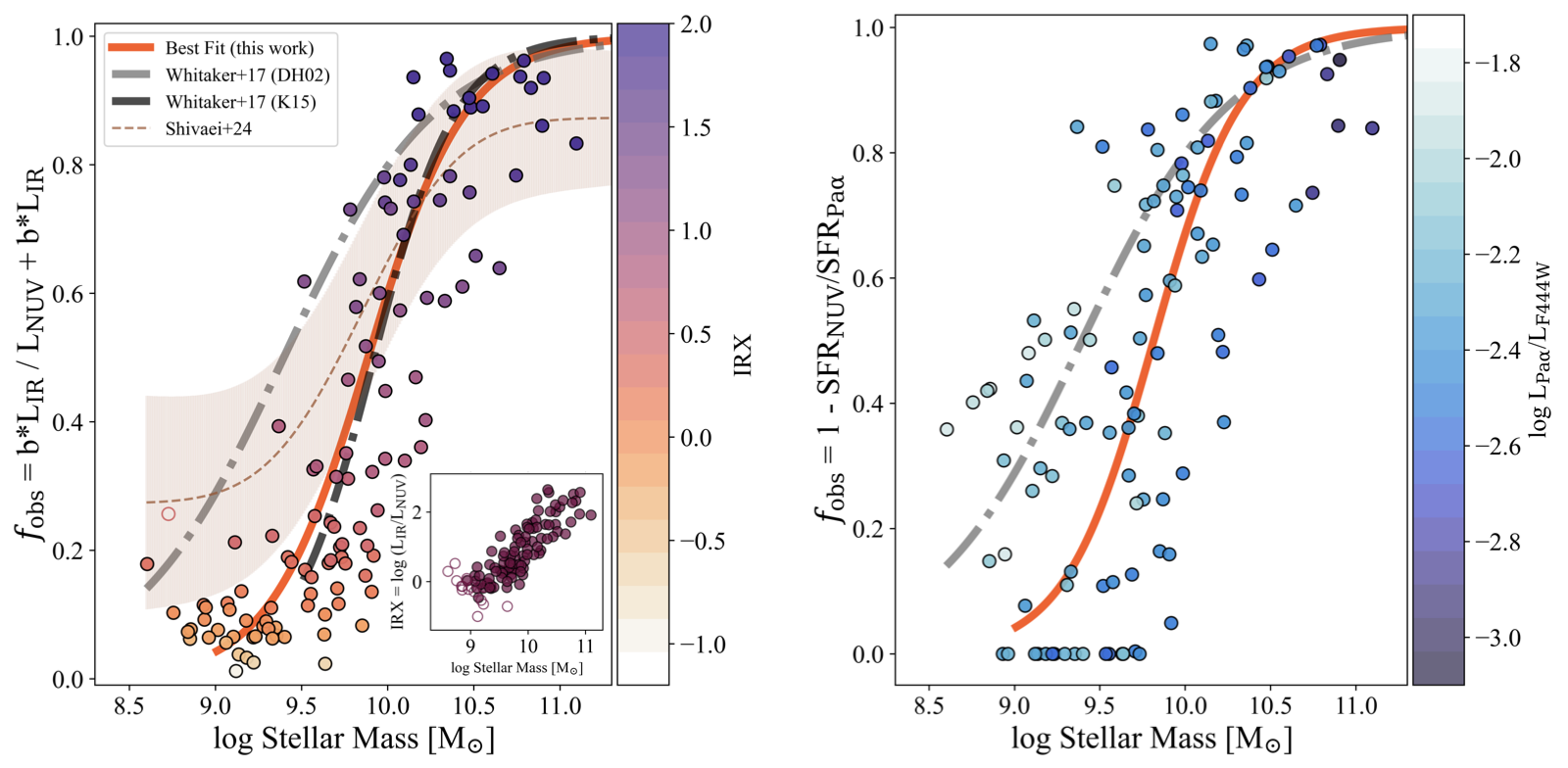}
    \caption{(left) The obscured fraction, $f_{\rm obs}$, as a function of stellar mass derived from the (uncorrected) NUV and $L^{\rm F1800W}_{\rm IR}$ via our derived composite SFR calibration (Eqn~\ref{eqn:uv_mir_sfrs}) for MS galaxies.  The dash dot gray and black lines show the fitted relation from \citet{whitaker2017} based on MIPS stacking plus the \citet{dale2002} and \citet{kirkpatrick2015} IR templates to derive $L_{\rm IR}$, respectively. The dashed line and shaded region show the fitted relation from \citet{shivaei2024}, based on SED fitting including MIRI broadband photometry.  Our best fit is shown as the orange solid line and is in good agreement with \citet{whitaker2017} when using the \citet{kirkpatrick2015} templates. The points are color-coded by IRX, which is also shown as a function of stellar mass in the inset. (right) Same as the left panel, only with $f_{\rm obs}$ derived using the SFR$_{\rm UV}$ and SFR$_{\Paa}$.  Sources with SFR$_{\rm UV}>$SFR$_{\Paa}$ have had their $f_{\rm obs}$ set to zero (see Section~\ref{sec:uv_excess}). The points are color-coded by the ratio of the \Paa\ to F444W luminosities, as a proxy for the \Paa\ EWs.}
    \label{fig:fobs}
\end{figure*}

\subsubsection{The obscured fraction and further verification of our template assumption}

Building on our composite SFR calibration presented in the previous section, in Figure~\ref{fig:fobs} (left) we show the obscured SFR fraction, defined as $f_{\rm obs}\equiv b_x L_{\rm IR} / L_{\rm NUV} + b_x L_{\rm IR}$, as a function of stellar mass.  As in previous studies, we find high $f_{\rm obs}$ at high mass, followed by a steep decline with decreasing mass.  However, our fall off is sharper than the main relation presented in \citet{whitaker2017} based on MIPS stacking.  As shown in that work, this is related to the dust emission template used. \citet{whitaker2017} compared their assumed template, a log-averaged IR template from \citet[][see also \citet{wuyts2008, wuyts2011}]{dale2002}, to alternative templates from \citet{kirkpatrick2015} and \citet{magdis2012}, finding the latter produced a steeper drop off (shown as the black dash-dot line in Figure~\ref{fig:fobs} [left]).  In this work, we have assumed the log $L_{\rm IR}=11.25$ template from \citet{rieke2009}, which is similar to the \citet{kirkpatrick2015} template and produces a similar best-fit relation.  

Which assumption is correct?  In Figure~\ref{fig:fobs} (right), we test these templates by deriving the $f_{\rm obs}$ independently from our UV and \Paa\ information, where $f_{\rm obs}\equiv 1 - \mathrm{SFR}^{\rm uncorr}_{\rm NUV} / \mathrm{SFR}_{\Paa}$.  SFR$^{\rm uncorr}_{\rm NUV}$ is derived using the NUV calibration from \citet{kennicutt2012} and is \textit{not} corrected for dust attenuation\footnote{We note that we find qualitatively the same result using FUV with the \citet{reddy2023a} calibration. We show the NUV here as it is better constrained by our dataset.}.  The result is that this independent measure, to first order, supports both template choices with significant scatter.  

However, it is important to remember here (and throughout this study) that the UV (and IR) and \Paa\ can trace very different star formation timescales \citep{calzetti2007a}.  Recent simulations have found that the UV in galaxies with constant SFHs will tend to trace similar timescales as the hydrogen recombination lines \citep[$\sim10$ Myr;][]{floresvelazquez2021} due to the continued renewal of O-type stars that can outshine older populations.  Stochastic SFHs, on the other hand, can produce UV that traces longer timescales ($\sim100$ Myr), averaging over the rise and fall of bursty star formation.  To test this, in Figure~\ref{fig:fobs} (right), we color code our sources by the \Paa\ equivalent width (EW, approximated as \Paa\ divided by the F444W flux), where larger EWs may indicate more bursty SF \citep[e.g.,][]{tran2020}.  At $\logM\lesssim9.5$, the sources that favor the original \citet{whitaker2017} fit are those with the highest EWs.  We propose that these sources are the most likely to overestimate $f_{\rm obs}$ when comparing UV and \Paa\ due to the timescale mismatch and tentatively conclude that this test supports the steeper decline and our template choice.  $f_{\rm obs}$ and the global dust properties of our sample are discussed further in Section~\ref{sec:calibration}. The UV properties of these high EW \Paa\ emitters are discussed further in Section~\ref{sec:uv_excess}.

\subsubsection{Comparison to UV- and SED fitting-based SFRs}

\begin{figure*}[htb!]
    \centering
    \includegraphics[width=0.95\textwidth]{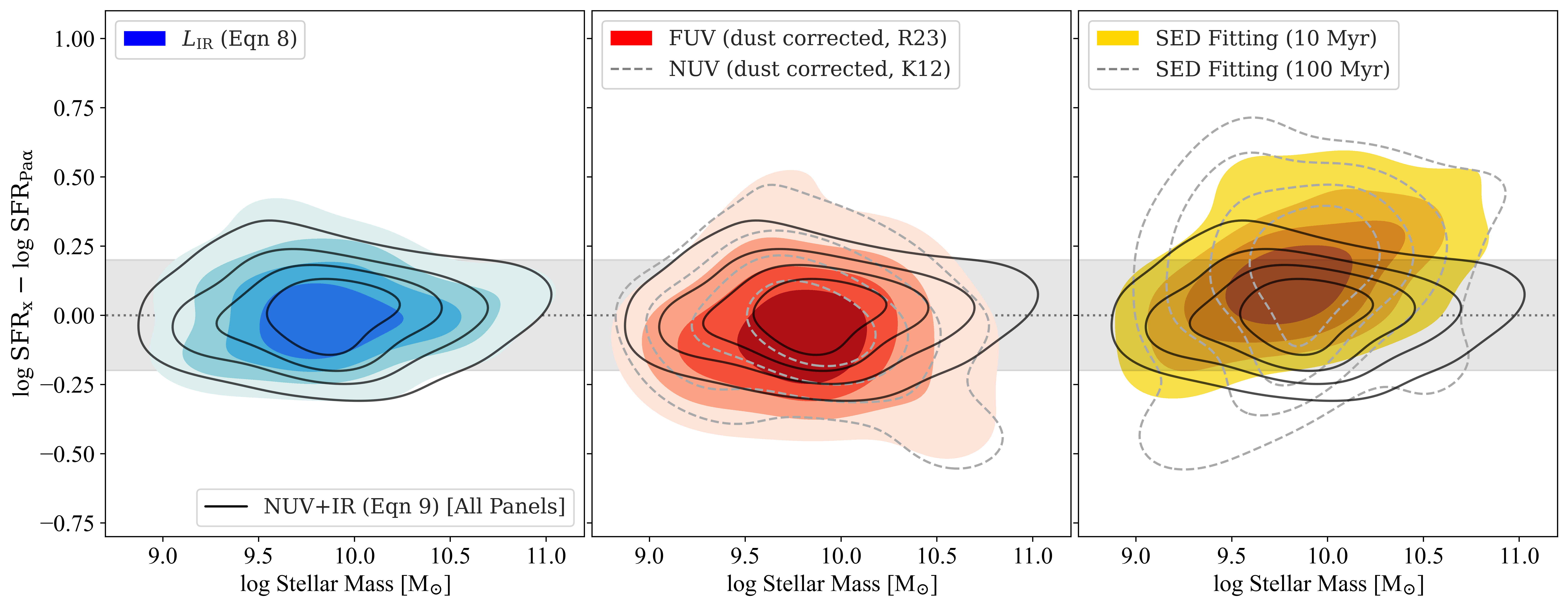}
    \caption{The residuals between SFRs calculated through different methods and the fiducial \Paa\ SFRs.  In all panels the residuals for the UV+IR composite SFR calibration (Eqn~\ref{eqn:uv_mir_sfrs}, Table~\ref{tbl:fit3}) are shown as solid black contours. (left) The blue color contours show the $L_{\rm IR}$-based SFR (Eqn~\ref{eqn:sfr_mir2}, Table~\ref{tbl:fit2}). (middle) The red color and dashed line contours show dust corrected UV-based SFRs from the FUV \citep{reddy2023} and NUV \citep{kennicutt2012}.  (right) The yellow color and dashed line contours show the best-fit SFRs from {\tt Prospector} over timescales of 10 and 100 Myr.} 
    \label{fig:scatter}
\end{figure*}

In Figure~\ref{fig:scatter}, we again show the residuals for MS galaxies between our fiducial \Paa\ SFRs for our $L_{\rm IR}$ and UV+IR composite SFR calibrations (Table~\ref{tbl:fit2}--\ref{tbl:fit3}) and compare them to residuals for dust-corrected UV-based SFRs and the SFRs reported our SED fitting (Section~\ref{sec:sed_fitting}).  For the UV, we correct for dust by determining $A_{\rm UV}$ from the best-fit attenuation curves from SED fitting.  We verify that this correction is in good agreement with applying the average SMC attenuation curve from \citet{gordon2024}, which has been found to well describe the UV attenuation in \Paa\ emitters \citep{reddy2023a}.  We show two UV-based SFR calibrations: 1) for rest $1600$ \AA, we apply the metallicity-dependent FUV-SFR conversion from \citet{reddy2023} and 2) at rest $2300$ \AA the NUV-SFR conversion from \citet{kennicutt2012}. We find that the UV-based SFRs have similar scatter to the $L_{\rm IR}$-based SFRs ($\sim0.2-0.3$ dex).  The FUV shows signs of being slightly systematically underestimated by $\sim0.05$ dex, potentially related to the assumed attenuation corrections. In the rightmost panel, we additionally show the SFRs derived from SED fitting \citep[e.g.,][]{leja2019a} over timescales of 10 and 100 Myr. The former is comparable to the timescales traced by \Paa\ and the latter with the UV and IR, modulo variations in SFH \citep[e.g.,][]{floresvelazquez2021}.  Both appear overestimated (by $\sim0.1$ dex) with the SFR$_{\rm 10 Myr}$ estimates being more overestimated at higher masses and having an overall scatter of 0.3 dex.  The scatter for SFR$_{\rm 100 Myr}$ is comparable to the UV and IR-based measurements, all of which exceed the scatter for the composite UV+IR SFRs ($\sim0.15$ dex).   

\section{Discussion}\label{sec:disc}

In this work, we have combined the unprecedented sensitivity of JWST/MIRI imaging with NIRCam's access to the near-infrared \Paa\ emission line, the gold standard in SFR indicators, to assess the use of the mid-IR as a SFR tracer in typical, main sequence galaxies at cosmic noon.  We find that the rest-frame $8 \mu$m luminosity $-$ as probed by the MIRI F1800W and F2100W bands over $1\lesssim z \lesssim1.7$ $-$ declines steeply at SFR$_{\Paa}\lesssim8\,\Msun$ yr$^{-1}$, such that $L_{\rm 8\mu m, rest}\propto$ SFR$_{\rm \Paa}^{2.2\pm0.3}$ (Section~\ref{sec:paa_lmir}). This is in sharp contrast to the unity-slope relation between the mid-IR and SFR in high mass, high metallicity galaxies at  $1\lesssim z\lesssim3$ \citep[e.g.,][]{rujopakarn2013, shipley2016}.

Fitting our full calibration sample of MS galaxies with a broken power-law with a unity-slope above $8\,\Msun$ yr$^{-1}$ results in a well behaved $L_{\rm 8\mu m, rest}$-SFR two-component function with a scatter of $0.33$ dex (Figure~\ref{fig:lmir}).  Likely contributing to this scatter are differences in the transmission curves of the MIRI filters and potential contributions from the rapidly changing spectral features over our redshift range (Figure~\ref{fig:dist}), which we have not controlled for in this first analysis. An addition source of scatter may be the combination of multiple sources of emission (i.e. PAH, hot dust continuum, stellar continuum, and/or AGN) contributing to the MIRI photometry, which we will not attempt to disentangle in this study.  Inverting this relation, we provided purely empirical SFR calibrations based on the observed F1800W and F2100W photometry (Eqn~\ref{eqn:sfr_mir}, Table~\ref{tbl:fit1}).

In Section~\ref{sec:uv_ir}, we expand our calibration to use  MIRI single-band photometry to predict the total infrared luminosity, $L_{\rm IR}$, based on dust emission templates.  We find that a single representative template can produce an $L_{\rm IR}$-SFR relation with slightly lower scatter than
the $L_{\rm 8\mu m, rest}$-SFR relation that makes no assumptions about the SED shape; the implications of this are discussed further below.  We presented the SFR calibrations using $L_{\rm IR}$ derived from single-band MIRI photometry in F1280W, F1500W, F1800W, and F2100W in Table~\ref{tbl:fit2}.  These calibrations are two-component and assume a breakpoint in SFR or, equivalently, stellar mass.  The latter assumption we review as part of a broader discussion on the $L_{\rm IR}$-SFR relation in MS galaxies in the next section.  From the single-band $L_{\rm IR}$ predictions, we further construct a composite SFR indicator \citep[e.g.,][]{kennicutt2012} combining the UV based on SED fitting constrained by HST photometry with the $L_{\rm IR}$, calibrated against the \Paa\ SFRs (Eqn~\ref{eqn:uv_mir_sfrs}, Figure~\ref{fig:lir1}).  Like previous composite SFR indicators, we find our data is well fit by a linear combination of UV+IR, with a tight scatter of $<0.2$ dex.  

These results demonstrate that, although the mid-IR does display different behavior in less massive and less metal-rich MS galaxies at cosmic noon, single-band imaging of the rest frame $8\mu$m or $6\mu\,$m (Section~\ref{sec:application}) spectral regions with JWST/MIRI can be used as a robust SFR indicator down to log $\logM\sim9$.   
Our parameterization of the composite UV + IR SFR follows the well-established literature and indicates that the energy balance arguments underlying that functional form still hold. 

In the following section, we examine the behavior we're observing in the mid-IR in terms of the PAH-specific and global dust properties of our galaxies.  Then in Section~\ref{sec:scatter}, we take a closer look at how our calibrations perform in terms of non-MS populations and outliers.

\subsection{Photometric SFR Calibration at Rest $8\,\mu$m from Local to Cosmic Noon Galaxies}\label{sec:calibration}

Given the relative sensitivity and spatial resolution compared to the far-IR, mid-IR SFR tracers have been well-studied in both local and high mass, high SFR galaxy populations up to cosmic noon.  The rest-frame $8\,\mu$m spectral region has been particularly targeted due to the convenient coverage by Spitzer IRAC Ch4 locally and MIPS24 at $z\sim2$ \citep{diaz-santos2008}.  Broadband measurements, as used in this work, combine PAH emission with continuum emission from hot dust, older  stellar populations, and AGN, if present. PAHs scale with the total dust emission, though with a scatter that depends on metallicity, ionization state, the SFR surface density, and age distribution of the stellar population \citep[e.g.,][]{helou2001,  alonso-herrero2004, engelbracht2005, engelbracht2008, madden2006, draine2007, smith2007, galliano2008, elbaz2011, magdis2013, egorov2023, pedrini2024}  For integrated studies of massive, luminous galaxies without AGN, this variation is minimal \citep[up to ULIRGs;][see Section~\ref{sec:ulirgs}]{pope2008, wu2010, fiolet2010, pope2013, shipley2016} and thus the rest $8\mu$m can be used as an (obscured) SFR tracer \citep{alonso-herrero2006, farrah2007, pope2008, treyer2010, pope2013, rujopakarn2013, kennicutt2009, shipley2016}, with the obscured SFR component found to provide a large portion ($\gtrsim80\%$ at log $\logM\gtrsim 10$) of the total SFR, roughly independent of redshift up to $z\sim2.5$ \citep{whitaker2017, zimmerman2024}.  

In this work, we have shown a steep decrease in the $8\mu$m emission of MS galaxies at $z\sim1.3$ below SFR$\,\sim8\,\Msun$ yr$^{-1}$, a departure from the behavior of previous IR-based SFR indicators.  For our SFR calibrations to be applicable over a wide redshift range, we need to now understand what drives this break and decline and whether our calibration needs to evolve with redshift.  We can easily rule out that SFR is the fundamental property setting the break; unity-slope relations between $L_{\rm 7.7PAH}$ and SFR have been shown over the range $\sim0.3-30\,\Msun$ yr$^{-1}$ in more massive galaxies at $z\sim0.4$ \citep{shipley2016}.  The next logical option is stellar mass and, specifically, its correlation with metallicity.  Studies of metal-poor HII regions and low-$Z$ local galaxies \citep{calzetti2007a, engelbracht2005, engelbracht2008, smith2007, hunt2010, remy-ruyer2015, chastenet2019, aniano2020} and $z\sim2$ \citep{shivaei2024} galaxies have found deficits in the rest-frame $8\mu$m, often attributed to a decrease in PAH abundance relative to $L_{\rm IR}$ at sub-solar metallicities.   

However, stellar mass is also positively correlated with global dust properties, such as dust obscuration. In the next two sections, we consider whether the behavior we see in the $L_{\rm 8\mu m}$-SFR relation is better described by a decrease in PAH abundance specifically, or a lower (global) dust obscuration, and discuss the implications for our calibration. 

\subsubsection{MIR dust emission at low stellar mass (log $\logM<10$) and metallicity}

The deficit of PAHs at low metallicity is well established in the literature, though the details remain unclear.  Some studies find that the PAH abundance drops sharply at $\sim0.2-0.3\,\Zsun$ \citep{engelbracht2005, engelbracht2008, smith2007, draine2007, chastenet2019, shim2023} while others find a more continuous decline at $\sim0.5-0.7\,\Zsun$ \citep{galliano2008, remy-ruyer2015, aniano2020, whitcomb2024, shivaei2024}.  In this work, we see a decline in $L_{8\mu m, rest}$ beginning at SFR$=8\,\Msun$ yr$^{-1}$, which for a MS galaxy at $z\sim1.3$ corresponds to log $M/\Msun\sim10.1$.  Adopting the mass-metallicity relation at this redshift from \citet{topping2021} and the O3N2-metallicity calibration from \citep{bian2018}, this corresponds to a metallicity of $\sim0.7\Zsun$.  This appears consistent with studies that find a continuous decline. 

However, it's important to remember that this decline may not effect all of the PAH features equally and we are particularly interested in the 6.2 and 7.7$\mu$m PAHs.  A lower PAH abundance is often attributed to either photodestruction of small grains in e.g. hard radiation fields or inhibited large grain growth due to lack of metals.  Using Spitzer IRS spectroscopy, \citet{smith2007} observed PAH feature ratios consistent with power shifting from long to short wavelength PAH bands with decreasing metallicity, indicating that large grain growth is inhibited at low $Z$.  The 6.2 and 7.7$\mu$m features, comprised of relatively smaller grains, were less effected \citep[see also][]{hunt2010, sandstrom2012}.  In three nearby spirals, \citet{whitcomb2024} confirmed that while the general trend is of decreasing total PAH luminosity with decreasing metallicity, the most rapid decrease is in the large grain $17\mu$m PAH feature, with the $7.7\mu$m PAH showing a more gradual decrease and the $6.2\mu$m PAH actually showing a modest increase. 

A decrease in the abundance of the grains that create the $7.7\mu$m emission complex would imply that our breakpoint and decline is tied directly to metallicity.  As a result, it would evolve with redshift following the mass-metallicity relation.  Our analysis, however, suggests this is not the case.  In Section~\ref{sec:uv_ir}, we derived a UV+$L_{\rm IR}$ SFR calibration using the common functional form (Eqn~\ref{eqn:uv_mir_sfrs}), which is built on the assumption of energy balance between UV photons associated with star formation and dust attenuation. Crucially, we derived the $L_{\rm IR}$ using a single IR template with a fixed $L_{\rm 8\mu m, rest}$ to $L_{\rm IR}$ ratio (Section~\ref{sec:mir_sfrs}).  If the 7.7PAH to $L_{\rm IR}$ ratio was decreasing smoothly, we would expect to systematically underestimate the $L_{\rm IR}$ by factors of $\gtrsim2$ at the low mass end of our calibration sample.  This would violate energy balance and not result in the linear UV+IR relation we derive.

\subsubsection{Global dust emission at low mass: the obscuration fraction sets the breakpoint}\label{sec:breakpoint}

\begin{figure*}[htb!]
    \centering
    \includegraphics[width=0.9\textwidth]{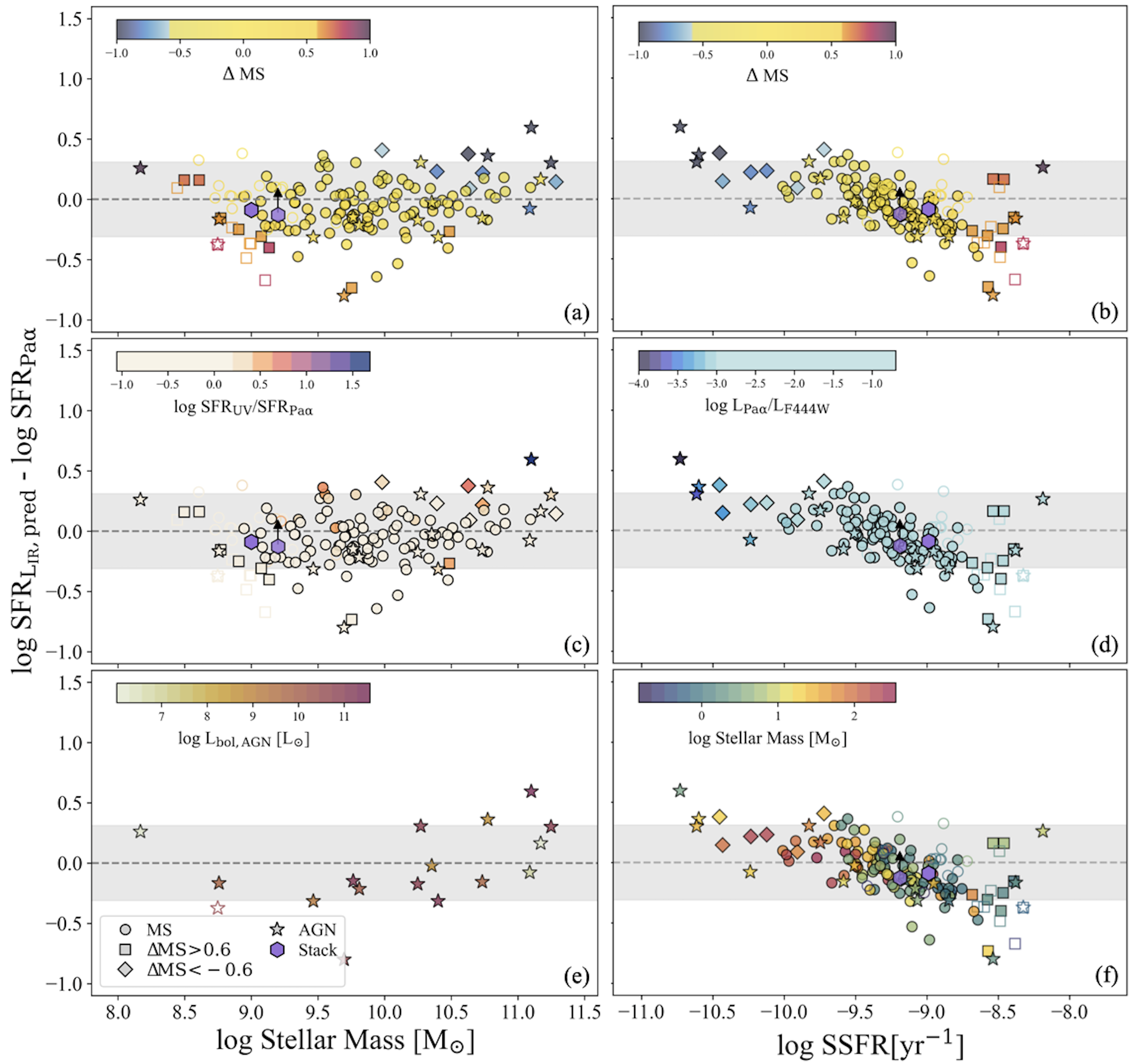}
    \caption{The residuals in dex between the predicted SFR from our $L_{\rm IR}$-based SFR calibration (Eqn~\ref{eqn:sfr_mir2}, Table~\ref{tbl:fit2}) as a function of stellar mass (left panels) and SSFR (right panels).  Solid (open) symbols are detections (SNR$\,<3$) in F1800W.  Circles, squares, diamonds, and stars show MS, above MS ($\Delta$MS$>0.6$), below MS ($\Delta$MS$<-0.6$), and AGN, respectively.  The shaded regions show a scatter of 0.3 dex, determined using the MS galaxy sample.  Different galaxy properties and potential sources of scatter are highlighted in different panels: (a-b) galaxies are colored accounting to whether they are on, above, or below the MS; (c) highlights sources with excess UV defined as SFR$_{\rm 2300\AA}$/SFR$_{\Paa}>2$ (Section~\ref{sec:uv_excess}); (d) highlights sources with low \Paa\ to continuum luminosities, measured in the F277W filter (Section~\ref{sec:evolved}); (e) shows AGN and their bolometric luminosities (Section~\ref{sec:agn}); and (f) shows the trend with SSFR while highlighting the stellar mass.  } 
    \label{fig:scatter_lir}
\end{figure*}

Given the discussion above, we next consider the global dust properties $-$ dominated by large grains in thermal equilibrium $-$ at low metallicity.  Global dust properties are, of course, also sensitive to metallicity, which will lower the dust-to-gas ratio and cause changes of the radiation field, which affects the dust equilibrium temperature.  A curious thing, then, is that it has been established that the obscuration fraction, $f_{\rm obs}$, at fixed stellar mass has weak or no evolution with redshift up to $z\sim3$ \citep{bouwens2016, whitaker2017, mclure2018, shapley2022}. Recent cosmological simulations \citep{zimmerman2024} suggest that $f_{\rm obs}$ actually does increase with redshift (at fixed mass), which could potentially counter the effects of decreasing metallicity and create the illusion of weak or no evolution.  This increase is attributed to changing star-dust geometry \citep{zimmerman2024}, as the well characterized $M_{\rm dust}-M_{\star}$ relation and gas-to-dust ratio-metallicity relation doesn't seem to evolve with redshift \citep{popping2022, zhang2023}.   

In Figure~\ref{fig:fobs} (left), we show the $f_{\rm obs}$ derived from our UV+IR calibration as a function of stellar mass.  As established in previous works \citep[e.g.,][]{whitaker2017}, we see an asymptote toward $f_{\rm obs}\sim0.8-1$ at high masses, followed by a steep decline with large scatter at fixed stellar mass. Via the colorbar and inset, we show the infrared excess, IRX$\,\equiv L_{\rm IR}/L_{\rm UV}$, for our sample.  IRX is known to be tightly correlated with stellar mass \citep[e.g.,][]{bouwens2016} and simulations confirm that IRX is a strong proxy for the effective UV opacity, which is set by the column density of dust relative to young stars \citep{popping2017, narayanan2018, liang2021c}.  Together IRX and dust mass, also tightly correlated with stellar mass, set $f_{\rm obs}$.

The drop in $L_{\rm 8\mu m, rest}$ following the drop in the global $f_{\rm obs}$ naturally sets a non-evolving breakpoint in stellar mass through the $f_{\rm obs}$-$M_{\star}$ relation up to $z\sim3$.  Global dust properties driving this trend as opposed to a drop in the 7.7$\,\mu$m (or $6.2\,\mu$m) PAH abundance is consistent with our assumption of a single dust template and our UV+IR composite SFR, which obeys energy balance.  It further supports our use of the $6.2\,\mu$m PAH to extend the redshift range for MIRI-based SFRs to $\sim3$ and suggests that our calibration is not overly sensitive the fraction of the 6.2 or 7.7$\mu$m captured within a given MIRI filter. As such, we conclude that our MIRI-based SFR calibration (Eqn~\ref{eqn:sfr_mir2}, Table~\ref{tbl:fit2}) can be applied to MS galaxies over a wide redshift range.

\subsection{Understanding the Outliers and Broader Applicability of MIRI-based SFRs}\label{sec:scatter}

In Figures~\ref{fig:scatter_lir} and \ref{fig:scatter_uv_lir}, we show the residuals of our predicted SFRs relative to SFR$_{\rm Pa\alpha}$ from our $L_{\rm IR}$ (Eqn~\ref{eqn:sfr_mir2}) and UV+IR (Eqn~\ref{eqn:uv_mir_sfrs}) calibrations, respectively.  Predicted SFRs are calculated using the F1800W photometry and the FUV from our SED fitting.   These residuals are shown as a function of stellar mass (left) and the instantaneous SSFR (right), calculated using SFR$_{\rm Pa\alpha}$.  The top two panels (a-b) display the full population $-$ MS, above/below MS, and AGN $-$ while the remaining panels (c-f) highlight different populations and potential sources of scatter. 

\subsubsection{(UV+)MIRI-based SFRs in high SSFR and Starburst galaxies}\label{sec:starburst}

In Figure~\ref{fig:scatter_lir}, the residuals between our $L_{\rm IR}$-based SFRs and the fiducial \Paa\ SFRs overall show weak to no trend with stellar mass (panel a) but some dependence on SSFR (panel b), with our SFR calibrations tending to over-predict at low SSFR and under-predict at high SSFR.  Starburst galaxies, defined as $\delta$MS$>0.6$, mostly fall below the scatter we have measured for our MS population.  In Figure~\ref{fig:scatter_uv_lir} (panel b), we see that this trend in the residuals with SSFR is much less obvious among MS galaxies; however, starbursts mostly remain outliers.

SSFR has been found to track the average radiation field, even on galaxy-integrated scales, and is anti-correlated with both $q_{\rm PAH}$ \citep{chastenet2025} and IRX (Figure~\ref{fig:fobs}).  Interestingly, the colorbar in Figure~\ref{fig:scatter_lir} (panel f) shows that the high SSFR and starburst outliers range in mass, and thus metallicity, which is consistent with the lack of strong trend in the residual outliers with mass and emphasizes the role of the SSFR and by proxy the radiation field in setting the $L_{\rm 8\mu m, rest}$/\LIR \citep[e.g.,][]{elbaz2011, mcnulty2026}.  As some of this trend with SSFR weakens and some (namely in the starburst population) remains when the UV is added into the predicted SFR, we posit that both a decrease in PAH abundance due to high radiation fields \citep[e.g.,][]{baron2025} and a change in global dust properties via star-dust geometry could be contributing to the intrinsic MS scatter and this outlier population.

\subsubsection{High predicted (UV+)MIRI-based SFRs in galaxies: UV excess?}\label{sec:uv_excess}

\begin{figure*}[htb!]
    \centering
    \includegraphics[width=0.9\textwidth]{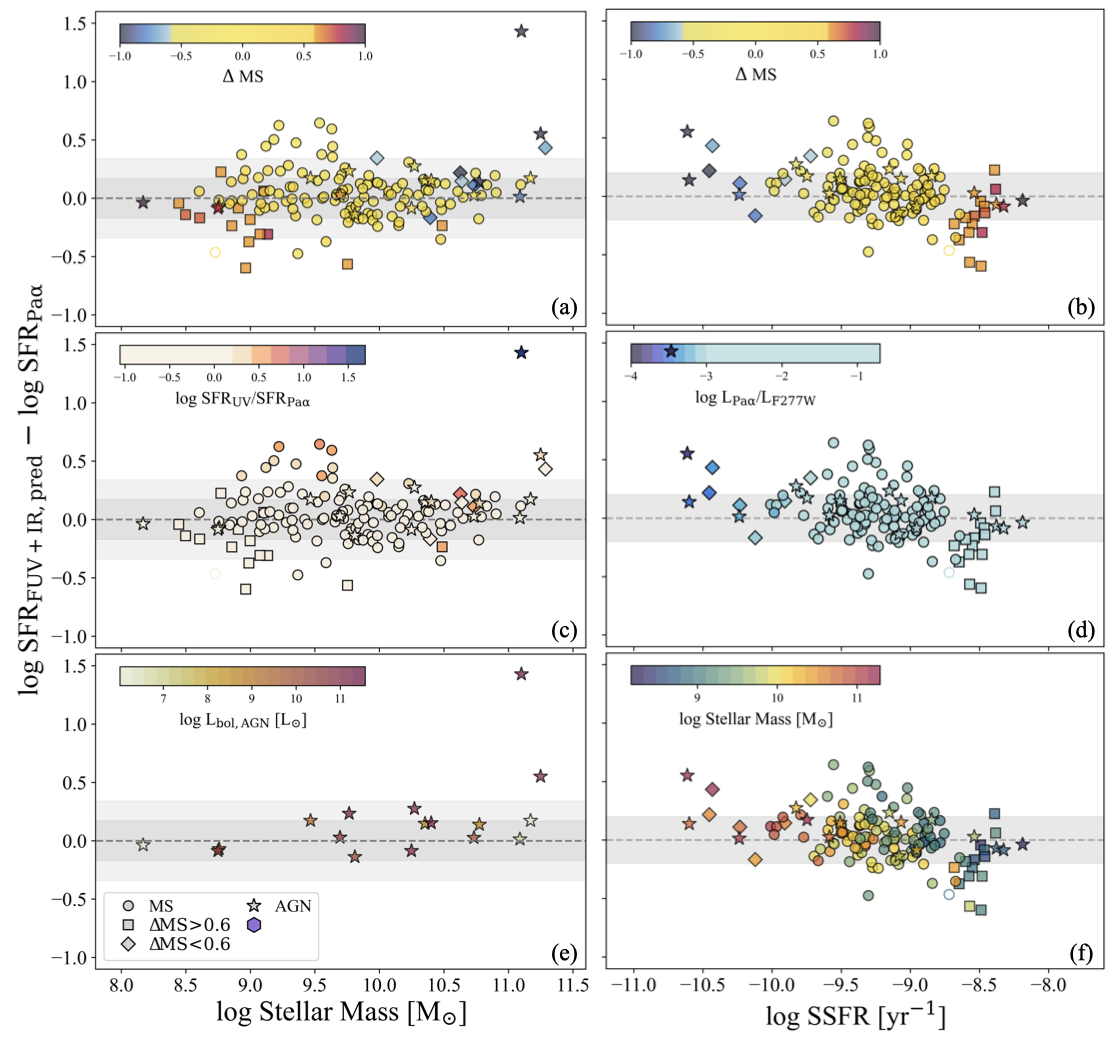}
    \caption{As in Figure~\ref{fig:scatter_lir}, except showing the residuals for the UV+IR based SFRs (Eqn~\ref{eqn:sfr_mir2}, Table~\ref{tbl:fit3}).} 
    \label{fig:scatter_uv_lir}
\end{figure*}

A comparison of Figures~\ref{fig:scatter_lir} and \ref{fig:scatter_uv_lir} highlights a particular set of outliers in our MS calibration sample at log $\logM\sim9-9.5$ that have elevated UV+IR SFRs. In panel c in these two figures, we explore this by looking at the ratio of the SFR derived solely from the \textit{uncorrected} UV, e.g. presumably tracing the unobscured star formation component, versus the \Paa\ SFR.  To derive the UV SFR, we adopt the metallicity-dependent calibration presented in \citet{reddy2023}, which is based on the same models as the their \Paa\ SFR calibration.  We find that these outliers have high UV-based SFRs, $\sim2-5\times$ the \Paa\ SFR\footnote{We only consider UV SFR excesses above \Paa\ SFRs of $>2\times$ to be conservative against calibration uncertainties.}.

A possible explanation is provided by recent simulations with FIRE-2 \citep{floresvelazquez2021}, which found that the FUV in galaxies with constant SFHs can trace short timescales ($\sim10$ Myr) similar to hydrogen recombination lines \citep[$\sim5-10$ Myr;][]{hao2011, floresvelazquez2021} due to the continued renewal of O-type stars that can outshine older populations.  In galaxies with stochastic SFHs, however, the UV can stay elevated for longer timescales \citep[$\gtrsim100$ Myr; see also][]{murphy2011a, kennicutt2012, calzetti2013} following the lull after a burst.   This is consistent with studies finding increased scatter between UV- and Balmer line-based SFRs in low mass galaxies \citep{dominguez2015, guo2016, emami2019, atek2022, cleri2022, reddy2023a, reddy2025}, which are increasingly thought to have stochastic SFHs \citep{navarro-carrera2026}.  This highlights that the timescales and other factors that may influence UV-based SFR calibrations need to be considered, particular as JWST has opened up access to the UV in high-$z$, low mass populations.

\subsubsection{High predicted  (UV+)MIRI-based SFRs in galaxies below the MS: contributions from evolved stellar populations?}\label{sec:evolved}

In the previous section, we discussed galaxies where the predicted UV SFR is higher than the \Paa\ SFR,  potentially owing to the different timescales being traced and stochastic SFHs.  Similarly, the timescale over which the IR traces SFR can have a complex relationship with SFH \citep[][and references therein]{calzetti2025} and, importantly, the IR can additionally include contributions from evolved stellar populations largely unrelated to recent star formation \citep[e.g.,][]{crocker2013, calapa2014, fumagalli2014, hao2011}.
Low $L_{\rm 8\mu m}$ to $L_{\rm \Paa}$ ratios have been found to be associated with smaller \Paa\ EWs in local, resolved studies, signaling older stellar ages \citep{diaz-santos2008}. In Figures~\ref{fig:scatter_lir} and \ref{fig:scatter_uv_lir} [panels a, b], we see that galaxies below the MS ($\Delta$MS$<-0.6$) $--$ which we note do not satisfy UVJ quiescent galaxy selection and so are not classified as quenched $--$ tend to inhabit the upper end of our scatter, though only one is a $>2\sigma$ outlier in the UV+IR SFRs.  In panel d, we show that low SSFR sources in our sample have  
low \Paa\ equivalent widths, $3-10\times$ lower than the median ratio.  This suggests our overestimated  predicted SFRs in low SSFR sources are boosted by contributions from evolved stars in the IR (or AGN, see Section~\ref{sec:agn}).  However, our subsample of sub-MS galaxies is generally within our scatter and small, including only 6 galaxies.  Among MS galaxies, we cannot rule out excess IR emission from evolved populations contributing to our overall scatter 
and indeed this could contribute to the correlation we see between $L_{\rm IR}$-based SFRs and SSFR in Figure~\ref{fig:scatter_lir}.

\subsubsection{(UV+)MIRI-based SFRs in AGN}\label{sec:agn}

In Section~\ref{sec:mir_sfrs}, we removed AGN from our calibration sample as AGN emission can contaminate the \Paa\ emission line \citep[e.g.,][]{sun2025} as well as the UV and mid-IR continuum.  For moderately luminous AGN, however, the host contribution may still dominate and in highly obscured AGN, the contributions to the UV, optical, and mid-IR can be minimal \citep{kirkpatrick2015, lyu2024, kirkpatrick2023}.  In these cases, our calibration may still provide a first order estimate of the star formation rate.

In Figures~\ref{fig:scatter_lir} and \ref{fig:scatter_uv_lir} (panel e), we examine the residuals between our $L_{\rm IR}$- and UV+IR-based SFR calibrations and the \Paa\ SFRs for previously identified AGN (Section~\ref{sec:sample}).  We find that most of our AGN fall in the expected scatter for non-AGN MS calibration sample.  We have two significant ($>2\sigma$) outliers with high predicted UV+IR SFRs (Figure~\ref{fig:scatter_uv_lir}).  Both are luminous X-ray AGN \citep{lyu2022a} in massive hosts. The most extreme outlier was additionally identified in \citet{sun2025} as having a broad component to its \Paa\ line.  Their selection in the X-ray and optical indicates they are unobscured AGN, which is consistent with their UV+IR SFRs being significantly more overestimated than their $L_{\rm IR}$-based SFRs. In our sample, five additional X-ray-selected AGN in more intermediate mass hosts fall within our calibration's scatter.

We additionally have one outlier with a low $L_{\rm IR}$-based SFR.  This AGN was identified in the mid-IR \citep{lyu2024} and via its X-ray to radio ratio \citep{alberts2020, lyu2022a} and has an AGN bolometric luminosity of log $L_{\rm AGN}/\Lsun=10.8$.  Its UV+IR SFR, however, is again within our scatter. At a more intermediate mass (log $M_{\star}/\Msun=9.8$), this AGN may have a host weak in the IR and/or the AGN could be contributing to the \Paa\ and UV.  Disentangling these is beyond the scope of this work.

In general, our SFR calibration provides SFR estimates in AGN candidates that are still within the scatter derived from the non-AGN calibration sample for the bulk of these AGN (14 out of 17).  Given this, we suggest that our calibration can be applied to AGN-hosting galaxies and that comparing $L_{\rm IR}$-based and UV+IR SFRs can help identify outliers.

\section{$\Paa$ and IR-based SFRs in Luminous Infrared Galaxies}\label{sec:ulirgs}

MIRI's capabilities now give us unprecedented coverage of the mid-IR beyond the local Universe, allowing us to individually detect galaxies where previously we had to rely on stacking.  In the previous sections, we calibrated MIRI- and UV+IR-based SFR indicators against the \Paa\ emission line in MS galaxies at cosmic noon.  Our sample was comprised of galaxies with low- to moderate-infrared luminosities (log $L_{\rm IR}/\Lsun\sim9-11.5$), a result of the relatively small MIRI FOV compared to the surveying power of e.g. Spitzer. As such, we relied on previous studies to parameterize our calibration at higher luminosities and SFRs.  However, as MIRI (and grism) surveys accumulate more area \citep[for example, the PRIMER and MINERVA MIRI surveys will cover over 250 arcmin$^2$, see][]{muzzin2025}, we will be able to examine the bright populations previously probed during the Spitzer era with the finer detail provided by MIRI's sensitivity and resolution.  Here we combine data from previous missions like Spitzer and Herschel with  new JWST observations to discuss \Paa\ and IR-based SFRs in galaxies with log $L_{\rm IR}/\Lsun\gtrsim11-13$.
%, termed LIRGs and ULIRGs. 

\subsection{PAH behavior at high luminosities}
\label{PAHhighz}

\begin{figure}[htb!]
    \centering
    \includegraphics[width=0.48\textwidth]{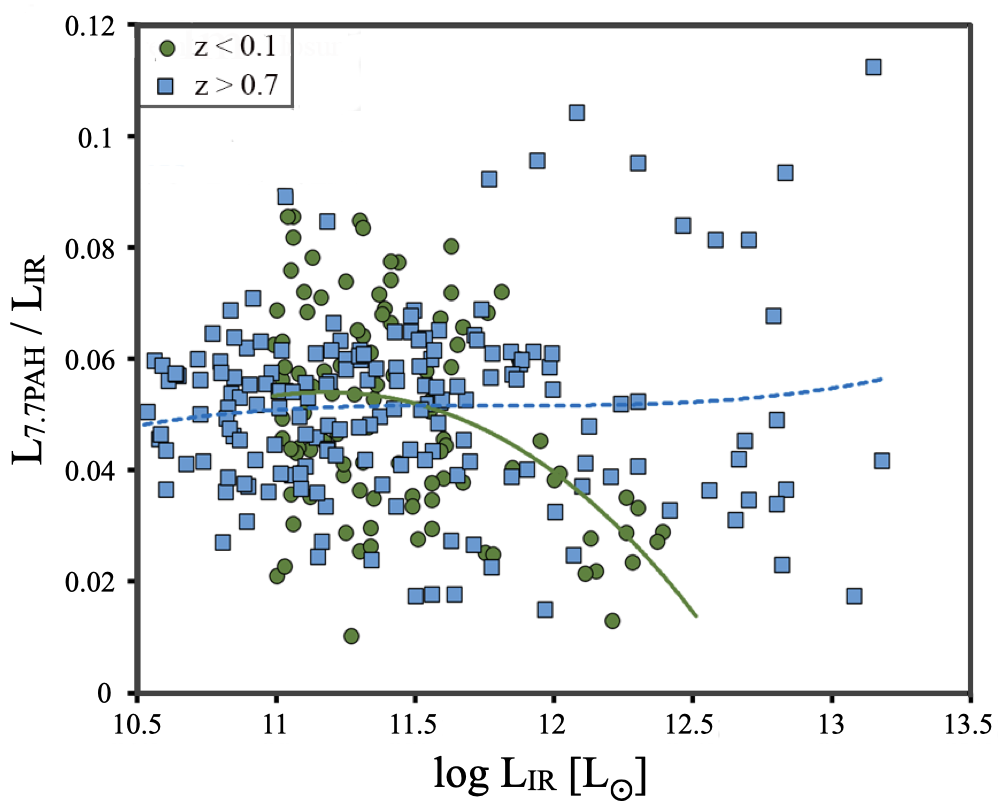}
    \caption{Ratio of the 7.7 $\mu$m PAH luminosity to the total infrared luminosity. The green dots are for galaxies at z $<$ 0.1, and the green solid line is a 3rd order fit to them. The blue squares are for galaxies at $z > 0.7$ and the blue dashed line is a 3rd order fit. }
    \label{fig:pahltir}
\end{figure}

The ratio of PAHs to the total infrared luminosity is known to drop sharply at high $L_{\rm IR}$, starting at a threshold of log $L_{\rm IR}/\Lsun\sim12$ locally 
\citep{shipley2016}.  This implies that PAHs may no longer track the obscured SFR at these luminosities.  However, quantifying and explaining this drop has historically been challenging due to potential contamination from heavily obscured AGN.
New analyses with JWST now allow us to revisit this issue based on improved tools for quantifying the contribution from AGN to the mid-IR in these extreme galaxies. In \citet{rieke2025}, NIRSpec and MIRI IFU observations of four local ULIRGs put strong upper limits on high excitation AGN lines such as [NeVI], indicating that AGN in local ULIRGs can be so deeply embedded and/or their accretion so intermittent that the AGN is not contributing to the mid-IR spectrum.  An examination of the dynamics and black hole masses of a larger ULIRG sample found a large fraction have low Eddington ratios.  Taken together, this indicates that, contrary to previous models, the mid-IR in many ULIRGs is powered by star formation, allowing us to examine the PAH deficit in this context.

Based on the analysis in \citet{rieke2025} outlined above, we re-examine the PAH abundance in local and high-z LIRGs and ULIRGs to assess the mid-IR as a SFR indicator. In Figure~\ref{fig:pahltir}, we show the $L_{\rm 7.7PAH}$/$L_{\rm IR}$ ratio from the GOALS sample \citep[][]{stierwalt2014} at $z<0.1$. The 7.7$\,\mu$m PAH fluxes were derived from Spitzer IRS spectra and are equivalent to those measured using the standard PAH modeling tool PAHFIT \citep{smith2007}.  We take the total infrared luminosities derived from IRAS photometry from \citet{sanders2003}. We note \citet{u2012} demonstrated that adding longer wavelength data does not significantly modify the $L_{\rm IR}$ estimates. We discard measurements of close pairs, where the Spitzer/IRS spectra and the IRAS photometry may not refer to the same region/galaxy. To be conservative, we also discard cases where an AGN has plausibly affected the PAH features, based on the full SED.  

Our compilation of local galaxies agrees with the trends found in previous studies \cite[e.g.,][]{shipley2016, lai2020}.  Namely, in local ULIRGs, the average  $L_{\rm 7.7PAH}$/$L_{\rm IR}$ ratio is about half the average value for the $ 11 < \mathrm{log}~L_{\rm IR}/\Lsun < 11.5$ sample. This drop apparently starts at log $L_{\rm IR}/\Lsun \sim 11.5$ and becomes dramatic for log $L_{\rm IR}/\Lsun > 12$, albeit with significant scatter. The behavior over the full range of luminosity appears to be relatively well behaved, making it tempting to fit the trend\footnote{
For $z < 0.1$, 
\begin{equation}\label{eqn:ulirgs}
    \frac{L_{\rm 7.7PAH}}{L_{\rm IR}} = -0.0025 * w^3 + 0.0646* w^2 - 0.5058*w + 1.131
\end{equation}
\noindent
% and for $z > 0.7$,
% \begin{equation}
%     \frac{L_{\rm PAH}}{L_{\rm IR}} = 0.0018 * w^3 - 0.0626* w^2 + 0.7380*w  - 2.8466
% \end{equation}
% \noindent
where $w = \log L_{\rm IR}/\Lsun$. }
as a correction term for local ULIRG measurements.

This behavior is known to be substantially reduced at even modest redshift \citep{rujopakarn2013} with a roughly constant ratio reported up to log $L_{\rm IR}\sim12.5$ at $z\gtrsim0.7$ \citep{shipley2016}.  To construct an improved high redshift comparison sample, we proceed as follows, with results plotted in Figure~\ref{fig:pahltir}. For the lower luminosities, log $L_{\rm IR}/\Lsun\lesssim11.5$, we again take advantage of SMILES over a larger redshift range, adopting values for $L_{\rm 7.7PAH}$ and \LIR\ from \citet{shivaei2024}.  These were estimated through full UV to IR SED fitting, assuming energy balance. We note that the priors used in \citet{shivaei2024} may be too restrictive to fully model the mid-IR dust \citep{mcnulty2026} and only a small fraction of sources had far-IR constraints on the $L_{\rm IR}$.  As such, we do not attempt to interpret the scatter in the measurement of the ratio, we only assess the qualitative trend in the $L_{\rm 7.7PAH}$/\LIR\ ratio with increasing infrared luminosity.  To compare this and the local sample with a consistent calibration, we renormalize such that the \citet{shivaei2024} sample's average ratio matches the best-fit average for local galaxies from \citet[][their Eqn~21]{shipley2016}.

For the higher luminosities, we take values from the literature thought to be from star formation-dominated galaxies \citep{sajina2007, rigby2008, huang2009, pope2013, shipley2016}. All PAH measurements were made in a way consistent with the PAHFIT approach.  Those reported relative to the local continuum were corrected by a factor of 1.7 as found compared to a spline-based decomposition \citep[e.g.,][]{pope2013}  There are a number of other samples with Spitzer IRS spectra, but they were not included if they used the PAHs as their primary means to identify AGN \citep[e.g.,][]{Dasyra2009, fiolet2010}. Nonetheless, there are likely some cases in our high luminosity sample with significant AGN power, which would lower their $L_{\rm 7.7PAH}$/\LIR\ ratio. Since Figure~\ref{fig:pahltir} illustrates that this sample on average has larger values of this ratio than do local high-luminosity galaxies, this contamination would have no effect on our overall conclusion.

As shown in Figure~\ref{fig:pahltir}, there is no clear drop in the ratio up to log $L_{\rm IR}/\Lsun \sim 13$.  In addition to measurement uncertainties, one caveat is that some of the apparently very high luminosity cases may be lower luminosity galaxies amplified by gravitational lensing \citep{spilker2016}. This  makes straightforward determination of their luminosities and star formation rates problematic, muddling the interpretation at the very highest luminosities.  As such, we conclude there is little evidence for PAH suppression in ULIRGs at cosmic noon, but more sources and more robust measurements are still needed.

\begin{deluxetable*}{lcccc|cc|cccc}\label{tbl:ulirgs}
\tabletypesize{\footnotesize}
\tablecaption{Star formation rates from infrared spectral features}
\tablehead{
\colhead{Galaxy} &
\colhead{$f_{\rm Pa\alpha}$} &
\colhead{$f_{\rm Pf\gamma}$} &
\colhead{$f_{\rm Br\alpha}$} &
\colhead{$f_{\rm 7.7PAH}$} &
\colhead{log $L_{\rm Pa\alpha}$} &
\colhead{log $L_{\rm Pa\alpha}$} &
\colhead{SFR$_{\rm Pa\alpha}$} &
\colhead{SFR$_{\rm 7.7PAH}$} &
\colhead{SFR$_{L_{\rm IR}}$} \\
\colhead{} &
\colhead{} &
\colhead{} &
\colhead{} &
\colhead{} &
\colhead{Fgnd.\ Screen} &
\colhead{Mixed} &
\colhead{} &
\colhead{} &
\colhead{} 
}
\startdata
Arp 220   & 15.16 & 9.23    & \nodata & $3.73\times10^{5}$ & 37.48 & 43.0  & 16   & 72  & 244 \\
IRAS 14378 &  2.76 & \nodata &  1.11   & 3612               &  5.77 &  5.99 & 41.5 & 27  & 212 \\
IRAS 17208 &  2.38 & \nodata &  1.67   & $2.42\times10^{5}$ & 10.59 & 14.4  & 39   & 65  & 369 \\
IRAS 23365 &  1.68 & \nodata &  0.76   & 4702               &  4.11 &  4.38 & 28   & 32  & 203 \\
\enddata
\tablecomments{Columns $2-5$: Flux densities are given in $10^{-17}$ W m$^{-2}$.  Hydrogen recombination line fluxes are from \citet{perna2024} for Arp 220 and \citet{rieke2025} for the IRAS sources.  The $7.7\,\mu$m PAH flux is from \citet{stierwalt2014}.  Columns $6-7$: extinction-corrected \Paa\ luminosities given in W for a foreground sceen and mixed geometry, respectively. Columns $8-10$: SFRs are given in $M_\odot$ yr$^{-1}$ (see Section~\ref{recombination} for details). }  
\end{deluxetable*}

\subsection{The behavior of \Paa \, (and $L_{\rm PAH}/L_{\rm IR}$) in local ULIRGs with JWST}
\label{recombination}

Throughout this work, we have assumed \Paa\ is a ``gold standard'' of SFR indicators, as initially described in detail in \citet{alonso-herrero2006} and \citet{calzetti2007a}.  These works found that at high luminosities, SFRs from IR increased faster than that from \Paa\, as quantified in \citet{rieke2009} for log $L_{\rm IR}\gtrsim$ 11.  Two scenarios emerged: \citet{alonso-herrero2006} ascribed this behavior to the dust in the most luminous local galaxies (which contain the densest star forming regions) being able to absorb a higher percentage of the ionizing photons from young stars. 
In this explanation, the IR gives a more reliable measure of the SFR than \Paa.  \citet{calzetti2007a} instead suggested that the dust is hotter in bright galaxies, increasing their output at $24\,\mu$m while \Paa\ continues to trace the SFR. 

With JWST, we can now revisit this issue.  In particular, access to multiple hydrogen recombination lines allows us to correct \Paa\ for extinction, which can become a significant factor in very luminous galaxies.  We look again at the four ULIRGs with JWST/NIRSpec IFU from \citet[][see also \cite{perna2024} and \cite{goldberg2024} for details on the data acquisition and line measurements for Arp 220]{rieke2025}. We correct \Paa\ for extinction using the Br$\alpha$ or Pfund $\gamma$ lines, assuming intrinsic ratios given by case B recombination \citep{hummer1987}\footnote{We assume T$_{e}$=10,000 K and N$_{e}=10^4$ cm$^{-3}$, which are roughly appropriate for the dense gas in a ULIRG nucleus.  The line ratios are slightly dependent on these parameters \citep{hummer1987} but this dependence is not enough to affect our conclusions.} and an extinction law with $R_{\rm V}=3.1$ from \citet{gordon2021}.  

We test two geometries.  The first is the common assumption that the obscured source is behind a foreground screen.  Although this model does not even rise to the level of minimal plausibility for ULIRGs, we list the extinction-corrected  \Paa\ in Table~\ref{tbl:ulirgs} for reference.  Secondly, we use a more plausible geometry where the emitting source and obscuring material are mixed.
In this case

\begin{equation}
F_{\mathrm{obs}} = F_{\mathrm{intrinsic}} \times \frac{1-e^{\tau}}{\tau}
\end{equation}
\noindent where we take $\tau$ to be proportional to the wavelength-dependent extinction in the law used in the screen geometry. As shown in the table, the difference in this case is minor, which is expected because $\tau$ at Pa$\alpha$ is relatively small, $\sim2-6$.
We note the full correction for extinction in ULIRGs is substantially more complex than this approximation \citep{donnan2024}. However, since our approach is equivalent to basing the recombination line flux on Br$\alpha$ (or Pf$\gamma$), even less sensitive to extinction with $\tau \lesssim1-1.5$ 
these complexities are unlikely to increase the estimated fluxes substantially. 

In Table~\ref{tbl:ulirgs}, we list the SFRs derived from the extinction-corrected \Paa\ based on mixed geometry and the conversion rate from \citet{rieke2009}.  Converting first to \Ha\ as in \citet{Calzetti2010} gives consistent results.  We further list the SFRs as derived from the $L_{\rm IR}$ from \citet{sanders2003, u2012}, using the conversion given in \citet{kennicutt2012}.  In all cases, the $L_{\rm IR}$-based SFRs are an order of magnitude larger than those derived from \Paa.

We show similar results for the 7.7 $\mu$m PAH-based SFRs from the results reported by \citet{stierwalt2014}. We have adopted their $L_{\rm PAH}$ and corrected for the relative contribution of the 7.7 $\mu$m feature alone, $\sim$ 56\%.  The estimated SFRs are similar to those from \Paa\, perhaps trending somewhat higher, but still far below the values from total infrared luminosity, as expected. If we apply this same mixed model using the derived values of $\tau$ from Pa$\alpha$ and Br$\alpha$ to the 7.7$\mu$m PAH feature, we find that the reduction in its intrinsic flux is of order a factor of 1.3 (1.8 for IRAS 17208). That is, extinction may contribute  to its reduced flux in these ULIRG but falls short of accounting for the entire effect. 

Returning to the two scenarios, we evaluate the possible explanations for the underestimates of the \Paa\ SFR in ULIRGs.  The critical region to test whether the dust temperature increases with increasing star formation is $\sim10-20\,\mu$m.  However, the shapes of the star formation-powered IR SEDs have been found to be remarkably independent of luminosity in this range \citep{rieke2009, magdis2012}, inconsistent with a change in dust temperature.  In comparison, the phenomenon of the dust in very dense regions competing for the ionizing photons is well understood \citep[e.g.,][]{inoue2001, dopita2003}.  Additional absorption of ionizing photons could also explain the PAH deficits as the small grain PAHs can be destroyed by photodissociation \citep[e.g.,][]{egorov2025}. Taken together, this suggests that long-wavelength IR measurements are the best proxy for SFR in these very infrared-luminous galaxies.

\subsection{Lessons to apply to cosmic noon}

In the two previous sections, we have confirmed the suppression of PAHs seen relative to the $L_{\rm IR}$ in local, highly luminous infrared galaxies \citep[Figure~\ref{fig:pahltir}; e.g.,][]{shipley2016}.  This behavior, however, appears to disappear even at modest redshifts, indicating that MIRI- and generally IR-based SFRs are robust in the most luminous infrared sources at cosmic noon. We additionally looked at the behavior of \Paa\ in four local ULIRGs $-$ taking advantage of JWST's coverage of multiple hydrogen recombination lines $-$ and found that extinction-corrected \Paa\ can underestimate the SFRs by up to an order of magnitude compared to $L_{\rm IR}$ based on long-wavelength IR data. As with the PAHs, however, this behavior may not be common at cosmic noon: in the local ULIRGs, the suppression of both \Paa\ and the PAHs is likely related to the star-dust geometry (as discussed in Section~\ref{sec:calibration}).  Locally, (U)LIRGs host very compact, dense starbursts, while star formation is thought to be more extended during cosmic noon \citep[e.g.,][]{rujopakarn2011}, which has been recently confirmed by a spatially resolved analysis with MIRI \citep{florian2025}.  As such, the MIRI-based SFRs derived in this paper, as well as \Paa\ SFRs, should be accurate and robust over a large range in luminosity and redshift, with the caveat that care must be taken when applying these measures to i.e. compact starbursts or cases with extremely high luminosity densities in their star forming regions.  The latter may be be an important population at $5<z<7$ \citep{derossi2018}, where extremely high densities of UV luminosity may be reached in star forming regions.  These conditions are predicted to be uncommon during cosmic noon (De Rossi et al., 2026, submitted).

\section{Conclusions}\label{sec:conclusions}

In this work, we have examined the behavior of the mid-IR emission in galaxies spanning four order of magnitude in total infrared luminosity ($9\lesssim L_{\rm IR}/\Lsun \lesssim 13$).  Our main analysis extends mid-IR-based SFR indicators into the regime previously limited to stacking analyses in the IR, namely MS galaxies at cosmic noon, based on MIRI photometry in the F1280W, F1500W, F1800W, and F2100W filters. To calibrate, we use fiducial SFRs from the extinction-corrected \Paa\ emission line, a gold standard SFR indicator \citep[e.g.][]{alonso-herrero2006, calzetti2007a}.  Our sample of 169 galaxies at $1<z<1.75$, cross-matched between the SMILES and FRESCO surveys, covers $9\lesssim\mathrm{log}\logM\lesssim11$, reaching a MS SFR of $\mathrm{SFR}\sim1$ $\Msun \mathrm{yr}^{-1}$, over an order of magnitude below the MIPS 24$\mu$m confusion limit at $z\sim1.3$.  Our SFR calibration efforts can be summarized as follows:

\begin{itemize}
    \item Using a clean calibration sample of MS galaxies (Section~\ref{sec:sample}), we examine the behavior of rest-frame $8\,\mu$m emission $-$ dominated by PAHs and probed by MIRI's F1800W and F2100W filters $-$ relative to the \Paa\ emission line.  We find the mid-IR luminosity steeply declines ($\alpha\sim2$) when compared to SFR$_{\Paa}$ below SFR$\sim8\,\Msun$ yr$^{-1}$. This is in sharp contrast to the unity relation between the (mid-)IR and SFR in massive galaxies at cosmic noon \citep[e.g.][]{rujopakarn2013, shipley2016}. We fit our full sample with a two-component, broken power-law (Eqn~\ref{eqn:pl}) with a variable faint end slope and bright end slope fixed to unity.  The latter is necessary based on our small sample size at the bright end, driven by the SMILES area. From this fit, we derive a SFR calibration based purely on F1800W and/or F2100W photometry (Eqn~\ref{eqn:sfr_mir}, Table~\ref{tbl:fit1}) with a scatter of $\sim0.3-0.4$ dex. 

    \item {\bf Our main MIRI-based SFR calibration} (Eqn~\ref{eqn:sfr_mir2}, Table~\ref{tbl:fit2}) builds on the previous analysis by introducing local dust emission templates parameterized by total IR luminosity from \citet{rieke2009} and scaling the single-band MIRI F1800W or F2100W photometry to measure $L_{\rm IR}$.  We find that using a single, representative template reduces the scatter of our SFR calibration to $\sim0.2-0.3$ dex, on par with previous IR-based SFR calibrations. Introducing multiple templates does not improve our calibration, likely due to the shape of the dust SED not being a simple function of $L_{\rm IR}$.  This implies that a fixed $8\mu$m to $L_{\rm IR}$ ratio is, to first order, a reasonable assumption for the galaxies in our sample.

    \item The redshift range over which the F1800W and/or F2100W filters cover the rest-frame $8\,\mu$m is $0.9\lesssim z \lesssim2$.  As such, we test extending our main MIRI-based SFR calibration into the rest-frame $6\,\mu$m region, dominated by the weaker 6.2PAH, using the F1280W and F1500W filters.   We find that the rest-frame 6 and $8\,\mu$m emission display the same behavior and conclude that MIRI photometry covering the rest-frame $6\,\mu$m can be used as a robust SFR proxy. This expands the range of our SFR calibrations up to $z\sim2.8$.

    \item Finally, we construct composite UV+$L_{\rm IR}$ SFR indicators (Eqn~\ref{eqn:uv_mir_sfrs}, Table~\ref{tbl:fit3}), using the typical formulation which assumes energy balance \citep{kennicutt2009, kennicutt2012}.  The relation between our UV+$L_{\rm IR}$ calibration and SFR$_{\Paa}$ is well described by a unity slope and tight scatter ($\sim0.15$ dex). This indicates that the assumption of energy balance holds over our full sample, despite the steep decline in the mid-IR emission.
\end{itemize}

From the above, we deduce that the mid-IR closely tracks the (global) dust obscuration fraction and thus the obscured SFR component.  The obscuration is known to be a strong function of stellar mass and to have only weak to no evolution with redshift up to $z\sim3$ \citep[e.g.,][]{whitaker2017}.  The implications are: 1) the rest-frame 6 and 8$\,\mu$m emission, dominated by PAHs,  
%the 6.2 and 7.7$\mu$m PAH abundances 
does not significantly change relative to $L_{\rm IR}$ over the mass and metallicity range of our sample. 2) The breakpoint in the mid-IR behavior at $\sim8\,\Msun$ yr$^{-1}$ is more fundamentally linked to stellar mass through its relation to the obscured fraction.  At this redshift, $8\,\Msun$ yr$^{-1}$ corresponds to log $\logM=10.1$ for MS galaxies \citep{leja2019a}. 3) Though our sample only spans $1<z<1.75$, the lack of significant redshift evolution of $f_{\rm obs}$ implies that our MIRI-based SFR calibrations and the breakpoint in the mid-IR behavior are valid up to $z\sim3$, above which the MIRI bands are not longer dominated by dust emission. 

Lastly, in Section~\ref{sec:ulirgs} we extended our analysis to the high luminosity regime of ULIRGs.  We compare the behavior of the $7.7\mu$m PAH vs $L_{\rm IR}$ between local and $z>0.7$ galaxies, confirming that the PAH deficits seen locally in ULIRGs are not evident at $z>0.7$ \citep[e.g.][]{shipley2016}, with the caveat that current samples of high-$z$ ULIRGs show significant scatter.  New JWST spectroscopy in four local ULIRGs further reveals that \Paa\ in heavily obscured galaxies underestimates the IR-based SFR.  This is likely due to dust absorption of ionizing photons in extreme ISM conditions.
%% IMPORTANT! The old "\acknowledgment" command has be depreciated. It was
%% not robust enough to handle our new dual anonymous review requirements and
%% thus been replaced with the acknowledgment environment. If you try to 
%% compile with \acknowledgment you will get an error print to the screen
%% and in the compiled pdf.
%% 
%% Also note that the akcnowlodgment environment does not support long amounts of text. If you have a lot of people and institutions to acknowledge, do not use this command. Instead, create a new \section{Acknowledgments}.
\begin{acknowledgments}
Acknowledgments: The authors thank Andras G\'aspar for his work on the empirical MIRI PSFs used in extracting photometry for this work. SA, GHR, and JL acknowledge support from the JWST Mid-Infrared Instrument (MIRI) Science Team Lead, grant 80NSSC18K0555, from NASA Goddard Space Flight Center to the University of Arizona.  ZJ, JMH, and CNAW acknowledge support from the NIRCam Science Team contract to the University of Arizona, NAS5-02105.  The research of CCW is supported by NOIRLab, which is managed by the Association of Universities for Research in Astronomy (AURA) under a cooperative agreement with the National Science Foundation. IS acknowledges funding from the European Research Council (ERC) DistantDust (Grant No.101117541) and the Atracc\'{i}on de Talento Grant No.2022-T1/TIC-20472 of the Comunidad de Madrid, Spain. This work is based on observations made with the NASA/ESA/CSA James Webb Space Telescope. The data were obtained from the Mikulski Archive for Space Telescopes at the Space Telescope Science Institute, which is operated by the Association of Universities for Research in Astronomy, Inc., under NASA contract NAS 5-03127 for JWST.
\end{acknowledgments}

%% To help institutions obtain information on the effectiveness of their 
%% telescopes the AAS Journals has created a group of keywords for telescope 
%% facilities.
%
%% Following the acknowledgments section, use the following syntax and the
%% \facility{} or \facilities{} macros to list the keywords of facilities used 
%% in the research for the paper.  Each keyword is check against the master 
%% list during copy editing.  Individual instruments can be provided in 
%% parentheses, after the keyword, but they are not verified.

\vspace{5mm}
\facilities{JWST, HST}

%% Similar to \facility{}, there is the optional \software command to allow 
%% authors a place to specify which programs were used during the creation of 
%% the manuscript. Authors should list each code and include either a
%% citation or url to the code inside ()s when available.
\software{JWST Calibration Pipeline \citep{bushouse2023}, astropy \citep{astropy:2013, astropy:2018, astropy:2022}, photutils \citep{bradley2025}, sep \citep{bertin1996, Barbary2016}, matplotlib \citep{Hunter:2007}, grizli \citep{brammer2018}}

%% Appendix material should be preceded with a single \appendix command.
%% There should be a \section command for each appendix. Mark appendix
%% subsections with the same markup you use in the main body of the paper.

%% Each Appendix (indicated with \section) will be lettered A, B, C, etc.
%% The equation counter will reset when it encounters the \appendix
%% command and will number appendix equations (A1), (A2), etc. The
%% Figure and Table counter will not reset.

%\appendix

%% For this sample we use BibTeX plus aasjournals.bst to generate the
%% the bibliography. The sample631.bib file was populated from ADS. To
%% get the citations to show in the compiled file do the following:
%%
%% pdflatex sample631.tex
%% bibtext sample631
%% pdflatex sample631.tex
%% pdflatex sample631.tex

\bibliography{main}{}

@article{brammer2018,
	title = {gbrammer/grizli: {Release} 2021},
	shorttitle = {gbrammer/grizli},
	url = {https://ui.adsabs.harvard.edu/abs/2021zndo...5012699B},
	doi = {10.5281/zenodo.5012699},
	urldate = {2025-03-14},
	journal = {Zenodo},
	publisher = {Zenodo},
	author = {Brammer, Gabe and Matharu, Jasleen},
	month = jun,
	year = {2021},
	note = {ADS Bibcode: 2021zndo...5012699B},
}

@article{bushouse2023,
  title = {{{JWST Calibration Pipeline}}},
  author = {Bushouse, Howard and Eisenhamer, Jonathan and Dencheva, Nadia and Davies, James and Greenfield, Perry and Morrison, Jane and Hodge, Phil and Simon, Bernie and Grumm, David and Droettboom, Michael and Slavich, Edward and Sosey, Megan and Pauly, Tyler and Miller, Todd and Jedrzejewski, Robert and Hack, Warren and Davis, David and Crawford, Steven and Law, David and Gordon, Karl and Regan, Michael and Cara, Mihai and MacDonald, Ken and Bradley, Larry and Shanahan, Clare and Jamieson, William and Teodoro, Mairan and Williams, Thomas and {Pena-Guerrero}, Maria},
  year = 2023,
  month = oct,
  journal = {Zenodo},
  publisher = {Zenodo},
  doi = {10.5281/zenodo.10022973},
  urldate = {2024-05-21}
}

@misc{illingworth2016,
  title = {The {{Hubble Legacy Fields}} ({{HLF-GOODS-S}}) v1.5 {{Data Products}}: {{Combining}} 2442 {{Orbits}} of {{GOODS-S}}/{{CDF-S Region ACS}} and {{WFC3}}/{{IR Images}}},
  shorttitle = {The {{Hubble Legacy Fields}} ({{HLF-GOODS-S}}) v1.5 {{Data Products}}},
  author = {Illingworth, Garth and Magee, Daniel and Bouwens, Rychard and Oesch, Pascal and Labbe, Ivo and {van Dokkum}, Pieter and Whitaker, Katherine and Holden, Bradford and Franx, Marijn and Gonzalez, Valentino},
  year = 2016,
  month = jun,
  journal = {arXiv e-prints},
  doi = {10.48550/arXiv.1606.00841},
  urldate = {2023-12-08}
}

@article{whitaker2019,
  title = {The {{Hubble Legacy Field GOODS-S Photometric Catalog}}},
  author = {Whitaker, Katherine E. and Ashas, Mohammad and Illingworth, Garth and Magee, Daniel and Leja, Joel and Oesch, Pascal and {van Dokkum}, Pieter and Mowla, Lamiya and Bouwens, Rychard and Franx, Marijn and Holden, Bradford and Labb{\'e}, Ivo and Rafelski, Marc and Teplitz, Harry and Gonzalez, Valentino},
  year = 2019,
  month = sep,
  journal = {ApJS},
  volume = {244},
  pages = {16},
  issn = {0067-0049},
  doi = {10.3847/1538-4365/ab3853},
  urldate = {2023-12-08}
}

@article{tacchella2022,
  title = {Fast, {{Slow}}, {{Early}}, {{Late}}: {{Quenching Massive Galaxies}} at z {$\sim$} 0.8},
  shorttitle = {Fast, {{Slow}}, {{Early}}, {{Late}}},
  author = {Tacchella, Sandro and Conroy, Charlie and Faber, S. M. and Johnson, Benjamin D. and Leja, Joel and Barro, Guillermo and Cunningham, Emily C. and Deason, Alis J. and Guhathakurta, Puragra and Guo, Yicheng and Hernquist, Lars and Koo, David C. and McKinnon, Kevin and Rockosi, Constance M. and Speagle, Joshua S. and {van Dokkum}, Pieter and Yesuf, Hassen M.},
  year = 2022,
  month = feb,
  journal = {ApJ},
  volume = {926},
  pages = {134},
  issn = {0004-637X},
  doi = {10.3847/1538-4357/ac449b},
  urldate = {2023-11-17}
}

@article{conroy2009,
  title = {The {{Propagation}} of {{Uncertainties}} in {{Stellar Population Synthesis Modeling}}. {{I}}. {{The Relevance}} of {{Uncertain Aspects}} of {{Stellar Evolution}} and the {{Initial Mass Function}} to the {{Derived Physical Properties}} of {{Galaxies}}},
  author = {Conroy, Charlie and Gunn, James E. and White, Martin},
  year = 2009,
  month = jul,
  journal = {ApJ},
  volume = {699},
  pages = {486--506},
  publisher = {IOP},
  issn = {0004-637X},
  doi = {10.1088/0004-637X/699/1/486},
  urldate = {2026-01-02}
}

@article{conroy2010,
  title = {The {{Propagation}} of {{Uncertainties}} in {{Stellar Population Synthesis Modeling}}. {{III}}. {{Model Calibration}}, {{Comparison}}, and {{Evaluation}}},
  author = {Conroy, Charlie and Gunn, James E.},
  year = 2010,
  month = apr,
  journal = {ApJ},
  volume = {712},
  pages = {833--857},
  publisher = {IOP},
  issn = {0004-637X},
  doi = {10.1088/0004-637X/712/2/833},
  urldate = {2026-01-02}
}

@ARTICLE{rodighiero2011,
       author = {{Rodighiero}, G. and {Daddi}, E. and {Baronchelli}, I. and {Cimatti}, A. and {Renzini}, A. and {Aussel}, H. and {Popesso}, P. and {Lutz}, D. and {Andreani}, P. and {Berta}, S. and {Cava}, A. and {Elbaz}, D. and {Feltre}, A. and {Fontana}, A. and {F{\"o}rster Schreiber}, N.~M. and {Franceschini}, A. and {Genzel}, R. and {Grazian}, A. and {Gruppioni}, C. and {Ilbert}, O. and {Le Floch}, E. and {Magdis}, G. and {Magliocchetti}, M. and {Magnelli}, B. and {Maiolino}, R. and {McCracken}, H. and {Nordon}, R. and {Poglitsch}, A. and {Santini}, P. and {Pozzi}, F. and {Riguccini}, L. and {Tacconi}, L.~J. and {Wuyts}, S. and {Zamorani}, G.},
        title = "{The Lesser Role of Starbursts in Star Formation at z = 2}",
      journal = {\apjl},
         year = 2011,
        month = oct,
       volume = {739},
       number = {2},
          eid = {L40},
        pages = {L40},
          doi = {10.1088/2041-8205/739/2/L40},
archivePrefix = {arXiv},
       eprint = {1108.0933},
 primaryClass = {astro-ph.CO},
       adsurl = {https://ui.adsabs.harvard.edu/abs/2011ApJ...739L..40R}
}

@ARTICLE{hayward2014,
       author = {{Hayward}, Christopher C. and {Lanz}, Lauranne and {Ashby}, Matthew L.~N. and {Fazio}, Giovanni and {Hernquist}, Lars and {Mart{\'\i}nez-Galarza}, Juan Rafael and {Noeske}, Kai and {Smith}, Howard A. and {Wuyts}, Stijn and {Zezas}, Andreas},
        title = "{The total infrared luminosity may significantly overestimate the star formation rate of quenching and recently quenched galaxies}",
      journal = {\mnras},
         year = 2014,
        month = dec,
       volume = {445},
       number = {2},
        pages = {1598-1604},
          doi = {10.1093/mnras/stu1843},
archivePrefix = {arXiv},
       eprint = {1402.0006},
 primaryClass = {astro-ph.GA},
       adsurl = {https://ui.adsabs.harvard.edu/abs/2014MNRAS.445.1598H}
}

@ARTICLE{scipy2020,
  author  = {Virtanen, Pauli and Gommers, Ralf and Oliphant, Travis E. and
            Haberland, Matt and Reddy, Tyler and Cournapeau, David and
            Burovski, Evgeni and Peterson, Pearu and Weckesser, Warren and
            Bright, Jonathan and {van der Walt}, St{\'e}fan J. and
            Brett, Matthew and Wilson, Joshua and Millman, K. Jarrod and
            Mayorov, Nikolay and Nelson, Andrew R. J. and Jones, Eric and
            Kern, Robert and Larson, Eric and Carey, C J and
            Polat, {\.I}lhan and Feng, Yu and Moore, Eric W. and
            {VanderPlas}, Jake and Laxalde, Denis and Perktold, Josef and
            Cimrman, Robert and Henriksen, Ian and Quintero, E. A. and
            Harris, Charles R. and Archibald, Anne M. and
            Ribeiro, Ant{\^o}nio H. and Pedregosa, Fabian and
            {van Mulbregt}, Paul and {SciPy 1.0 Contributors}},
  title   = {{{SciPy} 1.0: Fundamental Algorithms for Scientific
            Computing in Python}},
  journal = {Nature Methods},
  year    = {2020},
  volume  = {17},
  pages   = {261--272},
  adsurl  = {https://rdcu.be/b08Wh},
  doi     = {10.1038/s41592-019-0686-2},
}

@ARTICLE{wuyts2011,
       author = {{Wuyts}, Stijn and {F{\"o}rster Schreiber}, Natascha M. and {Lutz}, Dieter and {Nordon}, Raanan and {Berta}, Stefano and {Altieri}, Bruno and {Andreani}, Paola and {Aussel}, Herv{\'e} and {Bongiovanni}, Angel and {Cepa}, Jordi and {Cimatti}, Andrea and {Daddi}, Emanuele and {Elbaz}, David and {Genzel}, Reinhard and {Koekemoer}, Anton M. and {Magnelli}, Benjamin and {Maiolino}, Roberto and {McGrath}, Elizabeth J. and {P{\'e}rez Garc{\'\i}a}, Ana and {Poglitsch}, Albrecht and {Popesso}, Paola and {Pozzi}, Francesca and {Sanchez-Portal}, Miguel and {Sturm}, Eckhard and {Tacconi}, Linda and {Valtchanov}, Ivan},
        title = "{On Star Formation Rates and Star Formation Histories of Galaxies Out to z \raisebox{-0.5ex}\textasciitilde 3}",
      journal = {\apj},
         year = 2011,
        month = sep,
       volume = {738},
       number = {1},
          eid = {106},
        pages = {106},
          doi = {10.1088/0004-637X/738/1/106},
archivePrefix = {arXiv},
       eprint = {1106.5502},
 primaryClass = {astro-ph.CO},
       adsurl = {https://ui.adsabs.harvard.edu/abs/2011ApJ...738..106W}
}

@ARTICLE{bell2005,
       author = {{Bell}, Eric F. and {Papovich}, Casey and {Wolf}, Christian and {Le Floc'h}, Emeric and {Caldwell}, John A.~R. and {Barden}, Marco and {Egami}, Eiichi and {McIntosh}, Daniel H. and {Meisenheimer}, Klaus and {P{\'e}rez-Gonz{\'a}lez}, Pablo G. and {Rieke}, G.~H. and {Rieke}, M.~J. and {Rigby}, Jane R. and {Rix}, Hans-Walter},
        title = "{Toward an Understanding of the Rapid Decline of the Cosmic Star Formation Rate}",
      journal = {\apj},
         year = 2005,
        month = may,
       volume = {625},
       number = {1},
        pages = {23-36},
          doi = {10.1086/429552},
archivePrefix = {arXiv},
       eprint = {astro-ph/0502246},
 primaryClass = {astro-ph},
       adsurl = {https://ui.adsabs.harvard.edu/abs/2005ApJ...625...23B}
}

@ARTICLE{sun2025,
       author = {{Sun}, Yang and {Lyu}, Jianwei and {Rieke}, George H. and {Ji}, Zhiyuan and {Sun}, Fengwu and {Zhu}, Yongda and {Bunker}, Andrew J. and {Cargile}, Phillip A. and {Circosta}, Chiara and {D'Eugenio}, Francesco and {Egami}, Eiichi and {Hainline}, Kevin and {Helton}, Jakob M. and {Rinaldi}, Pierluigi and {Robertson}, Brant E. and {Scholtz}, Jan and {Shivaei}, Irene and {Stone}, Meredith A. and {Tacchella}, Sandro and {Williams}, Christina C. and {Willmer}, Christopher N.~A. and {Willott}, Chris},
        title = "{No Evidence for a Significant Evolution of M$_{{\textbullet}}${\textendash}M. Relation in Massive Galaxies up to z {\ensuremath{\sim}} 4}",
      journal = {\apj},
         year = 2025,
        month = jan,
       volume = {978},
       number = {1},
          eid = {98},
        pages = {98},
          doi = {10.3847/1538-4357/ad973b},
archivePrefix = {arXiv},
       eprint = {2409.06796},
 primaryClass = {astro-ph.GA},
       adsurl = {https://ui.adsabs.harvard.edu/abs/2025ApJ...978...98S}
}

@article{alberts2020,
  title = {Completing the {{Census}} of {{AGN}} in {{GOODS-S}}/{{HUDF}}: {{New Ultradeep Radio Imaging}} and {{Predictions}} for {{JWST}}},
  shorttitle = {Completing the {{Census}} of {{AGN}} in {{GOODS-S}}/{{HUDF}}},
  author = {Alberts, Stacey and Rujopakarn, Wiphu and Rieke, George H. and Jagannathan, Preshanth and Nyland, Kristina},
  year = 2020,
  month = oct,
  journal = {ApJ},
  volume = {901},
  number = {2},
  pages = {168},
  issn = {1538-4357},
  doi = {10.3847/1538-4357/abb1a0},
  urldate = {2021-12-16},
  langid = {english}
}

@article{alberts2024a,
  title = {{{SMILES Initial Data Release}}: {{Unveiling}} the {{Obscured Universe}} with {{MIRI Multiband Imaging}}},
  shorttitle = {{{SMILES Initial Data Release}}},
  author = {Alberts, Stacey and Lyu, Jianwei and Shivaei, Irene and Rieke, George H. and {P{\'e}rez-Gonz{\'a}lez}, Pablo G. and Bonaventura, Nina and Zhu, Yongda and Helton, Jakob M. and Ji, Zhiyuan and Morrison, Jane and Robertson, Brant E. and Stone, Meredith A. and Sun, Yang and Williams, Christina C. and Willmer, Christopher N. A.},
  year = {2024},
  month = dec,
  journal = {\apj},
  volume = {976},
  pages = {224},
  publisher = {IOP},
  issn = {0004-637X},
  doi = {10.3847/1538-4357/ad7396},
  urldate = {2024-12-20}
}

@article{alonso-herrero2004,
  title = {Obscured {{Star Formation}} in the {{Central Region}} of the {{Dwarf Galaxy NGC}} 5253},
  author = {{Alonso-Herrero}, Almudena and Takagi, Toshinobu and Baker, Andrew J. and Rieke, George H. and Rieke, Marcia J. and Imanishi, Masatoshi and Scoville, Nick Z.},
  year = {2004},
  month = sep,
  journal = {\apj},
  volume = {612},
  pages = {222--237},
  issn = {0004-637X},
  doi = {10.1086/422448},
  urldate = {2025-08-21}
}

@article{alonso-herrero2006,
  title = {Near-{{Infrared}} and {{Star}}-forming {{Properties}} of {{Local Luminous Infrared Galaxies}}},
  author = {Alonso-Herrero, Almudena and Rieke, George H. and Rieke, Marcia J. and Colina, Luis and Perez-Gonzalez, Pablo G. and Ryder, Stuart D.},
  year = {2006},
  month = oct,
  journal = {ApJ},
  volume = {650},
  number = {2},
  pages = {835--849},
  issn = {0004-637X, 1538-4357},
  doi = {10.1086/506958},
  urldate = {2021-11-10},
  langid = {english}
}

@article{aniano2020,
  title = {Modeling {{Dust}} and {{Starlight}} in {{Galaxies Observed}} by {{Spitzer}} and {{Herschel}}: {{The KINGFISH Sample}}},
  shorttitle = {Modeling {{Dust}} and {{Starlight}} in {{Galaxies Observed}} by {{Spitzer}} and {{Herschel}}},
  author = {Aniano, G. and Draine, B. T. and Hunt, L. K. and Sandstrom, K. and Calzetti, D. and Kennicutt, R. C. and Dale, D. A. and Galametz, M. and Gordon, K. D. and Leroy, A. K. and Smith, J. -D. T. and Roussel, H. and Sauvage, M. and Walter, F. and Armus, L. and Bolatto, A. D. and Boquien, M. and Crocker, A. and De Looze, I. and Donovan Meyer, J. and Helou, G. and Hinz, J. and Johnson, B. D. and Koda, J. and Miller, A. and Montiel, E. and Murphy, E. J. and Rela{\~n}o, M. and Rix, H. -W. and Schinnerer, E. and Skibba, R. and Wolfire, M. G. and Engelbracht, C. W.},
  year = {2020},
  month = feb,
  journal = {\apj},
  volume = {889},
  pages = {150},
  issn = {0004-637X},
  doi = {10.3847/1538-4357/ab5fdb},
  urldate = {2025-08-21}
}

@article{armus2007,
  title = {Observations of {{Ultraluminous Infrared Galaxies}} with the {{Infrared Spectrograph}} on the {{Spitzer Space Telescope}}. {{II}}. {{The IRAS Bright Galaxy Sample}}},
  author = {Armus, L. and Charmandaris, V. and {Bernard-Salas}, J. and Spoon, H. W. W. and Marshall, J. A. and Higdon, S. J. U. and Desai, V. and Teplitz, H. I. and Hao, L. and Devost, D. and Brandl, B. R. and Wu, Y. and Sloan, G. C. and Soifer, B. T. and Houck, J. R. and Herter, T. L.},
  year = {2007},
  month = feb,
  journal = {\apj},
  volume = {656},
  pages = {148--167},
  issn = {0004-637X},
  doi = {10.1086/510107},
  urldate = {2025-08-21}
}

@article{atek2022,
  title = {The Star Formation Burstiness and Ionizing Efficiency of Low-Mass Galaxies},
  author = {Atek, Hakim and Furtak, Lukas J. and Oesch, Pascal and {van Dokkum}, Pieter and Reddy, Naveen and Contini, Thierry and Illingworth, Garth and Wilkins, Stephen},
  year = {2022},
  month = apr,
  journal = {MNRAS},
  volume = {511},
  pages = {4464--4479},
  publisher = {OUP},
  issn = {0035-8711},
  doi = {10.1093/mnras/stac360},
  urldate = {2025-08-08}
}

@article{bacon2023,
  title = {The {{MUSE Hubble Ultra Deep Field}} Surveys: {{Data}} Release {{II}}},
  shorttitle = {The {{MUSE Hubble Ultra Deep Field}} Surveys},
  author = {Bacon, Roland and Brinchmann, Jarle and Conseil, Simon and Maseda, Michael and Nanayakkara, Themiya and Wendt, Martin and Bacher, Raphael and Mary, David and Weilbacher, Peter M. and Krajnovi{\'c}, Davor and Boogaard, Leindert and Bouch{\'e}, Nicolas and Contini, Thierry and Epinat, Beno{\^i}t and Feltre, Anna and Guo, Yucheng and Herenz, Christian and Kollatschny, Wolfram and Kusakabe, Haruka and Leclercq, Floriane and {Michel-Dansac}, L{\'e}o and Pello, Roser and Richard, Johan and Roth, Martin and Salvignol, Gregory and Schaye, Joop and Steinmetz, Matthias and Tresse, Laurence and Urrutia, Tanya and Verhamme, Anne and Vitte, Eloise and Wisotzki, Lutz and Zoutendijk, Sebastiaan L.},
  year = {2023},
  month = feb,
  journal = {A\&A},
  volume = {670},
  pages = {A4},
  issn = {0004-6361},
  doi = {10.1051/0004-6361/202244187},
  urldate = {2024-05-01}
}

@article{bakes1994,
  title = {The {{Photoelectric Heating Mechanism}} for {{Very Small Graphitic Grains}} and {{Polycyclic Aromatic Hydrocarbons}}},
  author = {Bakes, E. L. O. and Tielens, A. G. G. M.},
  year = {1994},
  month = jun,
  journal = {\apj},
  volume = {427},
  pages = {822},
  issn = {0004-637X},
  doi = {10.1086/174188},
  urldate = {2025-08-21}
}

@article{baron2025,
  title = {{{PHANGS-ML}}: {{The Universal Relation}} between {{PAH Band}} and {{Optical Line Ratios}} across {{Nearby Star-forming Galaxies}}},
  shorttitle = {{{PHANGS-ML}}},
  author = {Baron, Dalya and Sandstrom, Karin M. and Sutter, Jessica and Hassani, Hamid and Groves, Brent and Leroy, Adam K. and Schinnerer, Eva and Boquien, M{\'e}d{\'e}ric and Brazzini, Matilde and Chastenet, J{\'e}r{\'e}my and Dale, Daniel A. and Egorov, Oleg V. and Glover, Simon C. O. and Klessen, Ralf S. and Pathak, Debosmita and Rosolowsky, Erik and Bigiel, Frank and Chevance, M{\'e}lanie and Grasha, Kathryn and Hughes, Annie and {M{\'e}ndez-Delgado}, J. Eduardo and Pety, J{\'e}r{\^o}me and Williams, Thomas G. and Hannon, Stephen and Sarbadhicary, Sumit K.},
  year = {2025},
  month = jan,
  journal = {\apj},
  volume = {978},
  pages = {135},
  publisher = {IOP},
  issn = {0004-637X},
  doi = {10.3847/1538-4357/ad972a},
  urldate = {2025-07-14}
}

@article{barrera2023,
  title = {Formation of {{H2}} on Polycyclic Aromatic Hydrocarbons under Conditions of the {{ISM}}: An Ab Initio Molecular Dynamics Study},
  shorttitle = {Formation of {{H2}} on Polycyclic Aromatic Hydrocarbons under Conditions of the {{ISM}}},
  author = {Barrera, Nicol{\'a}s F. and Fuentealba, Patricio and Mu{\~n}oz, Francisco and G{\'o}mez, Tatiana and C{\'a}rdenas, Carlos},
  year = {2023},
  month = sep,
  journal = {MNRAS},
  volume = {524},
  pages = {3741--3748},
  issn = {0035-8711},
  doi = {10.1093/mnras/stad2106},
  urldate = {2025-08-18}
}

@article{bauschlicher1998,
  title = {The {{Reaction}} of {{Polycyclic Aromatic Hydrocarbon Cations}} with {{Hydrogen Atoms}}: {{The Astrophysical Implications}}},
  shorttitle = {The {{Reaction}} of {{Polycyclic Aromatic Hydrocarbon Cations}} with {{Hydrogen Atoms}}},
  author = {Bauschlicher, Jr., Charles W.},
  year = {1998},
  month = dec,
  journal = {\apj},
  volume = {509},
  pages = {L125-L127},
  issn = {0004-637X},
  doi = {10.1086/311782},
  urldate = {2025-08-26}
}

@article{bian2018,
  title = {``{{Direct}}'' {{Gas-phase Metallicity}} in {{Local Analogs}} of {{High-redshift Galaxies}}: {{Empirical Metallicity Calibrations}} for {{High-redshift Star-forming Galaxies}}},
  shorttitle = {``{{Direct}}'' {{Gas-phase Metallicity}} in {{Local Analogs}} of {{High-redshift Galaxies}}},
  author = {Bian, Fuyan and Kewley, Lisa J. and Dopita, Michael A.},
  year = {2018},
  month = jun,
  journal = {\apj},
  volume = {859},
  pages = {175},
  publisher = {IOP},
  issn = {0004-637X},
  doi = {10.3847/1538-4357/aabd74},
  urldate = {2024-11-12}
}

@article{boquien2021,
  title = {New-Generation Dust Emission Templates for Star-Forming Galaxies},
  author = {Boquien, M{\'e}d{\'e}ric and Salim, Samir},
  year = {2021},
  month = sep,
  journal = {A\&A},
  volume = {653},
  pages = {A149},
  issn = {0004-6361},
  doi = {10.1051/0004-6361/202140992},
  urldate = {2023-07-25}
}

@article{bouwens2016,
  title = {{{ALMA Spectroscopic Survey}} in the {{Hubble Ultra Deep Field}}: {{The Infrared Excess}} of {{UV-Selected}} z = 2-10 {{Galaxies}} as a {{Function}} of {{UV-Continuum Slope}} and {{Stellar Mass}}},
  shorttitle = {{{ALMA Spectroscopic Survey}} in the {{Hubble Ultra Deep Field}}},
  author = {Bouwens, Rychard J. and Aravena, Manuel and Decarli, Roberto and Walter, Fabian and {da Cunha}, Elisabete and Labb{\'e}, Ivo and Bauer, Franz E. and Bertoldi, Frank and Carilli, Chris and Chapman, Scott and Daddi, Emanuele and Hodge, Jacqueline and Ivison, Rob J. and Karim, Alex and Le Fevre, Olivier and Magnelli, Benjamin and Ota, Kazuaki and Riechers, Dominik and Smail, Ian R. and {van der Werf}, Paul and Weiss, Axel and Cox, Pierre and Elbaz, David and {Gonzalez-Lopez}, Jorge and Infante, Leopoldo and Oesch, Pascal and Wagg, Jeff and Wilkins, Steve},
  year = {2016},
  month = dec,
  journal = {\apj},
  volume = {833},
  pages = {72},
  issn = {0004-637X},
  doi = {10.3847/1538-4357/833/1/72},
  urldate = {2025-08-21}
}

@ARTICLE{bunker2023,
       author = {{Bunker}, Andrew J. and {Cameron}, Alex J. and {Curtis-Lake}, Emma and {Jakobsen}, Peter and {Carniani}, Stefano and {Curti}, Mirko and {Witstok}, Joris and {Maiolino}, Roberto and {D'Eugenio}, Francesco and {Looser}, Tobias J. and {Willott}, Chris and {Bonaventura}, Nina and {Hainline}, Kevin and {{\"U}bler}, Hannah and {Willmer}, Christopher N.~A. and {Saxena}, Aayush and {Smit}, Renske and {Alberts}, Stacey and {Arribas}, Santiago and {Baker}, William M. and {Baum}, Stefi and {Bhatawdekar}, Rachana and {Bowler}, Rebecca A.~A. and {Boyett}, Kristan and {Charlot}, Stephane and {Chen}, Zuyi and {Chevallard}, Jacopo and {Circosta}, Chiara and {DeCoursey}, Christa and {de Graaff}, Anna and {Egami}, Eiichi and {Eisenstein}, Daniel J. and {Endsley}, Ryan and {Ferruit}, Pierre and {Giardino}, Giovanna and {Hausen}, Ryan and {Helton}, Jakob M. and {Hviding}, Raphael E. and {Ji}, Zhiyuan and {Johnson}, Benjamin D. and {Jones}, Gareth C. and {Kumari}, Nimisha and {Laseter}, Isaac and {L{\"u}tzgendorf}, Nora and {Maseda}, Michael V. and {Nelson}, Erica and {Parlanti}, Eleonora and {Perna}, Michele and {Rauscher}, Bernard J. and {Rawle}, Tim and {Rix}, Hans-Walter and {Rieke}, Marcia and {Robertson}, Brant and {Rodr{\'\i}guez Del Pino}, Bruno and {Sandles}, Lester and {Scholtz}, Jan and {Sharpe}, Katherine and {Skarbinski}, Maya and {Stark}, Daniel P. and {Sun}, Fengwu and {Tacchella}, Sandro and {Topping}, Michael W. and {Villanueva}, Natalia C. and {Wallace}, Imaan E.~B. and {Williams}, Christina C. and {Woodrum}, Charity},
        title = "{JADES NIRSpec initial data release for the Hubble Ultra Deep Field: Redshifts and line fluxes of distant galaxies from the deepest JWST Cycle 1 NIRSpec multi-object spectroscopy}",
      journal = {\aap},
         year = 2024,
        month = oct,
       volume = {690},
          eid = {A288},
        pages = {A288},
          doi = {10.1051/0004-6361/202347094},
archivePrefix = {arXiv},
       eprint = {2306.02467},
 primaryClass = {astro-ph.GA},
       adsurl = {https://ui.adsabs.harvard.edu/abs/2024A&A...690A.288B}
}

@article{calapa2014,
  title = {The {{Heating}} of {{Mid-infrared Dust}} in the {{Nearby Galaxy M33}}: {{A Testbed}} for {{Tracing Galaxy Evolution}}},
  shorttitle = {The {{Heating}} of {{Mid-infrared Dust}} in the {{Nearby Galaxy M33}}},
  author = {Calapa, Marie D. and Calzetti, Daniela and Draine, Bruce T. and Boquien, M{\'e}d{\'e}ric and Kramer, Carsten and Xilouris, Manolis and Verley, Simon and Braine, Jonathan and Rela{\~n}o, Monica and {van der Werf}, Paul and Israel, Frank and Hermelo, Israel and Albrecht, Marcus},
  year = {2014},
  month = apr,
  journal = {\apj},
  volume = {784},
  pages = {130},
  publisher = {IOP},
  issn = {0004-637X},
  doi = {10.1088/0004-637X/784/2/130},
  urldate = {2025-08-04}
}

@article{calzetti2000,
  title = {The {{Dust Content}} and {{Opacity}} of {{Actively Star-forming Galaxies}}},
  author = {Calzetti, Daniela and Armus, Lee and Bohlin, Ralph C. and Kinney, Anne L. and Koornneef, Jan and {Storchi-Bergmann}, Thaisa},
  year = {2000},
  month = apr,
  journal = {\apj},
  volume = {533},
  pages = {682--695},
  issn = {0004-637X},
  doi = {10.1086/308692},
  urldate = {2023-11-17}
}

@article{calzetti2007a,
  title = {The {{Calibration}} of {{Mid-Infrared Star Formation Rate Indicators}}},
  author = {Calzetti, D. and Kennicutt, R. C. and Engelbracht, C. W. and Leitherer, C. and Draine, B. T. and Kewley, L. and Moustakas, J. and Sosey, M. and Dale, D. A. and Gordon, K. D. and Helou, G. X. and Hollenbach, D. J. and Armus, L. and Bendo, G. and Bot, C. and Buckalew, B. and Jarrett, T. and Li, A. and Meyer, M. and Murphy, E. J. and Prescott, M. and Regan, M. W. and Rieke, G. H. and Roussel, H. and Sheth, K. and Smith, J. D. T. and Thornley, M. D. and Walter, F.},
  year = {2007},
  month = sep,
  journal = {\apj},
  volume = {666},
  pages = {870--895},
  issn = {0004-637X},
  doi = {10.1086/520082},
  urldate = {2025-08-26}
}

@incollection{calzetti2013,
  title = {Star {{Formation Rate Indicators}}},
  booktitle = {Secular {{Evolution}} of {{Galaxies}}},
  author = {Calzetti, Daniela},
  year = {2013},
  month = oct,
  pages = {419},
  doi = {10.48550/arXiv.1208.2997},
  urldate = {2025-08-26}
}

@ARTICLE{calzetti2025,
       author = {{Calzetti}, Daniela and {Kennicutt}, Robert C. and {Adamo}, Angela and {Sandstrom}, Karin and {Dale}, Daniel A. and {Elmegreen}, Bruce and {Gallagher}, John S. and {Gregg}, Benjamin and {Bajaj}, Varun and {B{\"o}ker}, Torsten and {Bortolini}, Giacomo and {Boyer}, Martha and {Correnti}, Matteo and {De Looze}, Ilse and {Draine}, Bruce T. and {Duarte-Cabral}, Ana and {Faustino Vieira}, Helena and {Grasha}, Kathryn and {Hunt}, L.~K. and {Johnson}, Kelsey E. and {Klessen}, Ralf S. and {Krumholz}, Mark R. and {Lai}, Thomas S.-Y. and {Lapeer}, Drew and {Linden}, Sean T. and {Messa}, Matteo and {{\"O}stlin}, G{\"o}ran and {Pedrini}, Alex and {Rela{\~n}o}, M{\`o}nica and {Sabbi}, Elena and {Schinnerer}, Eva and {Skillman}, Evan and {Smith}, Linda J. and {Tosi}, Monica and {Walter}, Fabian and {Weinbeck}, Tony D.},
        title = "{Quantification of the Age Dependence of Mid-infrared Star Formation Rate Indicators}",
      journal = {\apj},
         year = 2025,
        month = oct,
       volume = {991},
       number = {2},
          eid = {198},
        pages = {198},
          doi = {10.3847/1538-4357/adfbe0},
archivePrefix = {arXiv},
       eprint = {2508.08451},
 primaryClass = {astro-ph.GA},
       adsurl = {https://ui.adsabs.harvard.edu/abs/2025ApJ...991..198C}
}

@article{cardelli1989,
  title = {The {{Relationship}} between {{Infrared}}, {{Optical}}, and {{Ultraviolet Extinction}}},
  author = {Cardelli, Jason A. and Clayton, Geoffrey C. and Mathis, John S.},
  year = {1989},
  month = oct,
  journal = {\apj},
  volume = {345},
  pages = {245},
  issn = {0004-637X},
  doi = {10.1086/167900},
  urldate = {2025-08-26}
}

@article{chastenet2019,
  title = {The {{Polycyclic Aromatic Hydrocarbon Mass Fraction}} on a 10 Pc {{Scale}} in the {{Magellanic Clouds}}},
  author = {Chastenet, J{\'e}r{\'e}my and Sandstrom, Karin and Chiang, I-Da and Leroy, Adam K. and Utomo, Dyas and Bot, Caroline and Gordon, Karl D. and Draine, Bruce T. and Fukui, Yasuo and Onishi, Toshikazu and Tsuge, Kisetsu},
  year = {2019},
  month = may,
  journal = {\apj},
  volume = {876},
  pages = {62},
  issn = {0004-637X},
  doi = {10.3847/1538-4357/ab16cf},
  urldate = {2025-08-21}
}

@article{chastenet2025,
  title = {The {{Resolved Behavior}} of {{Dust Mass}}, {{Polycyclic Aromatic Hydrocarbon Fraction}}, and {{Radiation Field}} in {$\sim$}800 {{Nearby Galaxies}}},
  author = {Chastenet, J{\'e}r{\'e}my and Sandstrom, Karin and Leroy, Adam K. and Bot, Caroline and Chiang, I-Da and Chown, Ryan and Gordon, Karl D. and Koch, Eric W. and Roussel, H{\'e}l{\`e}ne and Sutter, Jessica and Williams, Thomas G.},
  year = {2025},
  month = jan,
  journal = {\apj Suppl. Ser.},
  volume = {276},
  pages = {2},
  publisher = {IOP},
  issn = {0067-0049},
  doi = {10.3847/1538-4365/ad8a5c},
  urldate = {2025-07-14}
}

@article{chisholm2019,
  title = {Constraining the {{Metallicities}}, {{Ages}}, {{Star Formation Histories}}, and {{Ionizing Continua}} of {{Extragalactic Massive Star Populations}}},
  author = {Chisholm, J. and Rigby, J. R. and Bayliss, M. and Berg, D. A. and Dahle, H. and Gladders, M. and Sharon, K.},
  year = {2019},
  month = sep,
  journal = {\apj},
  volume = {882},
  pages = {182},
  issn = {0004-637X},
  doi = {10.3847/1538-4357/ab3104},
  urldate = {2025-08-21}
}

@article{cleri2022,
  title = {{{CLEAR}}: {{Paschen-$\beta$ Star Formation Rates}} and {{Dust Attenuation}} of {{Low-redshift Galaxies}}},
  shorttitle = {{{CLEAR}}},
  author = {Cleri, Nikko J. and Trump, Jonathan R. and Backhaus, Bren E. and Momcheva, Ivelina and Papovich, Casey and Simons, Raymond and Weiner, Benjamin and {Estrada-Carpenter}, Vicente and Finkelstein, Steven L. and Giavalisco, Mauro and Ji, Zhiyuan and Jung, Intae and Matharu, Jasleen and Martinez, Felix and Sturm, Megan R.},
  year = {2022},
  month = apr,
  journal = {\apj},
  volume = {929},
  pages = {3},
  issn = {0004-637X},
  doi = {10.3847/1538-4357/ac5a4c},
  urldate = {2023-08-07}
}

@article{crocker2013,
  title = {Quantifying {{Non-star-formation-associated}} 8 {$M$}m {{Dust Emission}} in {{NGC}} 628},
  author = {Crocker, Alison F. and Calzetti, Daniela and Thilker, David A. and Aniano, Gonzalo and Draine, Bruce T. and Hunt, Leslie K. and Kennicutt, Robert C. and Sandstrom, Karin and Smith, J. D. T.},
  year = {2013},
  month = jan,
  journal = {\apj},
  volume = {762},
  pages = {79},
  publisher = {IOP},
  issn = {0004-637X},
  doi = {10.1088/0004-637X/762/2/79},
  urldate = {2025-08-04}
}

@article{dale2002,
  title = {The {{Infrared Spectral Energy Distribution}} of {{Normal Star-forming Galaxies}}: {{Calibration}} at {{Far-Infrared}} and {{Submillimeter Wavelengths}}},
  shorttitle = {The {{Infrared Spectral Energy Distribution}} of {{Normal Star-forming Galaxies}}},
  author = {Dale, Daniel A. and Helou, George},
  year = {2002},
  month = sep,
  journal = {\apj},
  volume = {576},
  pages = {159--168},
  issn = {0004-637X},
  doi = {10.1086/341632},
  urldate = {2025-09-01}
}

@article{dannerbauer2005,
  title = {Follow-up near-Infrared Spectroscopy of Ultraluminous Infrared Galaxies Observed by {{ISO}}},
  author = {Dannerbauer, H. and Rigopoulou, D. and Lutz, D. and Genzel, R. and Sturm, E. and Moorwood, A. F. M.},
  year = {2005},
  month = oct,
  journal = {A\&A},
  volume = {441},
  pages = {999--1010},
  issn = {0004-6361},
  doi = {10.1051/0004-6361:20052812},
  urldate = {2025-08-21}
}

@article{derossi2018,
  title = {The {{Far-infrared Emission}} of the {{First Massive Galaxies}}},
  author = {De Rossi, Maria Emilia and Rieke, George H. and Shivaei, Irene and Bromm, Volker and Lyu, Jianwei},
  year = {2018},
  month = dec,
  journal = {\apj},
  volume = {869},
  pages = {4},
  issn = {0004-637X},
  doi = {10.3847/1538-4357/aaebf8},
  urldate = {2023-08-09}
}

@ARTICLE{deugenio2023,
       author = {{D'Eugenio}, Francesco and {Cameron}, Alex J. and {Scholtz}, Jan and {Carniani}, Stefano and {Willott}, Chris J. and {Curtis-Lake}, Emma and {Bunker}, Andrew J. and {Parlanti}, Eleonora and {Maiolino}, Roberto and {Willmer}, Christopher N.~A. and {Jakobsen}, Peter and {Robertson}, Brant E. and {Johnson}, Benjamin D. and {Tacchella}, Sandro and {Cargile}, Phillip A. and {Rawle}, Tim and {Arribas}, Santiago and {Chevallard}, Jacopo and {Curti}, Mirko and {Egami}, Eiichi and {Eisenstein}, Daniel J. and {Kumari}, Nimisha and {Looser}, Tobias J. and {Rieke}, Marcia J. and {Rodr{\'\i}guez Del Pino}, Bruno and {Saxena}, Aayush and {{\"U}bler}, Hannah and {Venturi}, Giacomo and {Witstok}, Joris and {Baker}, William M. and {Bhatawdekar}, Rachana and {Bonaventura}, Nina and {Boyett}, Kristan and {Charlot}, Stephane and {Danhaive}, A. Lola and {Hainline}, Kevin N. and {Hausen}, Ryan and {Helton}, Jakob M. and {Ji}, Xihan and {Ji}, Zhiyuan and {Jones}, Gareth C. and {Juod{\v{z}}balis}, Ignas and {Maseda}, Michael V. and {P{\'e}rez-Gonz{\'a}lez}, Pablo G. and {Perna}, Michele and {Pusk{\'a}s}, D{\'a}vid and {Shivaei}, Irene and {Silcock}, Maddie S. and {Simmonds}, Charlotte and {Smit}, Renske and {Sun}, Fengwu and {Villanueva}, Natalia C. and {Williams}, Christina C. and {Zhu}, Yongda},
        title = "{JADES Data Release 3: NIRSpec/Microshutter Assembly Spectroscopy for 4000 Galaxies in the GOODS Fields}",
      journal = {\apjs},
         year = 2025,
        month = mar,
       volume = {277},
       number = {1},
          eid = {4},
        pages = {4},
          doi = {10.3847/1538-4365/ada148},
archivePrefix = {arXiv},
       eprint = {2404.06531},
 primaryClass = {astro-ph.GA},
       adsurl = {https://ui.adsabs.harvard.edu/abs/2025ApJS..277....4D}
}

@article{diamond-stanic2010,
  title = {The {{Effect}} of {{Active Galactic Nuclei}} on the {{Mid-infrared Aromatic Features}}},
  author = {{Diamond-Stanic}, Aleksandar M. and Rieke, George H.},
  year = {2010},
  month = nov,
  journal = {\apj},
  volume = {724},
  pages = {140--153},
  issn = {0004-637X},
  doi = {10.1088/0004-637X/724/1/140},
  urldate = {2025-09-01}
}

@article{diaz-santos2008,
  title = {Understanding the 8 {$M$}m versus {{Pa$\alpha$ Relationship}} on {{Subarcsecond Scales}} in {{Luminous Infrared Galaxies}}},
  author = {{D{\'i}az-Santos}, Tanio and {Alonso-Herrero}, Almudena and Colina, Luis and Packham, Christopher and Radomski, James T. and Telesco, Charles M.},
  year = {2008},
  month = sep,
  journal = {\apj},
  volume = {685},
  pages = {211--224},
  issn = {0004-637X},
  doi = {10.1086/588276},
  urldate = {2023-08-07}
}

@article{dole2004,
  title = {Confusion of {{Extragalactic Sources}} in the {{Mid-}} and {{Far-Infrared}}: {{Spitzer}} and {{Beyond}}},
  shorttitle = {Confusion of {{Extragalactic Sources}} in the {{Mid-}} and {{Far-Infrared}}},
  author = {Dole, H. and Rieke, G. H. and Lagache, G. and Puget, J. -L. and {Alonso-Herrero}, A. and Bai, L. and Blaylock, M. and Egami, E. and Engelbracht, C. W. and Gordon, K. D. and Hines, D. C. and Kelly, D. M. and Le Floc'h, E. and Misselt, K. A. and Morrison, J. E. and Muzerolle, J. and Papovich, C. and {P{\'e}rez-Gonz{\'a}lez}, P. G. and Rieke, M. J. and Rigby, J. R. and Neugebauer, G. and Stansberry, J. A. and Su, K. Y. L. and Young, E. T. and Beichman, C. A. and Richards, P. L.},
  year = {2004},
  month = sep,
  journal = {\apjs},
  volume = {154},
  pages = {93--96},
  publisher = {IOP},
  issn = {0067-0049},
  doi = {10.1086/422690},
  urldate = {2024-05-17}
}

@article{dole2006,
  title = {The Cosmic Infrared Background Resolved by {{Spitzer}}. {{Contributions}} of Mid-Infrared Galaxies to the Far-Infrared Background},
  author = {Dole, H. and Lagache, G. and Puget, J.-L. and Caputi, K. I. and {Fern{\'a}ndez-Conde}, N. and Le Floc'h, E. and Papovich, C. and {P{\'e}rez-Gonz{\'a}lez}, P. G. and Rieke, G. H. and Blaylock, M.},
  year = {2006},
  month = may,
  journal = {A\&A},
  volume = {451},
  number = {2},
  pages = {417},
  issn = {0004-6361},
  doi = {10.1051/0004-6361:20054446},
  urldate = {2022-06-21},
  langid = {english}
}

@article{dominguez2015,
  title = {Consequences of Bursty Star Formation on Galaxy Observables at High Redshifts},
  author = {Dom{\'i}nguez, Alberto and Siana, Brian and Brooks, Alyson M. and Christensen, Charlotte R. and Bruzual, Gustavo and Stark, Daniel P. and Alavi, Anahita},
  year = {2015},
  month = jul,
  journal = {MNRAS},
  volume = {451},
  pages = {839--848},
  issn = {0035-8711},
  doi = {10.1093/mnras/stv1001},
  urldate = {2025-08-26}
}

@article{draine2007,
  title = {Infrared {{Emission}} from {{Interstellar Dust}}. {{IV}}. {{The Silicate-Graphite-PAH Model}} in the {{Post-Spitzer Era}}},
  author = {Draine, B. T. and Li, Aigen},
  year = {2007},
  month = mar,
  journal = {\apj},
  volume = {657},
  pages = {810--837},
  issn = {0004-637X},
  doi = {10.1086/511055},
  urldate = {2022-05-01}
}

@article{draine2021,
  title = {Excitation of {{Polycyclic Aromatic Hydrocarbon Emission}}: {{Dependence}} on {{Size Distribution}}, {{Ionization}}, and {{Starlight Spectrum}} and {{Intensity}}},
  shorttitle = {Excitation of {{Polycyclic Aromatic Hydrocarbon Emission}}},
  author = {Draine, B. T. and Li, Aigen and Hensley, Brandon S. and Hunt, L. K. and Sandstrom, K. and Smith, J. -D. T.},
  year = {2021},
  month = aug,
  journal = {\apj},
  volume = {917},
  pages = {3},
  issn = {0004-637X},
  doi = {10.3847/1538-4357/abff51},
  urldate = {2023-01-18}
}

@article{egorov2023,
  title = {{{PHANGS-JWST First Results}}: {{Destruction}} of the {{PAH Molecules}} in {{H II Regions Probed}} by {{JWST}} and {{MUSE}}},
  shorttitle = {{{PHANGS-JWST First Results}}},
  author = {Egorov, Oleg V. and Kreckel, Kathryn and Sandstrom, Karin M. and Leroy, Adam K. and Glover, Simon C. O. and Groves, Brent and Kruijssen, J. M. Diederik and Barnes, {\relax Ashley}. T. and Belfiore, Francesco and Bigiel, F. and Blanc, Guillermo A. and Boquien, M{\'e}d{\'e}ric and Cao, Yixian and Chastenet, J{\'e}r{\'e}my and Chevance, M{\'e}lanie and Congiu, Enrico and Dale, Daniel A. and Emsellem, Eric and Grasha, Kathryn and Klessen, Ralf S. and Larson, Kirsten L. and Liu, Daizhong and Murphy, Eric J. and Pan, Hsi-An and Pessa, Ismael and Pety, J{\'e}r{\^o}me and Rosolowsky, Erik and Scheuermann, Fabian and Schinnerer, Eva and Sutter, Jessica and Thilker, David A. and Watkins, Elizabeth J. and Williams, Thomas G.},
  year = {2023},
  month = feb,
  journal = {\apj},
  volume = {944},
  pages = {L16},
  issn = {0004-637X},
  doi = {10.3847/2041-8213/acac92},
  urldate = {2025-08-21}
}

@ARTICLE{eisenstein2023,
       author = {{Eisenstein}, Daniel J. and {Willott}, Chris and {Alberts}, Stacey and {Arribas}, Santiago and {Bonaventura}, Nina and {Bunker}, Andrew J. and {Cameron}, Alex J. and {Carniani}, Stefano and {Charlot}, Stephane and {Curtis-Lake}, Emma and {D'Eugenio}, Francesco and {Ferruit}, Pierre and {Giardino}, Giovanna and {Hainline}, Kevin and {Hausen}, Ryan and {Jakobsen}, Peter and {Johnson}, Benjamin D. and {Maiolino}, Roberto and {Rauscher}, Bernard J. and {Rieke}, Marcia and {Rieke}, George and {Rix}, Hans-Walter and {Robertson}, Brant and {Stark}, Daniel P. and {Tacchella}, Sandro and {Williams}, Christina C. and {Willmer}, Christopher N.~A. and {Baker}, William M. and {Baum}, Stefi and {Bhatawdekar}, Rachana and {Boyett}, Kristan and {Chen}, Zuyi and {Chevallard}, Jacopo and {Circosta}, Chiara and {Curti}, Mirko and {Danhaive}, A. Lola and {DeCoursey}, Christa and {Endsley}, Ryan and {de Graaff}, Anna and {Dressler}, Alan and {Egami}, Eiichi and {Helton}, Jakob M. and {Hviding}, Raphael E. and {Ji}, Zhiyuan and {Jones}, Gareth C. and {Kumari}, Nimisha and {L{\"u}tzgendorf}, Nora and {Laseter}, Isaac and {Looser}, Tobias J. and {Lyu}, Jianwei and {Maseda}, Michael V. and {Nelson}, Erica and {Parlanti}, Eleonora and {Perna}, Michele and {Pusk{\'a}s}, D{\'a}vid and {Rawle}, Tim and {Rodr{\'\i}guez Del Pino}, Bruno and {Rujopakarn}, Wiphu and {Sandles}, Lester and {Saxena}, Aayush and {Scholtz}, Jan and {Sharpe}, Katherine and {Shivaei}, Irene and {Silcock}, Maddie S. and {Simmonds}, Charlotte and {Skarbinski}, Maya and {Smit}, Renske and {Stone}, Meredith and {Suess}, Katherine A. and {Sun}, Fengwu and {Tang}, Mengtao and {Topping}, Michael W. and {{\"U}bler}, Hannah and {Villanueva}, Natalia C. and {Wallace}, Imaan E.~B. and {Whitler}, Lily and {Witstok}, Joris and {Woodrum}, Charity},
        title = "{Overview of the JWST Advanced Deep Extragalactic Survey (JADES)}",
      journal = {\apjs},
         year = 2026,
        month = mar,
       volume = {283},
       number = {1},
          eid = {6},
        pages = {6},
          doi = {10.3847/1538-4365/ae3163},
archivePrefix = {arXiv},
       eprint = {2306.02465},
 primaryClass = {astro-ph.GA},
       adsurl = {https://ui.adsabs.harvard.edu/abs/2026ApJS..283....6E}
}

@article{elbaz2005,
  title = {Observational Evidence for the Presence of {{PAHs}} in Distant {{Luminous Infrared Galaxies}} Using {{ISO}} and {{Spitzer}}},
  author = {Elbaz, D. and Le Floc'h, E. and Dole, H. and Marcillac, D.},
  year = {2005},
  month = apr,
  journal = {A\&A},
  volume = {434},
  pages = {L1-L4},
  issn = {0004-6361},
  doi = {10.1051/0004-6361:200500095},
  urldate = {2025-08-21}
}

@article{elbaz2011,
  title = {{{GOODS-Herschel}}: An Infrared Main Sequence for Star-Forming Galaxies},
  shorttitle = {{{GOODS-Herschel}}},
  author = {Elbaz, D. and Dickinson, M. and Hwang, H. S. and {D{\'i}az-Santos}, T. and Magdis, G. and Magnelli, B. and Le Borgne, D. and Galliano, F. and Pannella, M. and Chanial, P. and Armus, L. and Charmandaris, V. and Daddi, E. and Aussel, H. and Popesso, P. and Kartaltepe, J. and Altieri, B. and Valtchanov, I. and Coia, D. and Dannerbauer, H. and Dasyra, K. and Leiton, R. and Mazzarella, J. and Alexander, D. M. and Buat, V. and Burgarella, D. and Chary, R. -R. and Gilli, R. and Ivison, R. J. and Juneau, S. and Le Floc'h, E. and Lutz, D. and Morrison, G. E. and Mullaney, J. R. and Murphy, E. and Pope, A. and Scott, D. and Brodwin, M. and Calzetti, D. and Cesarsky, C. and Charlot, S. and Dole, H. and Eisenhardt, P. and Ferguson, H. C. and F{\"o}rster Schreiber, N. and Frayer, D. and Giavalisco, M. and Huynh, M. and Koekemoer, A. M. and Papovich, C. and Reddy, N. and Surace, C. and Teplitz, H. and Yun, M. S. and Wilson, G.},
  year = {2011},
  month = sep,
  journal = {A\&A},
  volume = {533},
  pages = {A119},
  issn = {0004-6361},
  doi = {10.1051/0004-6361/201117239},
  urldate = {2024-05-01}
}

@article{emami2019,
  title = {A {{Closer Look}} at {{Bursty Star Formation}} with {{L H$\alpha$}} and {{L UV Distributions}}},
  author = {Emami, Najmeh and Siana, Brian and Weisz, Daniel R. and Johnson, Benjamin D. and Ma, Xiangcheng and {El-Badry}, Kareem},
  year = {2019},
  month = aug,
  journal = {\apj},
  volume = {881},
  pages = {71},
  issn = {0004-637X},
  doi = {10.3847/1538-4357/ab211a},
  urldate = {2025-08-21}
}

@article{engelbracht2005,
  title = {Metallicity {{Effects}} on {{Mid-Infrared Colors}} and the 8 {$M$}m {{PAH Emission}} in {{Galaxies}}},
  author = {Engelbracht, C. W. and Gordon, K. D. and Rieke, G. H. and Werner, M. W. and Dale, D. A. and Latter, W. B.},
  year = {2005},
  month = jul,
  journal = {\apj},
  volume = {628},
  pages = {L29-L32},
  issn = {0004-637X},
  doi = {10.1086/432613},
  urldate = {2025-08-21}
}

@article{engelbracht2008,
  title = {Metallicity {{Effects}} on {{Dust Properties}} in {{Starbursting Galaxies}}},
  author = {Engelbracht, C. W. and Rieke, G. H. and Gordon, K. D. and Smith, J. -D. T. and Werner, M. W. and Moustakas, J. and Willmer, C. N. A. and Vanzi, L.},
  year = {2008},
  month = may,
  journal = {\apj},
  volume = {678},
  pages = {804--827},
  issn = {0004-637X},
  doi = {10.1086/529513},
  urldate = {2025-08-21}
}

@article{farrah2007,
  title = {High-{{Resolution Mid-Infrared Spectroscopy}} of {{Ultraluminous Infrared Galaxies}}},
  author = {Farrah, D. and {Bernard-Salas}, J. and Spoon, H. W. W. and Soifer, B. T. and Armus, L. and Brandl, B. and Charmandaris, V. and Desai, V. and Higdon, S. and Devost, D. and Houck, J.},
  year = {2007},
  month = sep,
  journal = {\apj},
  volume = {667},
  pages = {149--169},
  issn = {0004-637X},
  doi = {10.1086/520834},
  urldate = {2025-08-21}
}

@article{fiolet2010,
  title = {Mid-Infrared Spectroscopy of {{Spitzer-selected}} Ultra-Luminous Starbursts at z {\textasciitilde} 2},
  author = {Fiolet, N. and Omont, A. and Lagache, G. and Bertincourt, B. and Fadda, D. and Baker, A. J. and Beelen, A. and Berta, S. and Boulanger, F. and Farrah, D. and Kov{\'a}cs, A. and Lonsdale, C. and Owen, F. and Polletta, M. and Shupe, D. and Yan, L.},
  year = {2010},
  month = dec,
  journal = {A\&A},
  volume = {524},
  pages = {A33},
  issn = {0004-6361},
  doi = {10.1051/0004-6361/201015504},
  urldate = {2025-09-01}
}

@article{floresvelazquez2021,
  title = {The Time-Scales Probed by Star Formation Rate Indicators for Realistic, Bursty Star Formation Histories from the {{FIRE}} Simulations},
  author = {Flores Vel{\'a}zquez, Jos{\'e} A. and Gurvich, Alexander B. and {Faucher-Gigu{\`e}re}, Claude-Andr{\'e} and Bullock, James S. and Starkenburg, Tjitske K. and Moreno, Jorge and Lazar, Alexandres and Mercado, Francisco J. and Stern, Jonathan and Sparre, Martin and Hayward, Christopher C. and Wetzel, Andrew and {El-Badry}, Kareem},
  year = {2021},
  month = mar,
  journal = {MNRAS},
  volume = {501},
  pages = {4812--4824},
  publisher = {OUP},
  issn = {0035-8711},
  doi = {10.1093/mnras/staa3893},
  urldate = {2025-08-14}
}

@article{foley2018,
  title = {Molecular Hydrogen Formation on Interstellar {{PAHs}} through {{Eley-Rideal}} Abstraction Reactions},
  author = {Foley, Nolan and Cazaux, S. and Egorov, D. and Boschman, L. M. P. V. and Hoekstra, R. and Schlath{\"o}lter, T.},
  year = {2018},
  month = sep,
  journal = {MNRAS},
  volume = {479},
  pages = {649--656},
  issn = {0035-8711},
  doi = {10.1093/mnras/sty1528},
  urldate = {2025-08-21}
}

@article{galliano2008,
  title = {Stellar {{Evolutionary Effects}} on the {{Abundances}} of {{Polycyclic Aromatic Hydrocarbons}} and {{Supernova-Condensed Dust}} in {{Galaxies}}},
  author = {Galliano, Fr{\'e}d{\'e}ric and Dwek, Eli and Chanial, Pierre},
  year = {2008},
  month = jan,
  journal = {\apj},
  volume = {672},
  pages = {214--243},
  publisher = {IOP},
  issn = {0004-637X},
  doi = {10.1086/523621},
  urldate = {2024-10-11}
}

@article{garcia-bernete2022,
  title = {A High Angular Resolution View of the {{PAH}} Emission in {{Seyfert}} Galaxies Using {{JWST}}/{{MRS}} Data},
  author = {{Garc{\'i}a-Bernete}, I. and Rigopoulou, D. and {Alonso-Herrero}, A. and Donnan, F. R. and Roche, P. F. and {Pereira-Santaella}, M. and Labiano, A. and {Peralta de Arriba}, L. and Izumi, T. and Ramos Almeida, C. and Shimizu, T. and H{\"o}nig, S. and {Garc{\'i}a-Burillo}, S. and Rosario, D. J. and Ward, M. J. and Bellocchi, E. and Hicks, E. K. S. and Fuller, L. and Packham, C.},
  year = {2022},
  month = oct,
  journal = {A\&A},
  volume = {666},
  pages = {L5},
  issn = {0004-6361},
  doi = {10.1051/0004-6361/202244806},
  urldate = {2024-06-19}
}

@article{gardner2023,
  title = {The {{James Webb Space Telescope Mission}}},
  author = {Gardner, Jonathan P. and Mather, John C. and Abbott, Randy and Abell, James S. and Abernathy, Mark and Abney, Faith E. and Abraham, John G. and Abraham, Roberto and {Abul-Huda}, Yasin M. and Acton, Scott and Adams, Cynthia K. and Adams, Evan and Adler, David S. and Adriaensen, Maarten and Aguilar, Jonathan Albert and Ahmed, Mansoor and Ahmed, Nasif S. and Ahmed, Tanjira and Albat, R{\"u}deger and Albert, Lo{\"i}c and Alberts, Stacey and Aldridge, David and Allen, Mary Marsha and Allen, Shaune S. and Altenburg, Martin and Altunc, Serhat and Alvarez, Jose Lorenzo and {\'A}lvarez-M{\'a}rquez, Javier and {Alves de Oliveira}, Catarina and Ambrose, Leslie L. and Anandakrishnan, Satya M. and Andersen, Gregory C. and Anderson, Harry James and Anderson, Jay and Anderson, Kristen and Anderson, Sara M. and Aprea, Julio and Archer, Benita J. and Arenberg, Jonathan W. and Argyriou, Ioannis and Arribas, Santiago and Artigau, {\'E}tienne and Arvai, Amanda Rose and Atcheson, Paul and Atkinson, Charles B. and Averbukh, Jesse and Aymergen, Cagatay and Bacinski, John J. and Baggett, Wayne E. and Bagnasco, Giorgio and Baker, Lynn L. and Balzano, Vicki Ann and Banks, Kimberly A. and Baran, David A. and Barker, Elizabeth A. and Barrett, Larry K. and Barringer, Bruce O. and Barto, Allison and Bast, William and Baudoz, Pierre and Baum, Stefi and Beatty, Thomas G. and Beaulieu, Mathilde and Bechtold, Kathryn and Beck, Tracy and Beddard, Megan M. and Beichman, Charles and Bellagama, Larry and Bely, Pierre and Berger, Timothy W. and Bergeron, Louis E. and Bernier, Antoine-Darveau and Bertch, Maria D. and Beskow, Charlotte and Betz, Laura E. and Biagetti, Carl P. and Birkmann, Stephan and Bjorklund, Kurt F. and Blackwood, James D. and Blazek, Ronald Paul and Blossfeld, Stephen and Bluth, Marcel and Boccaletti, Anthony and Boegner, Jr., Martin E. and Bohlin, Ralph C. and Boia, John Joseph and B{\"o}ker, Torsten and Bonaventura, N. and Bond, Nicholas A. and Bosley, Kari Ann and Boucarut, Rene A. and Bouchet, Patrice and Bouwman, Jeroen and Bower, Gary and Bowers, Ariel S. and Bowers, Charles W. and Boyce, Leslye A. and Boyer, Christine T. and Boyer, Martha L. and Boyer, Michael and Boyer, Robert and Bradley, Larry D. and Brady, Gregory R. and Brandl, Bernhard R. and Brannen, Judith L. and Breda, David and Bremmer, Harold G. and Brennan, David and Bresnahan, Pamela A. and Bright, Stacey N. and Broiles, Brian J. and Bromenschenkel, Asa and Brooks, Brian H. and Brooks, Keira J. and Brown, Bob and Brown, Bruce and Brown, Thomas M. and Bruce, Barry W. and Bryson, Jonathan G. and Bujanda, Edwin D. and Bullock, Blake M. and Bunker, A. J. and Bureo, Rafael and Burt, Irving J. and Bush, James Aaron and Bushouse, Howard A. and Bussman, Marie C. and Cabaud, Olivier and Cale, Steven and Calhoon, Charles D. and Calvani, Humberto and Canipe, Alicia M. and Caputo, Francis M. and Cara, Mihai and Carey, Larkin and Case, Michael Eli and Cesari, Thaddeus and Cetorelli, Lee D. and Chance, Don R. and Chandler, Lynn and Chaney, Dave and Chapman, George N. and Charlot, S. and Chayer, Pierre and Cheezum, Jeffrey I. and Chen, Bin and Chen, Christine H. and Cherinka, Brian and Chichester, Sarah C. and Chilton, Zachary S. and Chittiraibalan, Dharini and Clampin, Mark and Clark, Charles R. and Clark, Kerry W. and Clark, Stephanie M. and Claybrooks, Edward E. and Cleveland, Keith A. and Cohen, Andrew L. and Cohen, Lester M. and Col{\'o}n, Knicole D. and Coleman, Benee L. and Colina, Luis and Comber, Brian J. and Comeau, Thomas M. and Comer, Thomas and Conde Reis, Alain and Connolly, Dennis C. and Conroy, Kyle E. and Contos, Adam R. and Contreras, James and Cook, Neil J. and Cooper, James L. and Cooper, Rachel Aviva and Correia, Michael F. and Correnti, Matteo and Cossou, Christophe and Costanza, Brian F. and Coulais, Alain and Cox, Colin R. and Coyle, Ray T. and Cracraft, Misty M. and Crew, Keith A. and Curtis, Gary J. and Cusveller, Bianca and Da Costa Maciel, Cleyciane and Dailey, Christopher T. and Daugeron, Fr{\'e}d{\'e}ric and Davidson, Greg S. and Davies, James E. and Davis, Katherine Anne and Davis, Michael S. and Day, Ratna and {de Chambure}, Daniel and {de Jong}, Pauline and De Marchi, Guido and Dean, Bruce H. and Decker, John E. and Delisa, Amy S. and Dell, Lawrence C. and Dellagatta, Gail and Dembinska, Franciszka and Demosthenes, Sandor and Dencheva, Nadezhda M. and Deneu, Philippe and DePriest, William W. and Deschenes, Jeremy and Dethienne, Nathalie and Detre, {\"O}rs Hunor and Diaz, Rosa Izela and Dicken, Daniel and DiFelice, Audrey S. and Dillman, Matthew and Disharoon, Maureen O. and Dixon, William V. and Doggett, Jesse B. and Dominguez, Keisha L. and Donaldson, Thomas S. and {Doria-Warner}, Cristina M. and Santos, Tony Dos and Doty, Heather and Douglas, Jr., Robert E. and Doyon, Ren{\'e} and Dressler, Alan and Driggers, Jennifer and Driggers, Phillip A. and Dunn, Jamie L. and DuPrie, Kimberly C. and Dupuis, Jean and Durning, John and Dutta, Sanghamitra B. and Earl, Nicholas M. and Eccleston, Paul and Ecobichon, Pascal and Egami, Eiichi and Ehrenwinkler, Ralf and Eisenhamer, Jonathan D. and Eisenhower, Michael and Eisenstein, Daniel J. and El Hamel, Zaky and Elie, Michelle L. and Elliott, James and Elliott, Kyle Wesley and Engesser, Michael and Espinoza, N{\'e}stor and Etienne, Odessa and Etxaluze, Mireya and Evans, Leah and Fabreguettes, Luce and Falcolini, Massimo and Falini, Patrick R. and Fatig, Curtis and Feeney, Matthew and Feinberg, Lee D. and Fels, Raymond and Ferdous, Nazma and Ferguson, Henry C. and Ferrarese, Laura and Ferreira, Marie-H{\'e}l{\'e}ne and Ferruit, Pierre and Ferry, Malcolm and Filippazzo, Joseph Charles and Firre, Daniel and Fix, Mees and Flagey, Nicolas and Flanagan, Kathryn A. and Fleming, Scott W. and Florian, Michael and Flynn, James R. and Foiadelli, Luca and Fontaine, Mark R. and Fontanella, Erin Marie and Forshay, Peter Randolph and Fortner, Elizabeth A. and Fox, Ori D. and Framarini, Alexandro P. and Francisco, John I. and Franck, Randy and Franx, Marijn and Franz, David E. and Friedman, Scott D. and Friend, Katheryn E. and Frost, James R. and Fu, Henry and Fullerton, Alexander W. and Gaillard, Lionel and Galkin, Sergey and Gallagher, Ben and Galyer, Anthony D. and Garc{\'i}a Mar{\'i}n, Macarena and Gardner, Lisa E. and Garland, Dennis and Garrett, Bruce Albert and Gasman, Danny and G{\'a}sp{\'a}r, Andr{\'a}s and Gastaud, Ren{\'e} and Gaudreau, Daniel and Gauthier, Peter Timothy and Geers, Vincent and Geithner, Paul H. and Gennaro, Mario and Gerber, John and Gereau, John C. and Giampaoli, Robert and Giardino, Giovanna and Gibbons, Paul C. and Gilbert, Karoline and Gilman, Larry and Girard, Julien H. and Giuliano, Mark E. and Gkountis, Konstantinos and Glasse, Alistair and Glassmire, Kirk Zachary and Glauser, Adrian Michael and Glazer, Stuart D. and Goldberg, Joshua and Golimowski, David A. and Gonzaga, Shireen P. and Gordon, Karl D. and Gordon, Shawn J. and Goudfrooij, Paul and Gough, Michael J. and Graham, Adrian J. and Grau, Christopher M. and Green, Joel David and Greene, Gretchen R. and Greene, Thomas P. and Greenfield, Perry E. and Greenhouse, Matthew A. and Greve, Thomas R. and Greville, Edgar M. and Grimaldi, Stefano and Groe, Frank E. and Groebner, Andrew and Grumm, David M. and Grundy, Timothy and G{\"u}del, Manuel and Guillard, Pierre and Guldalian, John and Gunn, Christopher A. and Gurule, Anthony and Gutman, Irvin Meyer and Guy, Paul D. and Guyot, Benjamin and Hack, Warren J. and Haderlein, Peter and Hagan, James B. and Hagedorn, Andria and Hainline, Kevin and Haley, Craig and Hami, Maryam and Hamilton, Forrest Clifford and Hammann, Jeffrey and Hammel, Heidi B. and Hanley, Christopher J. and Hansen, Carl August and Hardy, Bruce and Harnisch, Bernd and Harr, Michael Hunter and Harris, Pamela and Hart, Jessica Ann and Hartig, George F. and Hasan, Hashima and Hashim, Kathleen Marie and Hashimoto, Ryan and Haskins, Sujee J. and Hawkins, Robert Edward and Hayden, Brian and Hayden, William L. and Healy, Mike and Hecht, Karen and Heeg, Vince J. and Hejal, Reem and Helm, Kristopher A. and Hengemihle, Nicholas J. and Henning, Thomas and Henry, Alaina and Henry, Ronald L. and Henshaw, Katherine and Hernandez, Scarlin and Herrington, Donald C. and Heske, Astrid and Hesman, Brigette Emily and Hickey, David L. and Hilbert, Bryan N. and Hines, Dean C. and Hinz, Michael R. and Hirsch, Michael and Hitcho, Robert S. and Hodapp, Klaus and Hodge, Philip E. and Hoffman, Melissa and Holfeltz, Sherie T. and Holler, Bryan Jason and Hoppa, Jennifer Rose and Horner, Scott and Howard, Joseph M. and Howard, Richard J. and Huber, Jean M. and Hunkeler, Joseph S. and Hunter, Alexander and Hunter, David Gavin and Hurd, Spencer W. and Hurst, Brendan J. and Hutchings, John B. and Hylan, Jason E. and Ignat, Luminita Ilinca and Illingworth, Garth and Irish, Sandra M. and Isaacs, III, John C. and Jackson, Jr., Wallace C. and Jaffe, Daniel T. and Jahic, Jasmin and Jahromi, Amir and Jakobsen, Peter and James, Bryan and James, John C. and James, LeAndrea Rae and Jamieson, William Brian and Jandra, Raymond D. and Jayawardhana, Ray and Jedrzejewski, Robert and Jeffers, Basil S. and Jensen, Peter and Joanne, Egges and Johns, Alan T. and Johnson, Carl A. and Johnson, Eric L. and Johnson, Patricia and Johnson, Phillip Stephen and Johnson, Thomas K. and Johnson, Timothy W. and Johnstone, Doug and Jollet, Delphine and Jones, Danny P. and Jones, Gregory S. and Jones, Olivia C. and Jones, Ronald A. and Jones, Vicki and Jordan, Ian J. and Jordan, Margaret E. and Jue, Reginald and Jurkowski, Mark H. and Justis, Grant and Justtanont, Kay and Kaleida, Catherine C. and Kalirai, Jason S. and Kalmanson, Phillip Cabrales and Kaltenegger, Lisa and Kammerer, Jens and Kan, Samuel K. and Kanarek, Graham Childs and Kao, Shaw-Hong and Karakla, Diane M. and Karl, Hermann and Kassin, Susan A. and Kauffman, David D. and Kavanagh, Patrick and Kelley, Leigh L. and Kelly, Douglas M. and Kendrew, Sarah and Kennedy, Herbert V. and Kenny, Deborah A. and {Keski-Kuha}, Ritva A. and Keyes, Charles D. and Khan, Ali and Kidwell, Richard C. and Kimble, Randy A. and King, James S. and King, Richard C. and Kinzel, Wayne M. and Kirk, Jeffrey R. and Kirkpatrick, Marc E. and Klaassen, Pamela and Klingemann, Lana and Klintworth, Paul U. and Knapp, Bryan Adam and Knight, Scott and Knollenberg, Perry J. and Knutsen, Daniel Mark and Koehler, Robert and Koekemoer, Anton M. and Kofler, Earl T. and Kontson, Vicki L. and Kovacs, Aiden Rose and {Kozhurina-Platais}, Vera and Krause, Oliver and Kriss, Gerard A. and Krist, John and Kristoffersen, Monica R. and Krogel, Claudia and Krueger, Anthony P. and Kulp, Bernard A. and Kumari, Nimisha and Kwan, Sandy W. and Kyprianou, Mark and Labador, Aurora Gadiano and Labiano, {\'A}lvaro and Lafreni{\`e}re, David and Lagage, Pierre-Olivier and Laidler, Victoria G. and Laine, Benoit and Laird, Simon and Lajoie, Charles-Philippe and Lallo, Matthew D. and Lam, May Yen and LaMassa, Stephanie Marie and Lambros, Scott D. and Lampenfield, Richard Joseph and Lander, Matthew Ed and Langston, James Hutton and Larson, Kirsten and Larson, Melora and LaVerghetta, Robert Joseph and Law, David R. and Lawrence, Jon F. and Lee, David W. and Lee, Janice and Lee, Yat-Ning Paul and Leisenring, Jarron and Leveille, Michael Dunlap and Levenson, Nancy A. and Levi, Joshua S. and Levine, Marie B. and Lewis, Dan and Lewis, Jake and Lewis, Nikole and Libralato, Mattia and Lidon, Norbert and Liebrecht, Paula Louisa and Lightsey, Paul and Lilly, Simon and Lim, Frederick C. and Lim, Pey Lian and Ling, Sai-Kwong and Link, Lisa J. and Link, Miranda Nicole and Lipinski, Jamie L. and Liu, XiaoLi and Lo, Amy S. and Lobmeyer, Lynette and Logue, Ryan M. and Long, Chris A. and Long, Douglas R. and Long, Ilana D. and Long, Knox S. and {L{\'o}pez-Caniego}, Marcos and Lotz, Jennifer M. and {Love-Pruitt}, Jennifer M. and Lubskiy, Michael and Luers, Edward B. and Luetgens, Robert A. and Luevano, Annetta J. and Lui, Sarah Marie G. Flores and Lund, III, James M. and Lundquist, Ray A. and Lunine, Jonathan and L{\"u}tzgendorf, Nora and Lynch, Richard J. and MacDonald, Alex J. and MacDonald, Kenneth and Macias, Matthew J. and Macklis, Keith I. and Maghami, Peiman and Maharaja, Rishabh Y. and Maiolino, Roberto and Makrygiannis, Konstantinos G. and Malla, Sunita Giri and Malumuth, Eliot M. and Manjavacas, Elena and Marini, Andrea and Marrione, Amanda and Marston, Anthony and Martel, Andr{\'e} R. and Martin, Didier and Martin, Peter G. and Martinez, Kristin L. and Maschmann, Marc and Masci, Gregory L. and Masetti, Margaret E. and Maszkiewicz, Michael and Matthews, Gary and Matuskey, Jacob E. and McBrayer, Glen A. and McCarthy, Donald W. and McCaughrean, Mark J. and McClare, Leslie A. and McClare, Michael D. and McCloskey, John C. and McClurg, Taylore D. and McCoy, Martin and McElwain, Michael W. and McGregor, Roy D. and McGuffey, Douglas B. and McKay, Andrew G. and McKenzie, William K. and McLean, Brian and McMaster, Matthew and McNeil, Warren and De Meester, Wim and Mehalick, Kimberly L. and Meixner, Margaret and Mel{\'e}ndez, Marcio and Menzel, Michael P. and Menzel, Michael T. and Merz, Matthew and Mesterharm, David D. and Meyer, Michael R. and Meyett, Michele L. and Meza, Luis E. and Midwinter, Calvin and Milam, Stefanie N. and Miller, Jay Todd and Miller, William C. and Miskey, Cherie L. and Misselt, Karl and Mitchell, Eileen P. and Mohan, Martin and Montoya, Emily E. and Moran, Michael J. and Morishita, Takahiro and {Moro-Mart{\'i}n}, Amaya and Morrison, Debra L. and Morrison, Jane and Morse, Ernie C. and Moschos, Michael and Moseley, S. H. and Mosier, Gary E. and Mosner, Peter and Mountain, Matt and Muckenthaler, Jason S. and Mueller, Donald G. and Mueller, Migo and Muhiem, Daniella and M{\"u}hlmann, Prisca and Mullally, Susan Elizabeth and Mullen, Stephanie M. and Munger, Alan J. and Murphy, Jess and Murray, Katherine T. and Muzerolle, James C. and Mycroft, Matthew and Myers, Andrew and Myers, Carey R. and Myers, Fred Richard R. and Myers, Richard and Myrick, Kaila and Nagle, IV, Adrian F. and Nayak, Omnarayani and Naylor, Bret and Neff, Susan G. and Nelan, Edmund P. and Nella, John and Nguyen, Duy Tuong and Nguyen, Michael N. and Nickson, Bryony and Nidhiry, John Joseph and Niedner, Malcolm B. and {Nieto-Santisteban}, Maria and Nikolov, Nikolay K. and Nishisaka, Mary Ann and {Noriega-Crespo}, Alberto and Nota, Antonella and O'Mara, Robyn C. and Oboryshko, Michael and O'Brien, Marcus B. and Ochs, William R. and Offenberg, Joel D. and Ogle, Patrick Michael and Ohl, Raymond G. and Olmsted, Joseph Hamden and Osborne, Shannon Barbara and O'Shaughnessy, Brian Patrick and {\"O}stlin, G{\"o}ran and O'Sullivan, Brian and Otor, O. Justin and Ottens, Richard and Ouellette, Nathalie N. -Q. and Outlaw, Daria J. and Owens, Beverly A. and Pacifici, Camilla and Page, James Christophe and Paranilam, James G. and Park, Sang and Parrish, Keith A. and Paschal, Laura and Patapis, Polychronis and Patel, Jignasha and Patrick, Keith and Pattishall, Jr., Robert A. and Paul, Douglas William and Paul, Shirley J. and Pauly, Tyler Andrew and Pavlovsky, Cheryl M. and {Pe{\~n}a-Guerrero}, Maria and Pedder, Andrew H. and Peek, Matthew Weldon and Pelham, Patricia A. and Penanen, Konstantin and Perriello, Beth A. and Perrin, Marshall D. and Perrine, Richard F. and Perrygo, Chuck and Peslier, Muriel and Petach, Michael and Peterson, Karla A. and Pfarr, Tom and Pierson, James M. and Pietraszkiewicz, Martin and Pilchen, Guy and Pipher, Judy L. and Pirzkal, Norbert and Pitman, Joseph T. and Player, Danielle M. and Plesha, Rachel and Plitzke, Anja and Pohner, John A. and Poletis, Karyn Konstantin and Pollizzi, Joseph A. and Polster, Ethan and Pontius, James T. and Pontoppidan, Klaus and Porges, Susana C. and Potter, Gregg D. and Prescott, Stephen and Proffitt, Charles R. and Pueyo, Laurent and Quispe Neira, Irma Aracely and Radich, Armando and Rager, Reiko T. and Rameau, Julien and Ramey, Deborah D. and Ramos Alarcon, Rafael and Rampini, Riccardo and Rapp, Robert and Rashford, Robert A. and Rauscher, Bernard J. and Ravindranath, Swara and Rawle, Timothy and Rawlings, Tynika N. and Ray, Tom and Regan, Michael W. and Rehm, Brian and Rehm, Kenneth D. and Reid, Neill and Reis, Carl A. and Renk, Florian and Reoch, Tom B. and Ressler, Michael and Rest, Armin W. and Reynolds, Paul J. and Richon, Joel G. and Richon, Karen V. and Ridgaway, Michael and Riedel, Adric Richard and Rieke, George H. and Rieke, Marcia J. and Rifelli, Richard E. and Rigby, Jane R. and Riggs, Catherine S. and Ringel, Nancy J. and Ritchie, Christine E. and Rix, Hans-Walter and Robberto, Massimo and Robinson, Gregory L. and Robinson, Michael S. and Robinson, Orion and Rock, Frank W. and Rodriguez, David R. and {Rodr{\'i}guez del Pino}, Bruno and Roellig, Thomas and Rohrbach, Scott O. and Roman, Anthony J. and Romelfanger, Frederick J. and Romo, Jr., Felipe P. and Rosales, Jose J. and Rose, Perry and Roteliuk, Anthony F. and Roth, Marc N. and Rothwell, Braden Quinn and Rouzaud, Sylvain and Rowe, Jason and Rowlands, Neil and Roy, Arpita and Royer, Pierre and Rui, Chunlei and Rumler, Peter and Rumpl, William and Russ, Melissa L. and Ryan, Michael B. and Ryan, Richard M. and Saad, Karl and Sabata, Modhumita and Sabatino, Rick and Sabbi, Elena and Sabelhaus, Phillip A. and Sabia, Stephen and Sahu, Kailash C. and Saif, Babak N. and Salvignol, Jean-Christophe and {Samara-Ratna}, Piyal and Samuelson, Bridget S. and Sanders, Felicia A. and Sappington, Bradley and Sargent, B. A. and Sauer, Arne and Savadkin, Bruce J. and Sawicki, Marcin and Schappell, Tina M. and Scheffer, Caroline and Scheithauer, Silvia and Scherer, Ron and Schiff, Conrad and Schlawin, Everett and Schmeitzky, Olivier and Schmitz, Tyler S. and Schmude, Donald J. and Schneider, Analyn and Schreiber, J{\"u}rgen and {Schroeven-Deceuninck}, Hilde and Schultz, John J. and Schwab, Ryan and Schwartz, Curtis H. and Scoccimarro, Dario and Scott, John F. and Scott, Michelle B. and Seaton, Bonita L. and Seely, Bruce S. and Seery, Bernard and Seidleck, Mark and Sembach, Kenneth and Shanahan, Clare Elizabeth and Shaughnessy, Bryan and Shaw, Richard A. and Shay, Christopher Michael and Sheehan, Even and Sheth, Kartik and Shih, Hsin-Yi and Shivaei, Irene and Siegel, Noah and Sienkiewicz, Matthew G. and Simmons, Debra D. and Simon, Bernard P. and Sirianni, Marco and Sivaramakrishnan, Anand and Slade, Jeffrey E. and Sloan, G. C. and Slocum, Christine E. and Slowinski, Steven E. and Smith, Corbett T. and Smith, Eric P. and Smith, Erin C. and Smith, Koby and Smith, Robert and Smith, Stephanie J. and Smolik, John L. and Soderblom, David R. and Sohn, Sangmo Tony and Sokol, Jeff and Sonneborn, George and Sontag, Christopher D. and Sooy, Peter R. and Soummer, Remi and Southwood, Dana M. and Spain, Kay and Sparmo, Joseph and Speer, David T. and Spencer, Richard and Sprofera, Joseph D. and Stallcup, Scott S. and Stanley, Marcia K. and Stansberry, John A. and Stark, Christopher C. and Starr, Carl W. and Stassi, Diane Y. and Steck, Jane A. and Steeley, Christine D. and Stephens, Matthew A. and Stephenson, Ralph J. and Stewart, Alphonso C. and Stiavelli, Massimo and , Hervey Jr., Stockman and Strada, Paolo and Straughn, Amber N. and Streetman, Scott and Strickland, David Kendal and Strobele, Jingping F. and Stuhlinger, Martin and Stys, Jeffrey Edward and Such, Miguel and Sukhatme, Kalyani and Sullivan, Joseph F. and Sullivan, Pamela C. and Sumner, Sandra M. and Sun, Fengwu and Sunnquist, Benjamin Dale and Swade, Daryl Allen and Swam, Michael S. and Swenton, Diane F. and Swoish, Robby A. and Tam Litten, Oi In and Tamas, Laszlo and Tao, Andrew and Taylor, David K. and Taylor, Joanna M. and {te Plate}, Maurice and Van Tea, Mason and Teague, Kelly K. and Telfer, Randal C. and Temim, Tea and Texter, Scott C. and Thatte, Deepashri G. and Thompson, Christopher Lee and Thompson, Linda M. and Thomson, Shaun R. and Thronson, Harley and Tierney, C. M. and Tikkanen, Tuomo and Tinnin, Lee and Tippet, William Thomas and Todd, Connor William and Tran, Hien D. and Trauger, John and Trejo, Edwin Gregorio and Vinh Truong, Justin Hoang and Tsukamoto, Christine L. and Tufail, Yasir and Tumlinson, Jason and Tustain, Samuel and Tyra, Harrison and Ubeda, Leonardo and Underwood, Kelli and Uzzo, Michael A. and Vaclavik, Steven and Valenduc, Frida and Valenti, Jeff A. and Van Campen, Julie and {van de Wetering}, Inge and Van Der Marel, Roeland P. and {van Haarlem}, Remy and Vandenbussche, Bart and {van Dishoeck}, Ewine F. and Vanterpool, Dona D. and Vernoy, Michael R. and Vila Costas, Maria Bego{\~n}a and Volk, Kevin and Voorzaat, Piet and Voyton, Mark F. and Vydra, Ekaterina and Waddy, Darryl J. and Waelkens, Christoffel and Wahlgren, Glenn Michael and Walker, Jr., Frederick E. and Wander, Michel and Warfield, Christine K. and Warner, Gerald and Wasiak, Francis C. and Wasiak, Matthew F. and Wehner, James and Weiler, Kevin R. and Weilert, Mark and Weiss, Stanley B. and Wells, Martyn and Welty, Alan D. and Wheate, Lauren and Wheeler, Thomas P. and White, Christy L. and Whitehouse, Paul and Whiteleather, Jennifer Margaret and Whitman, William Russell and Williams, Christina C. and Willmer, Christopher N. A. and Willott, Chris J. and Willoughby, Scott P. and Wilson, Andrew and Wilson, Debra and Wilson, Donna V. and Windhorst, Rogier and Wislowski, Emily Christine and Wolfe, David J. and Wolfe, Michael A. and Wolff, Schuyler and Wondel, Amancio and Woo, Cindy and Woods, Robert T. and Worden, Elaine and Workman, William and Wright, Gillian S. and Wu, Carl and Wu, Chi-Rai and Wun, Dakin D. and Wymer, Kristen B. and Yadetie, Thomas and Yan, Isabelle C. and Yang, Keith C. and Yates, Kayla L. and Yeager, Christopher R. and Yerger, Ethan John and Young, Erick T. and Young, Gary and Yu, Gene and Yu, Susan and Zak, Dean S. and Zeidler, Peter and Zepp, Robert and Zhou, Julia and Zincke, Christian A. and Zonak, Stephanie and Zondag, Elisabeth},
  year = {2023},
  month = jun,
  journal = {PASP},
  volume = {135},
  pages = {068001},
  issn = {0004-6280},
  doi = {10.1088/1538-3873/acd1b5},
  urldate = {2024-02-19}
}

@article{gimenez-arteaga2022,
  title = {High-Resolution {{Hubble Space Telescope Imaging Survey}} of {{Local Star-forming Galaxies}}. {{I}}. {{Spatially Resolved Obscured Star Formation}} with {{H$\alpha$}} and {{Paschen-$\beta$ Recombination Lines}}},
  author = {{Gim{\'e}nez-Arteaga}, Clara and Brammer, Gabriel B. and Marchesini, Danilo and Colina, Luis and Bajaj, Varun and Brinch, Malte and Calzetti, Daniela and {Lange-Vagle}, Daniel and Murphy, Eric J. and Perna, Michele and {Piqueras-L{\'o}pez}, Javier and Snyder, Gregory F.},
  year = {2022},
  month = nov,
  journal = {\apj Suppl. Ser.},
  volume = {263},
  pages = {17},
  issn = {0067-0049},
  doi = {10.3847/1538-4365/ac958c},
  urldate = {2023-08-07}
}

@article{goldberg2024,
  title = {Search for {{High-excitation Emission Lines}} from a {{Quasar-scale Active Nucleus}} in {{Arp}} 220},
  author = {Goldberg, Charles E. and Buiten, Victorine A. and Rieke, George H. and {Alonso-Herrero}, Almudena and Paggi, A. and {van der Werf}, Paul and Stone, Meredith A. and Morrison, Jane E. and Alberts, Stacey and Dicken, Dan and Wright, Gillian},
  year = {2024},
  month = dec,
  journal = {\apj},
  volume = {977},
  pages = {55},
  issn = {0004-637X},
  doi = {10.3847/1538-4357/ad7eb0},
  urldate = {2025-09-01}
}

@article{gordon1997,
  title = {Dust in {{Starburst Galaxies}}},
  author = {Gordon, Karl D. and Calzetti, Daniela and Witt, Adolf N.},
  year = {1997},
  month = oct,
  journal = {\apj},
  volume = {487},
  pages = {625--635},
  issn = {0004-637X},
  doi = {10.1086/304654},
  urldate = {2025-08-26}
}

@article{guo2016,
  title = {The {{Bursty Star Formation Histories}} of {{Low-mass Galaxies}} at 0.4 {$<$} z {$<$} 1 {{Revealed}} by {{Star Formation Rates Measured From H$\beta$}} and {{FUV}}},
  author = {Guo, Yicheng and Rafelski, Marc and Faber, S. M. and Koo, David C. and Krumholz, Mark R. and Trump, Jonathan R. and Willner, S. P. and Amor{\'i}n, Ricardo and Barro, Guillermo and Bell, Eric F. and Gardner, Jonathan P. and Gawiser, Eric and Hathi, Nimish P. and Koekemoer, Anton M. and Pacifici, Camilla and {P{\'e}rez-Gonz{\'a}lez}, Pablo G. and Ravindranath, Swara and Reddy, Naveen and Teplitz, Harry I. and Yesuf, Hassen},
  year = {2016},
  month = dec,
  journal = {\apj},
  volume = {833},
  pages = {37},
  issn = {0004-637X},
  doi = {10.3847/1538-4357/833/1/37},
  urldate = {2025-08-26}
}

@article{hao2011,
  title = {Dust-Corrected {{Star Formation Rates}} of {{Galaxies}}. {{II}}. {{Combinations}} of {{Ultraviolet}} and {{Infrared Tracers}}},
  author = {Hao, Cai-Na and Kennicutt, Robert C. and Johnson, Benjamin D. and Calzetti, Daniela and Dale, Daniel A. and Moustakas, John},
  year = {2011},
  month = nov,
  journal = {\apj},
  volume = {741},
  pages = {124},
  issn = {0004-637X},
  doi = {10.1088/0004-637X/741/2/124},
  urldate = {2023-07-27}
}

@article{helou2001,
  title = {Evidence for the {{Heating}} of {{Atomic Interstellar Gas}} by {{Polycyclic Aromatic Hydrocarbons}}},
  author = {Helou, George and Malhotra, Sangeeta and Hollenbach, David J. and Dale, Daniel A. and Contursi, Alessandra},
  year = {2001},
  month = feb,
  journal = {\apj},
  volume = {548},
  pages = {L73-L76},
  issn = {0004-637X},
  doi = {10.1086/318916},
  urldate = {2025-08-21}
}

@article{huang2009,
  title = {Infrared {{Spectrograph Spectroscopy}} and {{Multi-Wavelength Study}} of {{Luminous Star-Forming Galaxies}} at z {$\simeq$} 1.9},
  author = {Huang, J. -S. and Faber, S. M. and Daddi, E. and Laird, E. S. and Lai, K. and Omont, A. and Wu, Y. and Younger, J. D. and Bundy, K. and Cattaneo, A. and Chapman, S. C. and Conselice, C. J. and Dickinson, M. and Egami, E. and Fazio, G. G. and Im, M. and Koo, D. and Le Floc'h, E. and Papovich, C. and Rigopoulou, D. and Smail, I. and Song, M. and {Van de Werf}, P. P. and Webb, T. M. A. and Willmer, C. N. A. and Willner, S. P. and Yan, L.},
  year = {2009},
  month = jul,
  journal = {\apj},
  volume = {700},
  pages = {183--198},
  issn = {0004-637X},
  doi = {10.1088/0004-637X/700/1/183},
  urldate = {2025-09-01}
}

@article{hummer1987,
  title = {Recombination-Line Intensities for Hydrogenic Ions - {{I}}. {{Case B}} Calculations for {{H I}} and {{He II}}.},
  author = {Hummer, D. G. and Storey, P. J.},
  year = {1987},
  month = feb,
  journal = {MNRAS},
  volume = {224},
  pages = {801--820},
  issn = {0035-8711},
  doi = {10.1093/mnras/224.3.801},
  urldate = {2025-09-01}
}

@article{hunt2010,
  title = {The {{Spitzer View}} of {{Low-Metallicity Star Formation}}. {{III}}. {{Fine-Structure Lines}}, {{Aromatic Features}}, and {{Molecules}}},
  author = {Hunt, Leslie K. and Thuan, Trinh X. and Izotov, Yuri I. and Sauvage, Marc},
  year = {2010},
  month = mar,
  journal = {\apj},
  volume = {712},
  pages = {164--187},
  issn = {0004-637X},
  doi = {10.1088/0004-637X/712/1/164},
  urldate = {2023-01-12}
}

@article{inami2018,
  title = {The {{AKARI}} 2.5--5 Micron Spectra of Luminous Infrared Galaxies in the Local {{Universe}}},
  author = {Inami, H. and Armus, L. and Matsuhara, H. and Charmandaris, V. and {D{\'i}az-Santos}, T. and Surace, J. and Stierwalt, S. and Ohyama, Y. and Howell, J. and Marshall, J. and Evans, A. S. and Linden, S. T. and Mazzarella, J.},
  year = {2018},
  month = sep,
  journal = {A\&A},
  volume = {617},
  pages = {A130},
  issn = {0004-6361, 1432-0746},
  doi = {10.1051/0004-6361/201833053},
  urldate = {2022-12-13},
  langid = {english}
}

@misc{ji2023,
  title = {{{JADES}} + {{JEMS}}: {{A Detailed Look}} at the {{Buildup}} of {{Central Stellar Cores}} and {{Suppression}} of {{Star Formation}} in {{Galaxies}} at {{Redshifts}} 3 {$<$} z {$<$} 4.5},
  shorttitle = {{{JADES}} + {{JEMS}}},
  author = {Ji, Zhiyuan and Williams, Christina C. and Tacchella, Sandro and Suess, Katherine A. and Baker, William M. and Alberts, Stacey and Bunker, Andrew J. and Johnson, Benjamin D. and Robertson, Brant and Sun, Fengwu and Eisenstein, Daniel J. and Rieke, Marcia and Maseda, Michael V. and Hainline, Kevin and Hausen, Ryan and Rieke, George and Willmer, Christopher N. A. and Egami, Eiichi and Shivaei, Irene and Carniani, Stefano and Charlot, Stephane and Chevallard, Jacopo and {Curtis-Lake}, Emma and Looser, Tobias J. and Maiolino, Roberto and Willott, Chris and Chen, Zuyi and Helton, Jakob M. and Lyu, Jianwei and Nelson, Erica and Bhatawdekar, Rachana and Boyett, Kristan and Sandles, Lester},
  year = {2023},
  month = may,
  journal = {arXiv e-prints},
  doi = {10.48550/arXiv.2305.18518},
  urldate = {2023-10-21}
}

@article{johnson2021,
  title = {Stellar {{Population Inference}} with {{Prospector}}},
  author = {Johnson, Benjamin D. and Leja, Joel and Conroy, Charlie and Speagle, Joshua S.},
  year = {2021},
  month = jun,
  journal = {\apj Suppl. Ser.},
  volume = {254},
  pages = {22},
  issn = {0067-0049},
  doi = {10.3847/1538-4365/abef67},
  urldate = {2023-12-16}
}

@article{kennicutt2007,
  title = {Star {{Formation}} in {{NGC}} 5194 ({{M51a}}). {{II}}. {{The Spatially Resolved Star Formation Law}}},
  author = {Kennicutt, Jr., Robert C. and Calzetti, Daniela and Walter, Fabian and Helou, George and Hollenbach, David J. and Armus, Lee and Bendo, George and Dale, Daniel A. and Draine, Bruce T. and Engelbracht, Charles W. and Gordon, Karl D. and Prescott, Moire K. M. and Regan, Michael W. and Thornley, Michele D. and Bot, Caroline and Brinks, Elias and {de Blok}, Erwin and {de Mello}, Dulia and Meyer, Martin and Moustakas, John and Murphy, Eric J. and Sheth, Kartik and Smith, J. D. T.},
  year = {2007},
  month = dec,
  journal = {\apj},
  volume = {671},
  pages = {333--348},
  issn = {0004-637X},
  doi = {10.1086/522300},
  urldate = {2025-08-21}
}

@article{kennicutt2009,
  title = {Dust-Corrected {{Star Formation Rates}} of {{Galaxies}}. {{I}}. {{Combinations}} of {{H$\alpha$}} and {{Infrared Tracers}}},
  author = {Kennicutt, Jr., Robert C. and Hao, Cai-Na and Calzetti, Daniela and Moustakas, John and Dale, Daniel A. and Bendo, George and Engelbracht, Charles W. and Johnson, Benjamin D. and Lee, Janice C.},
  year = {2009},
  month = oct,
  journal = {\apj},
  volume = {703},
  pages = {1672--1695},
  publisher = {IOP},
  issn = {0004-637X},
  doi = {10.1088/0004-637X/703/2/1672},
  urldate = {2025-07-29}
}

@article{kennicutt2012,
  title = {Star {{Formation}} in the {{Milky Way}} and {{Nearby Galaxies}}},
  author = {Kennicutt, Robert C. and Evans, Neal J.},
  year = {2012},
  month = sep,
  journal = {Annu. Rev. A\&A Vol 50 P531-608},
  volume = {50},
  pages = {531},
  issn = {0066-4146},
  doi = {10.1146/annurev-astro-081811-125610},
  urldate = {2022-03-30},
  langid = {english}
}

@article{kim2024a,
  title = {The {{Calibration}} of {{Polycyclic Aromatic Hydrocarbon Dust Emission}} as a {{Star Formation Rate Indicator}} in the {{AKARI NEP Survey}}},
  author = {Kim, Helen Kyung and Malkan, Matthew A. and Takagi, Toshinobu and Oi, Nagisa and Burgarella, Denis and Miyaji, Takamitsu and Shim, Hyunjin and Matsuhara, Hideo and Goto, Tomotsugu and Ohyama, Yoichi and Buat, Veronique and Kim, Seong Jin},
  year = {2024},
  month = oct,
  journal = {\apj},
  volume = {974},
  pages = {253},
  publisher = {IOP},
  issn = {0004-637X},
  doi = {10.3847/1538-4357/ad72e6},
  urldate = {2024-10-22}
}

@article{kirkpatrick2015,
  title = {The {{Role}} of {{Star Formation}} and an {{AGN}} in {{Dust Heating}} of z = 0.3-2.8 {{Galaxies}}. {{I}}. {{Evolution}} with {{Redshift}} and {{Luminosity}}},
  author = {Kirkpatrick, Allison and Pope, Alexandra and Sajina, Anna and Roebuck, Eric and Yan, Lin and Armus, Lee and {D{\'i}az-Santos}, Tanio and Stierwalt, Sabrina},
  year = {2015},
  month = nov,
  journal = {\apj},
  volume = {814},
  pages = {9},
  issn = {0004-637X},
  doi = {10.1088/0004-637X/814/1/9},
  urldate = {2022-05-01}
}

@misc{kirkpatrick2023,
  title = {{{CEERS Key Paper VII}}: {{JWST}}/{{MIRI Reveals}} a {{Faint Population}} of {{Galaxies}} at {{Cosmic Noon Unseen}} by {{Spitzer}}},
  shorttitle = {{{CEERS Key Paper VII}}},
  author = {Kirkpatrick, Allison and Yang, Guang and Bail, Aurelien Le and Troiani, Greg and Bell, Eric F. and Cleri, Nikko J. and Elbaz, David and Finkelstein, Steven L. and Hathi, Nimish P. and Hirschmann, Michaela and Holwerda, Benne W. and Kocevski, Dale D. and Lucas, Ray A. and McKinney, Jed and Papovich, Casey and {Perez-Gonzalez}, Pablo G. and {de la Vega}, Alexander and Bagley, Micaela B. and Daddi, Emanuele and Dickinson, Mark and Ferguson, Henry C. and Fontana, Adriano and Grazian, Andrea and Grogin, Norman A. and Haro, Pablo Arrabal and Kartaltepe, Jeyhan S. and Kewley, Lisa J. and Koekemoer, Anton M. and Lotz, Jennifer M. and Pentericci, Laura and Pirzkal, Nor and Ravindranath, Swara and Somerville, Rachel S. and Trump, Jonathan R. and Wilkins, Stephen M. and Yung, L. Y. Aaron},
  year = {2023},
  month = aug,
  number = {arXiv:2308.09750},
  eprint = {2308.09750},
  primaryclass = {astro-ph},
  publisher = {arXiv},
  doi = {10.48550/arXiv.2308.09750},
  urldate = {2023-08-22},
  archiveprefix = {arXiv}
}

@article{kodra2023,
  title = {Optimized {{Photometric Redshifts}} for the {{Cosmic Assembly Near-infrared Deep Extragalactic Legacy Survey}} ({{CANDELS}})},
  author = {Kodra, Dritan and Andrews, Brett H. and Newman, Jeffrey A. and Finkelstein, Steven L. and Fontana, Adriano and Hathi, Nimish and Salvato, Mara and Wiklind, Tommy and Wuyts, Stijn and Broussard, Adam and Chartab, Nima and Conselice, Christopher and Cooper, M. C. and Dekel, Avishai and Dickinson, Mark and Ferguson, Henry C. and Gawiser, Eric and Grogin, Norman A. and Iyer, Kartheik and Kartaltepe, Jeyhan and Kassin, Susan and Koekemoer, Anton M. and Koo, David C. and Lucas, Ray A. and Mantha, Kameswara Bharadwaj and McIntosh, Daniel H. and Mobasher, Bahram and Pacifici, Camilla and {P{\'e}rez-Gonz{\'a}lez}, Pablo G. and Santini, Paola},
  year = {2023},
  month = jan,
  journal = {\apj},
  volume = {942},
  pages = {36},
  issn = {0004-637X},
  doi = {10.3847/1538-4357/ac9f12},
  urldate = {2025-08-22}
}

@article{kriek2013,
  title = {The {{Dust Attenuation Law}} in {{Distant Galaxies}}: {{Evidence}} for {{Variation}} with {{Spectral Type}}},
  shorttitle = {The {{Dust Attenuation Law}} in {{Distant Galaxies}}},
  author = {Kriek, Mariska and Conroy, Charlie},
  year = {2013},
  month = sep,
  journal = {\apj},
  volume = {775},
  pages = {L16},
  issn = {0004-637X},
  doi = {10.1088/2041-8205/775/1/L16},
  urldate = {2025-08-21}
}

@article{kron1980,
  title = {Photometry of a Complete Sample of Faint Galaxies.},
  author = {Kron, R. G.},
  year = {1980},
  month = jun,
  journal = {\apj Suppl. Ser.},
  volume = {43},
  pages = {305--325},
  publisher = {IOP},
  issn = {0067-0049},
  doi = {10.1086/190669},
  urldate = {2024-05-24}
}

@article{lagache2005,
  title = {Dusty {{Infrared Galaxies}}: {{Sources}} of the {{Cosmic Infrared Background}}},
  shorttitle = {Dusty {{Infrared Galaxies}}},
  author = {Lagache, Guilaine and Puget, Jean-Loup and Dole, Herv{\'e}},
  year = {2005},
  month = sep,
  journal = {ARA\&A},
  volume = {43},
  number = {1},
  pages = {727},
  issn = {0066-4146},
  doi = {10.1146/annurev.astro.43.072103.150606},
  urldate = {2022-05-01},
  langid = {english}
}

@article{lai2020,
  title = {All the {{PAHs}}: {{An AKARI}}--{{Spitzer Cross-archival Spectroscopic Survey}} of {{Aromatic Emission}} in {{Galaxies}}},
  shorttitle = {All the {{PAHs}}},
  author = {Lai, Thomas S.-Y. and Smith, J. D. T. and Baba, Shunsuke and Spoon, Henrik W. W. and Imanishi, Masatoshi},
  year = {2020},
  month = dec,
  journal = {ApJ},
  volume = {905},
  number = {1},
  pages = {55},
  issn = {0004-637X, 1538-4357},
  doi = {10.3847/1538-4357/abc002},
  urldate = {2022-07-26},
  langid = {english}
}

@article{lai2022,
  title = {{{GOALS-JWST}}: {{Tracing AGN Feedback}} on the {{Star-forming Interstellar Medium}} in {{NGC}} 7469},
  shorttitle = {{{GOALS-JWST}}},
  author = {Lai, Thomas S.-Y. and Armus, Lee and U, Vivian and {D{\'i}az-Santos}, Tanio and Larson, Kirsten L. and Evans, Aaron and Malkan, Matthew A. and Appleton, Philip and Rich, Jeff and {M{\"u}ller-S{\'a}nchez}, Francisco and Inami, Hanae and Bohn, Thomas and McKinney, Jed and Finnerty, Luke and Law, David R. and Linden, Sean T. and Medling, Anne M. and Privon, George C. and Song, Yiqing and Stierwalt, Sabrina and {van der Werf}, Paul P. and {Barcos-Mu{\~n}oz}, Loreto and Smith, J. D. T. and Togi, Aditya and Aalto, Susanne and B{\"o}ker, Torsten and Charmandaris, Vassilis and Howell, Justin and Iwasawa, Kazushi and Kemper, Francisca and Mazzarella, Joseph M. and Murphy, Eric J. and Brown, Michael J. I. and Hayward, Christopher C. and Marshall, Jason and Sanders, David and Surace, Jason},
  year = {2022},
  month = dec,
  journal = {ApJL},
  volume = {941},
  number = {2},
  pages = {L36},
  issn = {2041-8205, 2041-8213},
  doi = {10.3847/2041-8213/ac9ebf},
  urldate = {2023-01-26},
  langid = {english}
}

@article{lee2015,
  title = {A {{Turnover}} in the {{Galaxy Main Sequence}} of {{Star Formation}} at {{M}} * {\textasciitilde} 1010 {{M}} {$\odot$} for {{Redshifts}} z {$<$} 1.3},
  author = {Lee, Nicholas and Sanders, D. B. and Casey, Caitlin M. and Toft, Sune and Scoville, N. Z. and Hung, Chao-Ling and Le Floc'h, Emeric and Ilbert, Olivier and Zahid, H. Jabran and Aussel, Herv{\'e} and Capak, Peter and Kartaltepe, Jeyhan S. and Kewley, Lisa J. and Li, Yanxia and Schawinski, Kevin and Sheth, Kartik and Xiao, Quanbao},
  year = {2015},
  month = mar,
  journal = {\apj},
  volume = {801},
  pages = {80},
  issn = {0004-637X},
  doi = {10.1088/0004-637X/801/2/80},
  urldate = {2025-08-26}
}

@article{lefloch2005,
  title = {Infrared {{Luminosity Functions}} from the {{Chandra Deep Field-South}}: {{The Spitzer View}} on the {{History}} of {{Dusty Star Formation}} at 0 {$<$}{\textasciitilde} z {$<$}{\textasciitilde} 1},
  shorttitle = {Infrared {{Luminosity Functions}} from the {{Chandra Deep Field-South}}},
  author = {Le Floc'h, Emeric and Papovich, Casey and Dole, Herv{\'e} and Bell, Eric F. and Lagache, Guilaine and Rieke, George H. and Egami, Eiichi and {P{\'e}rez-Gonz{\'a}lez}, Pablo G. and {Alonso-Herrero}, Almudena and Rieke, Marcia J. and Blaylock, Myra and Engelbracht, Charles W. and Gordon, Karl D. and Hines, Dean C. and Misselt, Karl A. and Morrison, Jane E. and Mould, Jeremy},
  year = {2005},
  month = oct,
  journal = {\apj},
  volume = {632},
  pages = {169--190},
  issn = {0004-637X},
  doi = {10.1086/432789},
  urldate = {2022-03-24}
}

@article{leja2019a,
  title = {An {{Older}}, {{More Quiescent Universe}} from {{Panchromatic SED Fitting}} of the {{3D-HST Survey}}},
  author = {Leja, Joel and Johnson, Benjamin D. and Conroy, Charlie and {van Dokkum}, Pieter and Speagle, Joshua S. and Brammer, Gabriel and Momcheva, Ivelina and Skelton, Rosalind and Whitaker, Katherine E. and Franx, Marijn and Nelson, Erica J.},
  year = {2019},
  month = jun,
  journal = {\apj},
  volume = {877},
  pages = {140},
  issn = {0004-637X},
  doi = {10.3847/1538-4357/ab1d5a},
  urldate = {2022-10-06}
}

@article{leja2019b,
  title = {How to {{Measure Galaxy Star Formation Histories}}. {{II}}. {{Nonparametric Models}}},
  author = {Leja, Joel and Carnall, Adam C. and Johnson, Benjamin D. and Conroy, Charlie and Speagle, Joshua S.},
  year = {2019},
  month = may,
  journal = {\apj},
  volume = {876},
  pages = {3},
  issn = {0004-637X},
  doi = {10.3847/1538-4357/ab133c},
  urldate = {2023-12-15}
}

@ARTICLE{leroy2022,
       author = {{Leroy}, Adam K. and {Bolatto}, Alberto D. and {Sandstrom}, Karin and {Rosolowsky}, Erik and {Barnes}, Ashley. T. and {Bigiel}, F. and {Boquien}, M{\'e}d{\'e}ric and {den Brok}, Jakob S. and {Cao}, Yixian and {Chastenet}, J{\'e}r{\'e}my and {Chevance}, M{\'e}lanie and {Chiang}, I-Da and {Chown}, Ryan and {Colombo}, Dario and {Ellison}, Sara L. and {Emsellem}, Eric and {Grasha}, Kathryn and {Henshaw}, Jonathan D. and {Hughes}, Annie and {Klessen}, Ralf S. and {Koch}, Eric W. and {Kim}, Jaeyeon and {Kreckel}, Kathryn and {Kruijssen}, J.~M. Diederik and {Larson}, Kirsten L. and {Lee}, Janice C. and {Levy}, Rebecca C. and {Lin}, Lihwai and {Liu}, Daizhong and {Meidt}, Sharon E. and {Pety}, J{\'e}r{\^o}me and {Querejeta}, Miguel and {Rubio}, M{\'o}nica and {Saito}, Toshiki and {Salim}, Samir and {Schinnerer}, Eva and {Sormani}, Mattia C. and {Sun}, Jiayi and {Thilker}, David A. and {Usero}, Antonio and {Vogel}, Stuart N. and {Watkins}, Elizabeth J. and {Whitcomb}, Cory M. and {Williams}, Thomas G. and {Wilson}, Christine D.},
        title = "{PHANGS-JWST First Results: A Global and Moderately Resolved View of Mid-infrared and CO Line Emission from Galaxies at the Start of the JWST Era}",
      journal = {\apjl},
         year = 2023,
        month = feb,
       volume = {944},
       number = {2},
          eid = {L10},
        pages = {L10},
          doi = {10.3847/2041-8213/acab01},
archivePrefix = {arXiv},
       eprint = {2212.09774},
 primaryClass = {astro-ph.GA},
       adsurl = {https://ui.adsabs.harvard.edu/abs/2023ApJ...944L..10L}
}

@article{li2020,
  title = {Spitzer's Perspective of Polycyclic Aromatic Hydrocarbons in Galaxies},
  author = {Li, Aigen},
  year = {2020},
  month = mar,
  journal = {Nat. Astron.},
  volume = {4},
  pages = {339--351},
  issn = {2397-3366},
  doi = {10.1038/s41550-020-1051-1},
  urldate = {2024-05-02}
}

@article{liang2021c,
  title = {The {{IRX-$\beta$}} Relation of High-Redshift Galaxies},
  author = {Liang, Lichen and Feldmann, Robert and Hayward, Christopher C. and Narayanan, Desika and {\c C}atmabacak, Onur and Kere{\v s}, Du{\v s}an and {Faucher-Gigu{\`e}re}, Claude-Andr{\'e} and Hopkins, Philip F.},
  year = {2021},
  month = apr,
  journal = {MNRAS},
  volume = {502},
  pages = {3210--3241},
  publisher = {OUP},
  issn = {0035-8711},
  doi = {10.1093/mnras/stab096},
  urldate = {2025-08-08}
}

@article{lorenz2025,
  title = {Measuring {{Emission Lines}} with {{JWST MegaScience Medium Bands}}: {{A New Window}} into {{Dust}} and {{Star Formation}} at {{Cosmic Noon}}},
  shorttitle = {Measuring {{Emission Lines}} with {{JWST MegaScience Medium Bands}}},
  author = {Lorenz, Brian and Suess, Katherine A. and Kriek, Mariska and Price, Sedona H. and Leja, Joel and Nelson, Erica and Atek, Hakim and Bezanson, Rachel and Brammer, Gabriel and Cutler, Sam E. and Dayal, Pratika and {de Graaff}, Anna and Greene, Jenny E. and Furtak, Lukas J. and Labb{\'e}, Ivo and Marchesini, Danilo and Maseda, Michael V. and Miller, Tim B. and Mintz, Abby and Mitsuhashi, Ikki and Pan, Richard and Porraz Barrera, Natalia and Wang, Bingjie and Weaver, John R. and Williams, Christina C. and Whitaker, Katherine E.},
  year = {2025},
  month = jul,
  journal = {\apj},
  volume = {988},
  pages = {L20},
  issn = {0004-637X},
  doi = {10.3847/2041-8213/ade887},
  urldate = {2025-08-26}
}

@ARTICLE{lyu2022a,
       author = {{Lyu}, Jianwei and {Alberts}, Stacey and {Rieke}, George H. and {Rujopakarn}, Wiphu},
        title = "{AGN Selection and Demographics in GOODS-S/HUDF from X-Ray to Radio}",
      journal = {\apj},
         year = 2022,
        month = dec,
       volume = {941},
       number = {2},
          eid = {191},
        pages = {191},
          doi = {10.3847/1538-4357/ac9e5d},
archivePrefix = {arXiv},
       eprint = {2209.06219},
 primaryClass = {astro-ph.GA},
       adsurl = {https://ui.adsabs.harvard.edu/abs/2022ApJ...941..191L}
}

@article{lyu2024,
  title = {Active {{Galactic Nuclei Selection}} and {{Demographics}}: {{A New Age}} with {{JWST}}/{{MIRI}}},
  shorttitle = {Active {{Galactic Nuclei Selection}} and {{Demographics}}},
  author = {Lyu, Jianwei and Alberts, Stacey and Rieke, George H. and Shivaei, Irene and {P{\'e}rez-Gonz{\'a}lez}, Pablo G. and Sun, Fengwu and Hainline, Kevin N. and Baum, Stefi and Bonaventura, Nina and Bunker, Andrew J. and Egami, Eiichi and Eisenstein, Daniel J. and Florian, Michael and Ji, Zhiyuan and Johnson, Benjamin D. and Morrison, Jane and Rieke, Marcia and Robertson, Brant and Rujopakarn, Wiphu and Tacchella, Sandro and Scholtz, Jan and Willmer, Christopher N. A.},
  year = {2024},
  month = may,
  journal = {\apj},
  volume = {966},
  pages = {229},
  publisher = {IOP},
  issn = {0004-637X},
  doi = {10.3847/1538-4357/ad3643},
  urldate = {2024-05-23}
}

@article{madau2014,
  title = {Cosmic {{Star Formation History}}},
  author = {Madau, Piero and Dickinson, Mark},
  year = {2014},
  month = aug,
  journal = {Annu. Rev. A\&A},
  volume = {52},
  number = {1},
  eprint = {1403.0007},
  pages = {415--486},
  issn = {0066-4146, 1545-4282},
  doi = {10.1146/annurev-astro-081811-125615},
  urldate = {2022-02-27},
  archiveprefix = {arXiv}
}

@article{madden2006,
  title = {{{ISM}} Properties in Low-Metallicity Environments},
  author = {Madden, S. C. and Galliano, F. and Jones, A. P. and Sauvage, M.},
  year = {2006},
  month = feb,
  journal = {A\&A},
  volume = {446},
  pages = {877--896},
  issn = {0004-6361},
  doi = {10.1051/0004-6361:20053890},
  urldate = {2025-08-21}
}

@article{magdis2013,
  title = {Mid- to Far-Infrared Properties of Star-Forming Galaxies and Active Galactic Nuclei},
  author = {Magdis, G. E. and Rigopoulou, D. and Helou, G. and Farrah, D. and Hurley, P. and {Alonso-Herrero}, A. and Bock, J. and Burgarella, D. and Chapman, S. and Charmandaris, V. and Cooray, A. and Dai, Y. Sophia and Dale, D. and Elbaz, D. and Feltre, A. and Hatziminaoglou, E. and Huang, J. -S. and Morrison, G. and Oliver, S. and Page, M. and Scott, D. and Shi, Y.},
  year = {2013},
  month = oct,
  journal = {A\&A},
  volume = {558},
  pages = {A136},
  issn = {0004-6361},
  doi = {10.1051/0004-6361/201322226},
  urldate = {2025-08-26}
}

@article{magnelli2011,
  title = {Evolution of the Dusty Infrared Luminosity Function from z = 0 to z = 2.3 Using Observations from {{Spitzer}}},
  author = {Magnelli, B. and Elbaz, D. and Chary, R. R. and Dickinson, M. and Le Borgne, D. and Frayer, D. T. and Willmer, C. N. A.},
  year = {2011},
  month = apr,
  journal = {A\&A},
  volume = {528},
  pages = {A35},
  issn = {0004-6361},
  doi = {10.1051/0004-6361/200913941},
  urldate = {2024-05-01}
}

@article{maiolino2019,
  title = {De Re Metallica: The Cosmic Chemical Evolution of Galaxies},
  shorttitle = {De Re Metallica},
  author = {Maiolino, R. and Mannucci, F.},
  year = {2019},
  month = feb,
  journal = {A\&A Rev.},
  volume = {27},
  pages = {3},
  issn = {0935-4956},
  doi = {10.1007/s00159-018-0112-2},
  urldate = {2023-12-15}
}

@article{marble2010,
  title = {An {{Aromatic Inventory}} of the {{Local Volume}}},
  author = {Marble, A. R. and Engelbracht, C. W. and {van Zee}, L. and Dale, D. A. and Smith, J. D. T. and Gordon, K. D. and Wu, Y. and Lee, J. C. and Kennicutt, R. C. and Skillman, E. D. and Johnson, L. C. and Block, M. and Calzetti, D. and Cohen, S. A. and Lee, H. and Schuster, M. D.},
  year = {2010},
  month = may,
  journal = {\apj},
  volume = {715},
  pages = {506--540},
  issn = {0004-637X},
  doi = {10.1088/0004-637X/715/1/506},
  urldate = {2023-07-27}
}

@article{mclure2018,
  title = {Dust Attenuation in 2 {$<$} z {$<$} 3 Star-Forming Galaxies from Deep {{ALMA}} Observations of the {{Hubble Ultra Deep Field}}},
  author = {McLure, R. J. and Dunlop, J. S. and Cullen, F. and Bourne, N. and Best, P. N. and Khochfar, S. and Bowler, R. A. A. and Biggs, A. D. and Geach, J. E. and Scott, D. and Micha{\l}owski, M. J. and Rujopakarn, W. and {van Kampen}, E. and Kirkpatrick, A. and Pope, A.},
  year = {2018},
  month = may,
  journal = {MNRAS},
  volume = {476},
  pages = {3991--4006},
  issn = {0035-8711},
  doi = {10.1093/mnras/sty522},
  urldate = {2025-08-26}
}

@article{murphy2001,
  title = {K-{{Band Spectroscopy}} of {{Ultraluminous Infrared Galaxies}}: {{The}} 2 {{JY Sample}}},
  shorttitle = {K-{{Band Spectroscopy}} of {{Ultraluminous Infrared Galaxies}}},
  author = {Murphy, Jr., T. W. and Soifer, B. T. and Matthews, K. and Armus, L. and Kiger, J. R.},
  year = {2001},
  month = jan,
  journal = {Astron. J.},
  volume = {121},
  pages = {97--127},
  issn = {0004-6256},
  doi = {10.1086/318031},
  urldate = {2025-08-26}
}

@article{murphy2011a,
  title = {Calibrating {{Extinction-free Star Formation Rate Diagnostics}} with 33 {{GHz Free-free Emission}} in {{NGC}} 6946},
  author = {Murphy, E. J. and Condon, J. J. and Schinnerer, E. and Kennicutt, R. C. and Calzetti, D. and Armus, L. and Helou, G. and Turner, J. L. and Aniano, G. and Beir{\~a}o, P. and Bolatto, A. D. and Brandl, B. R. and Croxall, K. V. and Dale, D. A. and Donovan Meyer, J. L. and Draine, B. T. and Engelbracht, C. and Hunt, L. K. and Hao, C. -N. and Koda, J. and Roussel, H. and Skibba, R. and Smith, J. -D. T.},
  year = {2011},
  month = aug,
  journal = {\apj},
  volume = {737},
  pages = {67},
  issn = {0004-637X},
  doi = {10.1088/0004-637X/737/2/67},
  urldate = {2025-08-21}
}

@article{narayanan2018,
  title = {The {{IRX-$\beta$}} Dust Attenuation Relation in Cosmological Galaxy Formation Simulations},
  author = {Narayanan, Desika and Dav{\'e}, Romeel and Johnson, Benjamin D. and Thompson, Robert and Conroy, Charlie and Geach, James},
  year = {2018},
  month = feb,
  journal = {MNRAS},
  volume = {474},
  pages = {1718--1736},
  issn = {0035-8711},
  doi = {10.1093/mnras/stx2860},
  urldate = {2025-08-25}
}

@misc{neufeld2024,
  title = {{{FRESCO}}: {{The Paschen-}}\${\textbackslash}alpha\$ {{Star Forming Sequence}} at {{Cosmic Noon}}},
  shorttitle = {{{FRESCO}}},
  author = {Neufeld, Chloe and {van Dokkum}, Pieter and Asali, Yasmeen and {Covelo-Paz}, Alba and Leja, Joel and Lin, Jamie and Matthee, Jorryt and Oesch, Pascal A. and Reddy, Naveen A. and Shivaei, Irene and Whitaker, Katherine E. and Wuyts, Stijn and Brammer, Gabriel and Marchesini, Danilo and Maseda, Michael V. and Naidu, Rohan P. and Nelson, Erica J. and Velichko, Anna and Weibel, Andrea and Xiao, Mengyuan},
  year = {2024},
  month = apr,
  journal = {arXiv e-prints},
  doi = {10.48550/arXiv.2404.10816},
  urldate = {2024-06-09}
}

@article{noll2009,
  title = {Analysis of Galaxy Spectral Energy Distributions from Far-{{UV}} to Far-{{IR}} with {{CIGALE}}: Studying a {{SINGS}} Test Sample},
  shorttitle = {Analysis of Galaxy Spectral Energy Distributions from Far-{{UV}} to Far-{{IR}} with {{CIGALE}}},
  author = {Noll, S. and Burgarella, D. and Giovannoli, E. and Buat, V. and Marcillac, D. and {Mu{\~n}oz-Mateos}, J. C.},
  year = {2009},
  month = dec,
  journal = {A\&A},
  volume = {507},
  pages = {1793--1813},
  issn = {0004-6361},
  doi = {10.1051/0004-6361/200912497},
  urldate = {2023-12-05}
}

@misc{oesch2023,
  title = {The {{JWST FRESCO Survey}}: {{Legacy NIRCam}}/{{Grism Spectroscopy}} and {{Imaging}} in the Two {{GOODS Fields}}},
  shorttitle = {The {{JWST FRESCO Survey}}},
  author = {Oesch, P. A. and Brammer, G. and Naidu, R. P. and Bouwens, R. J. and Chisholm, J. and Illingworth, G. D. and Matthee, J. and Nelson, E. and Qin, Y. and Reddy, N. and Shapley, A. and Shivaei, I. and {van Dokkum}, P. and Weibel, A. and Whitaker, K. and Wuyts, S. and {Covelo-Paz}, A. and Endsley, R. and Fudamoto, Y. and Giovinazzo, E. and {Herard-Demanche}, T. and Kerutt, J. and Kramarenko, I. and Labbe, I. and Leonova, E. and Lin, J. and Magee, D. and Marchesini, D. and Maseda, M. and Mason, C. and Matharu, J. and Meyer, R. A. and Neufeld, C. and Lyon, G. Prieto and Schaerer, D. and Sharma, R. and Shuntov, M. and Smit, R. and Stefanon, M. and Wyithe, J. S. B. and Xiao, M.},
  year = {2023},
  month = apr,
  number = {arXiv:2304.02026},
  eprint = {2304.02026},
  primaryclass = {astro-ph},
  publisher = {arXiv},
  urldate = {2023-04-06},
  archiveprefix = {arXiv},
  langid = {english}
}

@article{oke1983,
  title = {Secondary Standard Stars for Absolute Spectrophotometry.},
  author = {Oke, J. B. and Gunn, J. E.},
  year = {1983},
  month = mar,
  journal = {\apj},
  volume = {266},
  pages = {713--717},
  issn = {0004-637X},
  doi = {10.1086/160817},
  urldate = {2023-12-09}
}

@article{osterbrock1989,
  title = {Active Galactic Nuclei},
  author = {Osterbrock, Donald E.},
  year = {1989},
  month = dec,
  journal = {Ann. N. Y. Acad. Sci.},
  volume = {571},
  pages = {99--109},
  issn = {0077-8923},
  doi = {10.1111/j.1749-6632.1989.tb50500.x},
  urldate = {2025-08-21}
}

@article{pedrini2024,
  title = {{{FEAST}}: {{Feedback}} in {{Emerging extragAlactic Star ClusTers}}: {{JWST Spots Polycyclic Aromatic Hydrocarbon Destruction}} in {{NGC}} 628 during the {{Emerging Phase}} of {{Star Formation}}},
  shorttitle = {{{FEAST}}},
  author = {Pedrini, Alex and Adamo, Angela and Calzetti, Daniela and Bik, Arjan and Gregg, Benjamin and Linden, Sean T. and Bajaj, Varun and Ryon, Jenna E. and Ali, Ahmad A. and Bortolini, Giacomo and Correnti, Matteo and Elmegreen, Bruce G. and Elmegreen, Debra Meloy and Gallagher, John S. and Grasha, Kathryn and Gutermuth, Robert A. and Johnson, Kelsey E. and Melinder, Jens and Messa, Matteo and {\"O}stlin, G{\"o}ran and Sabbi, Elena and Smith, Linda J. and Tosi, Monica and Faustino Vieira, Helena},
  year = {2024},
  month = aug,
  journal = {\apj},
  volume = {971},
  pages = {32},
  publisher = {IOP},
  issn = {0004-637X},
  doi = {10.3847/1538-4357/ad534d},
  urldate = {2025-04-25}
}

@article{perna2024,
  title = {No Evidence of Active Galactic Nucleus Features in the Nuclei of {{Arp}} 220 from {{JWST}}/{{NIRSpec IFS}}},
  author = {Perna, Michele and Arribas, Santiago and Lamperti, Isabella and {Pereira-Santaella}, Miguel and Ulivi, Lorenzo and B{\"o}ker, Torsten and Maiolino, Roberto and Bunker, Andrew J. and Charlot, St{\'e}phane and Cresci, Giovanni and Rodr{\'i}guez Del Pino, Bruno and D'Eugenio, Francesco and {\"U}bler, Hannah and Fahrion, Katja and Ceci, Matteo},
  year = 2024,
  month = oct,
  journal = {A\&A},
  volume = {690},
  pages = {A171},
  issn = {0004-6361},
  doi = {10.1051/0004-6361/202450094},
  urldate = {2025-09-01}
}

@article{planckcollaboration2020,
  title = {Planck 2018 Results. {{I}}. {{Overview}} and the Cosmological Legacy of {{Planck}}},
  author = {{Planck Collaboration} and Aghanim, N. and Akrami, Y. and Arroja, F. and Ashdown, M. and Aumont, J. and Baccigalupi, C. and Ballardini, M. and Banday, A. J. and Barreiro, R. B. and Bartolo, N. and Basak, S. and Battye, R. and Benabed, K. and Bernard, J. -P. and Bersanelli, M. and Bielewicz, P. and Bock, J. J. and Bond, J. R. and Borrill, J. and Bouchet, F. R. and Boulanger, F. and Bucher, M. and Burigana, C. and Butler, R. C. and Calabrese, E. and Cardoso, J. -F. and Carron, J. and Casaponsa, B. and Challinor, A. and Chiang, H. C. and Colombo, L. P. L. and Combet, C. and Contreras, D. and Crill, B. P. and Cuttaia, F. and {de Bernardis}, P. and {de Zotti}, G. and Delabrouille, J. and Delouis, J. -M. and D{\'e}sert, F. -X. and Di Valentino, E. and Dickinson, C. and Diego, J. M. and Donzelli, S. and Dor{\'e}, O. and Douspis, M. and Ducout, A. and Dupac, X. and Efstathiou, G. and Elsner, F. and En{\ss}lin, T. A. and Eriksen, H. K. and Falgarone, E. and Fantaye, Y. and Fergusson, J. and {Fernandez-Cobos}, R. and Finelli, F. and Forastieri, F. and Frailis, M. and Franceschi, E. and Frolov, A. and Galeotta, S. and Galli, S. and Ganga, K. and {G{\'e}nova-Santos}, R. T. and Gerbino, M. and Ghosh, T. and {Gonz{\'a}lez-Nuevo}, J. and G{\'o}rski, K. M. and Gratton, S. and Gruppuso, A. and Gudmundsson, J. E. and Hamann, J. and Handley, W. and Hansen, F. K. and Helou, G. and Herranz, D. and Hildebrandt, S. R. and Hivon, E. and Huang, Z. and Jaffe, A. H. and Jones, W. C. and Karakci, A. and Keih{\"a}nen, E. and Keskitalo, R. and Kiiveri, K. and Kim, J. and Kisner, T. S. and Knox, L. and Krachmalnicoff, N. and Kunz, M. and {Kurki-Suonio}, H. and Lagache, G. and Lamarre, J. -M. and Langer, M. and Lasenby, A. and Lattanzi, M. and Lawrence, C. R. and Le Jeune, M. and Leahy, J. P. and Lesgourgues, J. and Levrier, F. and Lewis, A. and Liguori, M. and Lilje, P. B. and Lilley, M. and Lindholm, V. and {L{\'o}pez-Caniego}, M. and Lubin, P. M. and Ma, Y. -Z. and {Mac{\'i}as-P{\'e}rez}, J. F. and Maggio, G. and Maino, D. and Mandolesi, N. and Mangilli, A. and {Marcos-Caballero}, A. and Maris, M. and Martin, P. G. and Martinelli, M. and {Mart{\'i}nez-Gonz{\'a}lez}, E. and Matarrese, S. and Mauri, N. and McEwen, J. D. and Meerburg, P. D. and Meinhold, P. R. and Melchiorri, A. and Mennella, A. and Migliaccio, M. and Millea, M. and Mitra, S. and {Miville-Desch{\^e}nes}, M. -A. and Molinari, D. and Moneti, A. and Montier, L. and Morgante, G. and Moss, A. and Mottet, S. and M{\"u}nchmeyer, M. and Natoli, P. and {N{\o}rgaard-Nielsen}, H. U. and Oxborrow, C. A. and Pagano, L. and Paoletti, D. and Partridge, B. and Patanchon, G. and Pearson, T. J. and Peel, M. and Peiris, H. V. and Perrotta, F. and Pettorino, V. and Piacentini, F. and Polastri, L. and Polenta, G. and Puget, J. -L. and Rachen, J. P. and Reinecke, M. and Remazeilles, M. and Renault, C. and Renzi, A. and Rocha, G. and Rosset, C. and Roudier, G. and {Rubi{\~n}o-Mart{\'i}n}, J. A. and {Ruiz-Granados}, B. and Salvati, L. and Sandri, M. and Savelainen, M. and Scott, D. and Shellard, E. P. S. and Shiraishi, M. and Sirignano, C. and Sirri, G. and Spencer, L. D. and Sunyaev, R. and {Suur-Uski}, A. -S. and Tauber, J. A. and Tavagnacco, D. and Tenti, M. and Terenzi, L. and Toffolatti, L. and Tomasi, M. and Trombetti, T. and Valiviita, J. and Van Tent, B. and Vibert, L. and Vielva, P. and Villa, F. and Vittorio, N. and Wandelt, B. D. and Wehus, I. K. and White, M. and White, S. D. M. and Zacchei, A. and Zonca, A.},
  year = {2020},
  month = sep,
  journal = {A\&A},
  volume = {641},
  pages = {A1},
  issn = {0004-6361},
  doi = {10.1051/0004-6361/201833880},
  urldate = {2024-08-06}
}

@article{pope2008,
  title = {Mid-{{Infrared Spectral Diagnosis}} of {{Submillimeter Galaxies}}},
  author = {Pope, Alexandra and Chary, Ranga-Ram and Alexander, David M. and Armus, Lee and Dickinson, Mark and Elbaz, David and Frayer, David and Scott, Douglas and Teplitz, Harry},
  year = {2008},
  month = mar,
  journal = {\apj},
  volume = {675},
  pages = {1171--1193},
  issn = {0004-637X},
  doi = {10.1086/527030},
  urldate = {2025-08-21}
}

@article{pope2013,
  title = {Probing the {{Interstellar Medium}} of z {\textasciitilde} 1 {{Ultraluminous Infrared Galaxies}} through {{Interferometric Observations}} of {{CO}} and {{Spitzer Mid-infrared Spectroscopy}}},
  author = {Pope, Alexandra and Wagg, Jeff and Frayer, David and Armus, Lee and Chary, Ranga-Ram and Daddi, Emanuele and Desai, Vandana and Dickinson, Mark E. and Elbaz, David and Gabor, Jared and Kirkpatrick, Allison},
  year = {2013},
  month = aug,
  journal = {\apj},
  volume = {772},
  pages = {92},
  issn = {0004-637X},
  doi = {10.1088/0004-637X/772/2/92},
  urldate = {2022-12-22}
}

@article{popesso2023,
  title = {The Main Sequence of Star-Forming Galaxies across Cosmic Times},
  author = {Popesso, P. and Concas, A. and Cresci, G. and Belli, S. and Rodighiero, G. and Inami, H. and Dickinson, M. and Ilbert, O. and Pannella, M. and Elbaz, D.},
  year = {2023},
  month = feb,
  journal = {MNRAS},
  volume = {519},
  pages = {1526--1544},
  issn = {0035-8711},
  doi = {10.1093/mnras/stac3214},
  urldate = {2023-07-26}
}

@article{popping2017,
  title = {The Dust Content of Galaxies from z = 0 to z = 9},
  author = {Popping, Gerg{\"o} and Somerville, Rachel S. and Galametz, Maud},
  year = {2017},
  month = nov,
  journal = {MNRAS},
  volume = {471},
  pages = {3152--3185},
  publisher = {OUP},
  issn = {0035-8711},
  doi = {10.1093/mnras/stx1545},
  urldate = {2025-04-24}
}

@article{reddy2008,
  title = {Multiwavelength {{Constraints}} on the {{Cosmic Star Formation History}} from {{Spectroscopy}}: {{The Rest-Frame Ultraviolet}}, {{H$\alpha$}}, and {{Infrared Luminosity Functions}} at {{Redshifts}} 1.9 Lesssim z Lesssim 3.4},
  shorttitle = {Multiwavelength {{Constraints}} on the {{Cosmic Star Formation History}} from {{Spectroscopy}}},
  author = {Reddy, Naveen A. and Steidel, Charles C. and Pettini, Max and Adelberger, Kurt L. and Shapley, Alice E. and Erb, Dawn K. and Dickinson, Mark},
  year = {2008},
  month = mar,
  journal = {\apj Suppl. Ser.},
  volume = {175},
  pages = {48--85},
  publisher = {IOP},
  issn = {0067-0049},
  doi = {10.1086/521105},
  urldate = {2024-05-01}
}

@article{reddy2015,
  title = {The {{MOSDEF Survey}}: {{Measurements}} of {{Balmer Decrements}} and the {{Dust Attenuation Curve}} at {{Redshifts}} z \textasciitilde{} 1.4-2.6},
  shorttitle = {The {{MOSDEF Survey}}},
  author = {Reddy, Naveen A. and Kriek, Mariska and Shapley, Alice E. and Freeman, William R. and Siana, Brian and Coil, Alison L. and Mobasher, Bahram and Price, Sedona H. and Sanders, Ryan L. and Shivaei, Irene},
  year = 2015,
  month = jun,
  journal = {ApJ},
  volume = {806},
  pages = {259},
  publisher = {IOP},
  issn = {0004-637X},
  doi = {10.1088/0004-637X/806/2/259},
  urldate = {2026-01-05}
}

@article{reddy2018,
  title = {The {{HDUV Survey}}: {{A Revised Assessment}} of the {{Relationship}} between {{UV Slope}} and {{Dust Attenuation}} for {{High-redshift Galaxies}}},
  shorttitle = {The {{HDUV Survey}}},
  author = {Reddy, Naveen A. and Oesch, Pascal A. and Bouwens, Rychard J. and Montes, Mireia and Illingworth, Garth D. and Steidel, Charles C. and {van Dokkum}, Pieter G. and Atek, Hakim and Carollo, Marcella C. and Cibinel, Anna and Holden, Brad and Labb{\'e}, Ivo and Magee, Dan and Morselli, Laura and Nelson, Erica J. and Wilkins, Steve},
  year = {2018},
  month = jan,
  journal = {\apj},
  volume = {853},
  pages = {56},
  publisher = {IOP},
  issn = {0004-637X},
  doi = {10.3847/1538-4357/aaa3e7},
  urldate = {2025-08-08}
}

@article{reddy2020,
  title = {The {{MOSDEF Survey}}: {{The First Direct Measurements}} of the {{Nebular Dust Attenuation Curve}} at {{High Redshift}}*},
  shorttitle = {The {{MOSDEF Survey}}},
  author = {Reddy, Naveen A. and Shapley, Alice E. and Kriek, Mariska and Steidel, Charles C. and Shivaei, Irene and Sanders, Ryan L. and Mobasher, Bahram and Coil, Alison L. and Siana, Brian and Freeman, William R. and Azadi, Mojegan and Fetherolf, Tara and Leung, Gene and Price, Sedona H. and Zick, Tom},
  year = 2020,
  month = oct,
  journal = {ApJ},
  volume = {902},
  number = {2},
  pages = {123},
  issn = {0004-637X, 1538-4357},
  doi = {10.3847/1538-4357/abb674},
  urldate = {2026-01-05},
  langid = {english}
}

@article{reddy2022,
  title = {The {{Effects}} of {{Stellar Population}} and {{Gas Covering Fraction}} on the {{Emergent Ly$\alpha$ Emission}} of {{High-redshift Galaxies}}},
  author = {Reddy, Naveen A. and Topping, Michael W. and Shapley, Alice E. and Steidel, Charles C. and Sanders, Ryan L. and Du, Xinnan and Coil, Alison L. and Mobasher, Bahram and Price, Sedona H. and Shivaei, Irene},
  year = {2022},
  month = feb,
  journal = {\apj},
  volume = {926},
  pages = {31},
  issn = {0004-637X},
  doi = {10.3847/1538-4357/ac3b4c},
  urldate = {2024-01-05}
}

@misc{reddy2023,
  title = {The {{Impact}} of {{Star-Formation-Rate Surface Density}} on the {{Electron Density}} and {{Ionization Parameter}} of {{High-Redshift Galaxies}}},
  author = {Reddy, Naveen A. and Sanders, Ryan L. and Shapley, Alice E. and Topping, Michael W. and Kriek, Mariska and Coil, Alison L. and Mobasher, Bahram and Siana, Brian and Rezaee, Saeed},
  year = {2023},
  month = feb,
  journal = {arXiv e-prints},
  doi = {10.48550/arXiv.2302.10213},
  urldate = {2023-03-18}
}

@article{reddy2023a,
  title = {Paschen-Line {{Constraints}} on {{Dust Attenuation}} and {{Star Formation}} at z   1-3 with {{JWST}}/{{NIRSpec}}},
  author = {Reddy, Naveen A. and Topping, Michael W. and Sanders, Ryan L. and Shapley, Alice E. and Brammer, Gabriel},
  year = {2023},
  month = may,
  journal = {\apj},
  volume = {948},
  pages = {83},
  issn = {0004-637X},
  doi = {10.3847/1538-4357/acc869},
  urldate = {2023-07-25}
}

@misc{reddy2025,
  title = {The {{JWST}}/{{AURORA Survey}}: {{Multiple Balmer}} and {{Paschen Emission Lines}} for {{Individual Star-forming Galaxies}} at Z=1.5-4.4. {{I}}. {{A Diversity}} of {{Nebular Attenuation Curves}} and {{Evidence}} for {{Non-Unity Dust Covering Fractions}}},
  shorttitle = {The {{JWST}}/{{AURORA Survey}}},
  author = {Reddy, Naveen A. and Shapley, Alice E. and Sanders, Ryan L. and Topping, Michael W. and Ellis, Richard S. and Pettini, Max and Brammer, Gabriel and Cullen, Fergus and Forster Schreiber, Natascha M. and Khostovan, Ali A. and McLeod, Derek J. and McLure, Ross J. and Narayanan, Desika and Oesch, Pascal A. and Pahl, Anthony J. and Steidel, Charles C. and Berg, Danielle A.},
  year = {2025},
  month = jun,
  publisher = {arXiv},
  doi = {10.48550/arXiv.2506.17396},
  urldate = {2025-08-08}
}

@article{remy-ruyer2015,
  title = {Linking Dust Emission to Fundamental Properties in Galaxies: The Low-Metallicity Picture},
  shorttitle = {Linking Dust Emission to Fundamental Properties in Galaxies},
  author = {{R{\'e}my-Ruyer}, A. and Madden, S. C. and Galliano, F. and Lebouteiller, V. and Baes, M. and Bendo, G. J. and Boselli, A. and Ciesla, L. and Cormier, D. and Cooray, A. and Cortese, L. and De Looze, I. and {Doublier-Pritchard}, V. and Galametz, M. and Jones, A. P. and Karczewski, O. {\L}. and Lu, N. and Spinoglio, L.},
  year = {2015},
  month = oct,
  journal = {A\&A},
  volume = {582},
  pages = {A121},
  issn = {0004-6361},
  doi = {10.1051/0004-6361/201526067},
  urldate = {2025-08-26}
}

@article{rieke2004,
  title = {The {{Multiband Imaging Photometer}} for {{Spitzer}} ({{MIPS}})},
  author = {Rieke, G. H. and Young, E. T. and Engelbracht, C. W. and Kelly, D. M. and Low, F. J. and Haller, E. E. and Beeman, J. W. and Gordon, K. D. and Stansberry, J. A. and Misselt, K. A. and Cadien, J. and Morrison, J. E. and Rivlis, G. and Latter, W. B. and {Noriega-Crespo}, A. and Padgett, D. L. and Stapelfeldt, K. R. and Hines, D. C. and Egami, E. and Muzerolle, J. and {Alonso-Herrero}, A. and Blaylock, M. and Dole, H. and Hinz, J. L. and Le Floc'h, E. and Papovich, C. and {P{\'e}rez-Gonz{\'a}lez}, P. G. and Smith, P. S. and Su, K. Y. L. and Bennett, L. and Frayer, D. T. and Henderson, D. and Lu, N. and Masci, F. and Pesenson, M. and Rebull, L. and Rho, J. and Keene, J. and Stolovy, S. and Wachter, S. and Wheaton, W. and Werner, M. W. and Richards, P. L.},
  year = {2004},
  month = sep,
  journal = {\apj Suppl. Ser.},
  volume = {154},
  pages = {25--29},
  issn = {0067-0049},
  doi = {10.1086/422717},
  urldate = {2022-08-11}
}

@article{rieke2009,
  title = {Determining {{Star Formation Rates}} for {{Infrared Galaxies}}},
  author = {Rieke, G. H. and {Alonso-Herrero}, A. and Weiner, B. J. and {P{\'e}rez-Gonz{\'a}lez}, P. G. and Blaylock, M. and Donley, J. L. and Marcillac, D.},
  year = {2009},
  month = feb,
  journal = {\apj},
  volume = {692},
  pages = {556--573},
  issn = {0004-637X},
  doi = {10.1088/0004-637X/692/1/556},
  urldate = {2023-07-31}
}

@article{rieke2015,
  title = {The {{Mid-Infrared Instrument}} for the {{James Webb Space Telescope}}, {{I}}: {{Introduction}}},
  shorttitle = {The {{Mid-Infrared Instrument}} for the {{James Webb Space Telescope}}, {{I}}},
  author = {Rieke, G. H. and Wright, G. S. and B{\"o}ker, T. and Bouwman, J. and Colina, L. and Glasse, Alistair and Gordon, K. D. and Greene, T. P. and G{\"u}del, Manuel and Henning, {\relax Th}. and Justtanont, K. and Lagage, P. -O. and Meixner, M. E. and {N{\o}rgaard-Nielsen}, H. -U. and Ray, T. P. and Ressler, M. E. and {van Dishoeck}, E. F. and Waelkens, C.},
  year = {2015},
  month = jul,
  journal = {PASP},
  volume = {127},
  pages = {584},
  publisher = {IOP},
  issn = {0004-6280},
  doi = {10.1086/682252},
  urldate = {2024-05-01}
}

@misc{rieke2023,
  title = {{{JADES Initial Data Release}} for the {{Hubble Ultra Deep Field}}: {{Revealing}} the {{Faint Infrared Sky}} with {{Deep JWST NIRCam Imaging}}},
  shorttitle = {{{JADES Initial Data Release}} for the {{Hubble Ultra Deep Field}}},
  author = {Rieke, Marcia and {the JADES Collaboration}},
  year = {2023},
  month = jun,
  journal = {arXiv e-prints},
  doi = {10.48550/arXiv.2306.02466},
  urldate = {2023-07-19}
}

@misc{rieke2024,
  title = {The {{SMILES Mid-Infrared Survey}}},
  author = {Rieke, George and Alberts, Stacey and Shivaei, Irene and Lyu, Jianwei and Willmer, Christopher N. A. and {Perez-Gonzalez}, Pablo and Williams, Christina C.},
  year = {2024},
  month = jun,
  journal = {arXiv e-prints},
  doi = {10.48550/arXiv.2406.03518},
  urldate = {2024-08-06}
}

@article{rieke2025,
  title = {Low {{Accretion Rates}} in {{Black Holes}} in {{Late-stage Merger Ultraluminous Infrared Galaxies}}},
  author = {Rieke, George H. and Buiten, Victorine A. and Goldberg, Charles E. and Morrison, Jane and {van der Werf}, Paul and {Alonso-Herrero}, Almudena and Alberts, Stacey and Bonaventura, Nina and Ji, Zhiyuan and Lyu, Jianwei and Rinaldi, P. and Stone, Meredith A. and Sun, Yang and Zhu, Yongda},
  year = 2025,
  month = jul,
  journal = {ApJ},
  volume = {988},
  pages = {17},
  publisher = {IOP},
  issn = {0004-637X},
  doi = {10.3847/1538-4357/add2fd},
  urldate = {2026-03-20}
}

@misc{ronayne2023,
  title = {{{CEERS}}: 7.7 \$\{{\textbackslash}mu\}\$m {{PAH Star Formation Rate Calibration}} with {{JWST MIRI}}},
  shorttitle = {{{CEERS}}},
  author = {Ronayne, Kaila and Papovich, Casey and Yang, Guang and Shen, Lu and Dickinson, Mark and Kennicutt, Robert and Alavi, Anahita and Haro, Pablo Arrabal and Bagley, Micaela and Burgarella, Denis and Bail, Aur{\'e}lien Le and Bell, Eric and Cleri, Nikko and Cole, Justin and Costantin, Luca and {de la Vega}, Alexander and Daddi, Emanuele and Elbaz, David and Finkelstein, Steven and Grogin, Norman and Holwerda, Benne and Kartaltepe, Jeyhan and Kirkpatrick, Allison and Koekemoer, Anton and Lucas, Ray and Magnelli, Benjamin and Mobasher, Bahram and {Perez-Gonzalez}, Pablo and Prichard, Laura and Rafelski, Marc and Rodighiero, Giulia and Sunnquist, Ben and Teplitz, Harry and Wang, Xin and Windhorst, Rogier and Yung, L. Y. Aaron},
  year = {2023},
  month = oct,
  number = {arXiv:2310.07766},
  eprint = {2310.07766},
  primaryclass = {astro-ph},
  publisher = {arXiv},
  doi = {10.48550/arXiv.2310.07766},
  urldate = {2023-11-11},
  archiveprefix = {arXiv}
}

@article{rujopakarn2011,
  title = {Morphology and {{Size Differences Between Local}} and {{High-redshift Luminous Infrared Galaxies}}},
  author = {Rujopakarn, Wiphu and Rieke, George H. and Eisenstein, Daniel J. and Juneau, St{\'e}phanie},
  year = {2011},
  month = jan,
  journal = {\apj},
  volume = {726},
  pages = {93},
  publisher = {IOP},
  issn = {0004-637X},
  doi = {10.1088/0004-637X/726/2/93},
  urldate = {2024-08-06}
}

@article{rujopakarn2013,
  title = {Mid-Infrared {{Determination}} of {{Total Infrared Luminosity}} and {{Star Formation Rates}} of {{Local}} and {{High-redshift Galaxies}}},
  author = {Rujopakarn, W. and Rieke, G. H. and Weiner, B. J. and {P{\'e}rez-Gonz{\'a}lez}, P. and Rex, M. and Walth, G. L. and Kartaltepe, J. S.},
  year = {2013},
  month = apr,
  journal = {\apj},
  volume = {767},
  pages = {73},
  issn = {0004-637X},
  doi = {10.1088/0004-637X/767/1/73},
  urldate = {2022-03-25}
}

@article{sajina2007,
  title = {Spitzer {{Mid-Infrared Spectroscopy}} of {{Infrared Luminous Galaxies}} at Z{\textasciitilde}2. {{II}}. {{Diagnostics}}},
  author = {Sajina, Anna and Yan, Lin and Armus, Lee and Choi, Philip and Fadda, Dario and Helou, George and Spoon, Henrik},
  year = {2007},
  month = aug,
  journal = {\apj},
  volume = {664},
  pages = {713--737},
  issn = {0004-637X},
  doi = {10.1086/519446},
  urldate = {2025-08-26}
}

@article{sanders2003,
  title = {The {{IRAS Revised Bright Galaxy Sample}}},
  author = {Sanders, D. B. and Mazzarella, J. M. and Kim, D. -C. and Surace, J. A. and Soifer, B. T.},
  year = {2003},
  month = oct,
  journal = {Astron. J.},
  volume = {126},
  pages = {1607--1664},
  publisher = {IOP},
  issn = {0004-6256},
  doi = {10.1086/376841},
  urldate = {2024-05-21}
}

@article{sanders2017,
  title = {Biases in {{Metallicity Measurements}} from {{Global Galaxy Spectra}}: {{The Effects}} of {{Flux Weighting}} and {{Diffuse Ionized Gas Contamination}}},
  shorttitle = {Biases in {{Metallicity Measurements}} from {{Global Galaxy Spectra}}},
  author = {Sanders, Ryan L. and Shapley, Alice E. and Zhang, Kai and Yan, Renbin},
  year = {2017},
  month = dec,
  journal = {\apj},
  volume = {850},
  pages = {136},
  issn = {0004-637X},
  doi = {10.3847/1538-4357/aa93e4},
  urldate = {2025-08-21}
}

@article{sanders2021,
  title = {The {{MOSDEF Survey}}: {{The Evolution}} of the {{Mass-Metallicity Relation}} from z = 0 to z 3.3},
  shorttitle = {The {{MOSDEF Survey}}},
  author = {Sanders, Ryan L. and Shapley, Alice E. and Jones, Tucker and Reddy, Naveen A. and Kriek, Mariska and Siana, Brian and Coil, Alison L. and Mobasher, Bahram and Shivaei, Irene and Dav{\'e}, Romeel and Azadi, Mojegan and Price, Sedona H. and Leung, Gene and Freeman, William R. and Fetherolf, Tara and {de Groot}, Laura and Zick, Tom and Barro, Guillermo},
  year = {2021},
  month = jun,
  journal = {\apj},
  volume = {914},
  pages = {19},
  issn = {0004-637X},
  doi = {10.3847/1538-4357/abf4c1},
  urldate = {2023-07-26}
}

@article{shapley2022,
  title = {The {{MOSFIRE Deep Evolution Field Survey}}: {{Implications}} of the {{Lack}} of {{Evolution}} in the {{Dust Attenuation-Mass Relation}} to z   2},
  shorttitle = {The {{MOSFIRE Deep Evolution Field Survey}}},
  author = {Shapley, Alice E. and Sanders, Ryan L. and Salim, Samir and Reddy, Naveen A. and Kriek, Mariska and Mobasher, Bahram and Coil, Alison L. and Siana, Brian and Price, Sedona H. and Shivaei, Irene and Dunlop, James S. and McLure, Ross J. and Cullen, Fergus},
  year = {2022},
  month = feb,
  journal = {\apj},
  volume = {926},
  pages = {145},
  issn = {0004-637X},
  doi = {10.3847/1538-4357/ac4742},
  urldate = {2025-08-26}
}
\bibliographystyle{aasjournal}

%% This command is needed to show the entire author+affiliation list when
%% the collaboration and author truncation commands are used.  It has to
%% go at the end of the manuscript.
%\allauthors

%% Include this line if you are using the \added, \replaced, \deleted
%% commands to see a summary list of all changes at the end of the article.
%\listofchanges

\end{document}